\documentstyle[prd,aps,preprint]{revtex} 

\def\greaterthansquiggle{\raise.3ex\hbox{$>$\kern-.75em\lower1ex\hbox{$\sim$}}}
\def\lessthansquiggle{\raise.3ex\hbox{$<$\kern-.75em\lower1ex\hbox{$\sim$}}}
\newcommand{\bdi}{\begin{displaymath}}
\newcommand{\edi}{\end{displaymath}}
\newcommand{\bfi}{\begin{figure}}
\newcommand{\efi}{\end{figure}}

\newcommand{\beq}{\begin{equation}}
\newcommand{\eeq}{\end{equation}}

\newcommand{\gam}{\gamma_{\mu}}
\newcommand{\gan}{\gamma_{\nu}}

\newcommand{\gaN}{\gamma^{\nu}}
\newcommand{\gaf}{\gamma_{5}}
\newcommand{\emn}{\epsilon_{\mu\nu}}

\newcommand{\eML}{\epsilon^{\mu\lambda}}
\newcommand{\tr}{\mbox{tr}}

\newcommand{\beqa}{\begin{eqnarray}}
\newcommand{\eeqa}{\end{eqnarray}}
\newcommand{\no}{\nonumber}

\newcommand{\ra}{\rightarrow}

\newcommand{\wt}{\widetilde}
\newcommand{\wh}{\widehat}

\def\au{{\setbox0=\hbox{\lower1.36775ex%
\hbox{''}\kern-.05em}\dp0=.36775ex\hskip0pt\box0}}
\def\ao{{}\kern-.10em\hbox{``}}

\newcommand{\dsla}{\partial\hspace{-6pt} /  }  
\newcommand{\Asla}{A\hspace{-6.5pt}  /  } 
\newcommand{\Dsla}{D\hspace{-7.3pt}  /  }

\newcommand{\DDsla}{D\hspace{-5.83pt}  /  }
\newcommand{\ddsla}{\partial\hspace{-4.6pt} /  }

\newcommand{\AAsla}{A\hspace{-5pt}  /  }

\begin{document}

\setcounter{page}{0}
\def\footnoterule{\kern-3pt \hrule width\hsize \kern3pt}
\tighten
\title{MASSIVE SCHWINGER MODEL WITHIN MASS PERTURBATION THEORY\thanks
{This work is supported by a Schr\"odinger Stipendium of the Austrian FWF and
in part by funds provided by the U.S.
Department of Energy (D.O.E.) under cooperative 
research agreement \#DF-FC02-94ER40818.}}

\author{Christoph Adam\footnote{Email address: {\tt adam@ctp.mit.edu, 
adam@pap.univie.ac.at}}}

\address{Center for Theoretical Physics \\
Laboratory for Nuclear Science \\
and Department of Physics \\
Massachusetts Institute of Technology \\
Cambridge, Massachusetts 02139 \\
and \\
Inst. f. theoret. Physik d. Uni Wien \\
Boltzmanngasse 5, 1090 Wien, Austria \\
{~}}

\date{MIT-CTP-2618,~ {~~~~~} March 1997}
\maketitle
\thispagestyle{empty}

\begin{abstract}

In this article we give a detailed discussion 
of the mass perturbation theory of the
massive Schwinger model. After discussing some general features  
and briefly reviewing the exact solution of the massless case, we 
compute the vacuum energy density of the massive model and some related 
quantities. We derive the Feynman rules of mass perturbation theory and 
discuss the exact $n$-point functions  with the help of the
Dyson-Schwinger equations. Further we identify the stable and unstable
bound states of the theory and compute some bound-state masses and decay
widths. Finally we discuss scattering processes, where the resonances
and particle production thresholds of the model are properly taken into
account by our methods.

\medskip

%PACS-Numbers:   11.10, 11.80, 11.90

\end{abstract}
\vspace*{\fill}
%\begin{center}
%Submitted to: {\it Research Journal}
%\end{center}


\tableofcontents

%\newpage

\section{Introduction}

The massless and massive Schwinger models are two-dimensional QED with one
massless or massive fermion. Both models have been subject to intensive study
in the last decades. First of all, the exact solubility of the massless 
Schwinger model was discovered by J. Schwinger more than 30 years ago
\cite{Sc1}. Later on, Lowenstein and Swieca \cite{LS1} constructed a complete 
operator solution of the massless model. By carefully investigating the role 
of large gauge transformations they found that these large gauge transformations
have the effect of changing the vacuum ("instanton vacuum"), and therefore a
superposition of all these vacua ("$\theta$ vacuum") has to be introduced 
as a new, physical vacuum in order to render the theory sensible
(requirement of vacuum clustering).
In the usual terminology of gauge theories, the occurrence of instanton-like
gauge-field configurations can be traced back to the fact that the
boundary of Euclidean space-time is a $S^1$, and, further, that the
first homotopy group of the gauge group is nontrivial, $\Pi_1 (U(1))={\bf Z}$.

In addition, the chiral anomaly and Schwinger terms are present in these
models, and all these nontrivial field-theoretical features were one reason
for the rising interest in the Schwinger model \cite{DR1,Frish},
\cite{AAR} -- \cite{IP1}, \cite{ABH,Bai}.

However, there is another reason for the study of the Schwinger model,
namely its similarity to some aspects of QCD. All the above-mentioned
features are present in QCD, too. In addition, a fermion condensate is
formed both in QCD and in the massless and massive Schwinger models 
\cite{Jay,SW1,Adam,Sm1}, and
confinement is realized in the latter models in a quite understandable way.
In both models there are no fermions in the physical particle spectrum. 
The lowest physical state is a massive mesonic fermion-antifermion bound state,
the Schwinger boson. In the massless model this boson is free, and higher
states are trivial free $n$-particle states. In the massive model these
states turn into $n$-boson bound states, because the Schwinger boson is 
an interacting particle there  \cite{Co1,SMASS,BOUND,GBOUND}.

When confinement properties are tested with external charges, the two models
behave differently, too. In the massless model widely separated probe
charges are completely screened via vacuum polarization, and the potential
between the external charges approaches a constant.

For the massive model this screening takes place as long as the probe
charges $g$ are integer multiples of the fundamental fermion charge $e$.
When $g\ne ne$, the potential between the probe charges rises linearly
for large distances. So "screening" is realized in the massless model, whereas 
true confinement takes place in the massive model (this feature was
first discovered in \cite{CJS} within a bosonization approach, and is
further discussed e.g. in \cite{GKMS,AAR,HRS1}; see also Section 6).
Therefore, the Schwinger model was studied in order to get more insight into
the phenomena of quark trapping and confinement \cite{AAR,CKS,KS1}.

Instanton physics was studied, too, in the Schwinger model \cite{Sm3,Sm2}.

In addition, because of its simplicity, the Schwinger model is often used
for testing some new methods of QFT, like light-cone quantization
\cite{Vary} -- \cite{Els2}, 
semi-classical methods \cite{AAR,Gatt2,GKMS} or lattice
calculations \cite{Crew} -- \cite{Gatt3}.

As mentioned, the massless Schwinger model was first completely solved
within the operator formalism \cite{LS1}, and the operator approach and
the specific two-dimensional method of bosonization were mainly used in the
subsequent years \cite{CKS,KS1,CJS,Co1,Gro}. 
It took some time until a path integral approach to the
Schwinger model arose (mainly because of the problems with the
nontrivial vacuum structure) \cite{GS1}, \cite{NS1} -- \cite{RRS1}, 
\cite{Jay} -- \cite{DSW}, \cite{Adam} -- \cite{DSEQ}.

In this article all computations are based on the path integral formalism in 
flat Euclidean space-time.

This article is organized as follows: \\
After fixing our conventions we discuss some general features and the physical 
meaning of the vacuum angle $\theta$. Then we turn to the massless model,
because it will be the starting point for a mass perturbation theory. We
discuss the relevance of the instanton vacuum and briefly review the exact 
path integral solution of the massless model to the accuracy we need in the
sequel.

In a next step we compute the vacuum functional and vacuum energy
density of the massive model in
mass perturbation theory by a method that is analogous to the cluster
expansion of statistical physics. Further we prove the IR finiteness of
the mass perturbation theory and comment on the UV regularization. 
%(it may indeed be proven that the super-renormalizability of the model,
%that is obvious within conventional (electrical charge) perturbation
%theory, may be recovered in the framework of mass perturbation theory, see
%\cite{DIV}). 

With the help of the vacuum energy density 
we compute the fermion and photon condensates and the scalar and
topological susceptibilities  and generalize (to arbitrary order mass
perturbation theory)
some recent path integral computations \cite{GKMS} on screening 
and confinement in the massive Schwinger model.

Then we develop the Feynman rules for the mass perturbation theory, which
we need in the sequel. Actually, because of the chiral properties of the
model these Feynman rules acquire a matrix structure.

Further we compute the mass of the Schwinger boson by a direct application
of this mass perturbation theory.

From the equations of motion we derive, in Section 9, the Dyson-Schwinger 
equations of the model and use them to re-express the exact $n$-point
functions in a way that is more suitable for the subsequent discussion
of bound states, decays and scattering. From the exact two-point function
we will be able to infer the complete bound-state spectrum of the
model and find that it contains, in addition to the $n$-boson bound
states, some further bound states (Section 10).

Further, this exact two-point function provides information on the decay
widths of all the unstable bound states. As an illustration, we will compute 
the masses and decay widths of the lowest bound states.

At last, we discuss the scattering processes of the model, where we take 
properly into account all the resonances and higher production thresholds.
All this may be derived from a resummed mass perturbation theory, without
further approximations.

We will use the following conventions,
\beq
g_{\mu\nu}=\delta_{\mu\nu} \quad , \quad \epsilon_{01} =-i \quad ,\quad
\epsilon_{\mu\nu}\epsilon^{\nu\lambda}=g_\mu^\lambda
\eeq
\beq 
\gam\gaf =\emn \gaN \quad ,\quad {\rm tr }\gaf \gam\gan =-2\emn \quad ,\quad
\gaf =-i\gamma^0 \gamma^1
\eeq
\beq
\gamma^{0} =\left( \begin{array}{cc}0 & i \\ -i & 0 \end{array} \right)
\qquad ,\qquad
\gamma^{1} =\left( \begin{array}{cc}0 & 1 \\ 1 & 0 \end{array} \right)
\qquad ,\qquad
\gaf =\left( \begin{array}{cc}1 & 0 \\ 0 & -1 \end{array} \right)
\eeq
and we find it convenient to keep the notation $(x_0 ,x_1 )$ in Euclidean 
space-time. As a consequence, the dual field strength pseudoscalar
\beq
F=\frac{1}{2}\emn F^{\mu\nu} =\frac{1}{e}\Box\beta
\eeq
is imaginary ($\beta$ is defined in (6)). 
The Lagrangian of Euclidean QED$_2$ is
\beq
L=\bar \Psi (i\dsla -e\Asla +m)\Psi -\frac{1}{4}F_{\mu\nu}F^{\mu\nu}
\eeq
and we use the following representation for the gauge field
\beq
A_{\mu}(x)=\frac{1}{e}(\partial_{\mu}\alpha (x)+\epsilon_{\mu\nu}\partial^{\nu}
\beta (x)).
\eeq
The pure gauge part, $\alpha$, is unimportant for gauge invariant VEVs (as
we deal with throughout the article) and may therefore be set equal to
zero, $\alpha \equiv 0$ (Lorentz condition). Using the electric charge
$e$ (which has the dimension of mass in two space-time dimensions)
we define the Schwinger mass
\beq
\mu^2_0 =\frac{e^2}{\pi}
\eeq
which is the mass of the Schwinger boson in the massless Schwinger model.
Further we need the currents
\bdi
J_\mu =\bar\Psi\gam\Psi \quad  ,\quad  J^5_\mu =\bar\Psi\gam\gaf\Psi
=\emn J^\nu 
\edi
\bdi
S=\bar\Psi \Psi \quad ,\quad  P=\bar\Psi \gaf \Psi 
\edi
\beq
S_\pm =\bar\Psi P_\pm \Psi \quad ,\quad  P_\pm =\frac{1}{2}(1\pm \gaf )
\eeq
and define the Schwinger boson field, $\Phi$, by
\beq
J_\mu =: \frac{1}{\sqrt{\pi}}\emn \partial^\nu \Phi .
\eeq
On the quantized theory the following equations of motion hold (more
precisely, they hold on the physical subspace, see the remark after (103)),
\beq
\partial_\mu F^{\mu\nu} =eJ^\nu
\eeq
\beq
\partial^\mu J^5_\mu =\frac{e}{\pi}F-2imP,
\eeq
which are the Maxwell and anomaly equations. In addition, the Dirac equation
holds, but we do not need it.

We use the following Feynman propagators: \\ \\
massless scalar propagator
\beq
D_0 (x)=\frac{1}{4\pi}\ln (x^2 -i\epsilon )\quad ,\quad
\Box D_0 (x)=\delta (x)
\eeq
massless fermion propagator
\beq
G_0 (x)=\frac{x^{\mu}\gam}{2\pi (x^2 -i\epsilon )}=\dsla D_0 (x),
\eeq
massive scalar propagator
\beq
D_\mu (x)=-\frac{1}{2\pi}K_0 (\mu|x|) \quad ,\quad \wt D_\mu (p)=
\frac{-1}{p^2 +\mu^2}
\eeq
and $K_0$ is the McDonald function with the properties
\bdi
K_0 (z)\to -\gamma -\ln \frac{z}{2}
+o(z^2) \quad {\rm for}\quad  z\to 0
\edi
\beq
K_0 (z)\to \sqrt{\frac{\pi}{2z}}e^{-z} \quad {\rm for} \quad z\to \infty 
\eeq
where $\gamma$ is the Euler constant.

In the sequel we will have to distinguish between two types of VEVs, namely
VEVs with respect to the massless and massive Schwinger models. The former VEVs
are always denoted by $\langle\rangle_0$, the latter ones sometimes by
$\langle\rangle_m$ and sometimes without a subscript, $\langle\rangle$.

Here we should add a last comment on our Euclidean conventions. They are 
chosen in such a way that they are as similar as possible to their 1+1
Minkowski-space counterparts. However, for this advantage we have to pay
the prize that in the intermediate Euclidean computations some
quantities are unphysical (e.g. imaginary field strength $F$  and vacuum angle 
$\theta$).

\section{The $\theta$ vacuum}

Let us briefly recall the way the vacuum angle $\theta$ enters the theory.
The key observation is the existence of large gauge transformations
$G_i$ that do not leave the vacuum invariant and therefore give rise
to an infinite number of vacua,
\beq
G_1 |0\rangle =:|1\rangle \quad ,\quad G_n |0\rangle =(G_1 )^n |0\rangle
=:|n\rangle
\eeq
and $n$ is restricted to integer values by the requirement that gauge 
transformations must act uniquely on the matter fields (here the fermion).
As a consequence, gauge fields may exist that tend to different pure gauges
$G_{n_\pm}$ for the time 
$t\to \pm \infty$. They have instanton number $k=n_+ -n_-$,
and this instanton number may be computed from the Pontryagin index density
$\nu (x)$,
\beq
\nu =\int dx\, \nu (x) =-\frac{ie}{2\pi}\int dxF(x) =k.
\eeq
These instantons induce transitions between different, gauge equivalent
vacua (16). Therefore, a superposition of these vacua has to be introduced 
as a new, physical vacuum,
\beq
|\theta\rangle =\sum_{n=-\infty}^{\infty} e^{in\theta}|n\rangle \quad ,
\quad G_n |\theta\rangle =e^{-in\theta}|\theta\rangle
\eeq
\beq
\langle \theta' |\hat O |\theta \rangle =\delta (\theta -\theta' )
\sum_{k=-\infty}^\infty e^{ik\theta}\langle 0|\hat O |k\rangle
\eeq
and in (19) we used $\langle m|\hat O |n\rangle =\langle 0|\hat O 
|n-m\rangle $,
which holds for gauge invariant operators $\hat O $. Because all $\theta$
vacua are invariant with respect to large gauge transformations, 
up to a phase, (18),
the angle $\theta$ has to be included as a new physical parameter into
the theory.

From (19) we find that within the path integral approach the vacuum angle
$\theta$ may be included, e.g. for the vacuum functional of the massive
Schwinger model, like
\beqa
Z(m,\theta)&=&\sum_{k=-\infty}^{\infty}e^{ik\theta}\int DA_k^\mu D\bar\Psi
D\Psi e^{\int dx[\bar\Psi (i\ddsla -e\AAsla +m)\Psi -\frac{1}{4}F_{\mu\nu}
F^{\mu\nu}]} \\
&=&\sum_{k=-\infty}^{\infty}\int DA_k^\mu D\bar\Psi
D\Psi e^{\int dx[\bar\Psi (i\ddsla -e\AAsla +m)\Psi +\frac{1}{2}F^2 +
\frac{e\theta}{2\pi}F]}
\eeqa
(where we used (17)) and $A_k^\mu$ has instanton number $k$. Due to the anomaly
there is a third representation for $Z(m,\theta)$. By performing a constant
chiral transformation on the fermions,
\beq
\Psi \ra e^{i\beta\gaf}\Psi \quad ,\quad \bar\Psi \ra \bar\Psi e^{i\beta\gaf},
\eeq	
the change of the fermionic path integral causes the anomaly,
\beq
{\bf A} =-\frac{e}{\pi} \beta\int dxF,
\eeq
and in the action only the mass term is changed by this transformation.
Choosing $\beta =-\frac{\theta}{2}$, the $\theta F$ term is cancelled and we
find
\beq
Z(m,\theta)=\sum_{k=-\infty}^{\infty}\int DA_k^\mu D\bar\Psi
D\Psi e^{\int dx[\bar\Psi (i\ddsla -e\AAsla )\Psi +m\cos\theta\bar \Psi
\Psi +im\sin\theta\bar\Psi \gaf \Psi +\frac{1}{2}F^2 ]}
\eeq

Rewriting the $\theta$-dependent part of (24) like
\beq
m(\cos\theta S+i\sin\theta P)=m(e^{i\theta}S_+ +e^{-i\theta}S_- )
\eeq
we conclude that general VEVs will not be holomorphic in $me^{i\theta}$ but
depend on its complex conjugate, too, $\langle\rangle_m (me^{i\theta},
me^{-i\theta})$.

There is a nice physical interpretation of $\theta$ that was extensively
discussed in \cite{Co1}, which we want to present now. Ignoring the
fermions at the moment and treating $F$ as the fundamental field in (21),
we find the equation of motion
\beq
F=-\frac{e\theta}{2\pi}.
\eeq
So $\theta$ may be interpreted as a constant background electric field.

[{\em Remark}: There is a simple and very instructive way of discussing
this feature in the Schr\"odinger picture, which we want to describe
briefly (this was shown to us by R. Jackiw). The Hamiltonian of the
theory without fermion reads, in the gauge $A_0 =0$ (remember that $A_0$
has no conjugate momentum; $F=\partial_0 A_1$; we temporarily introduce
a finite length $L$ for the space direction $x_1$)
\bdi
H=\frac{1}{2}\int_0^L dx_1 F^2
\edi
The quantum theory is described in the Schr\"odinger picture by operators
$A_1 (x)$ and $\wh F(x)\equiv i\frac{\delta}{\delta A_1 (x)}$ acting on 
wave functionals $\Psi [A_1]$. The Gauss law for physical states simply
reads
\bdi
\partial_1 \wh F(x) \Psi [A_1]\equiv \partial_1 i\frac{\delta}{\delta
A_1 (x)}\Psi [A_1]=0
\edi
with the general solution
\bdi
\Psi [A_1]=f(a) \quad ,\quad a:=\int_0^L dx_1 A_1 (x)
\edi
where $f$ is an arbitrary function. So there remains only one degree of
freedom (the zero Fourier component of $A_1$). The Schr\"odinger equation
reads ($E$ \ldots energy)
\bdi
\frac{1}{2}\int_0^L dx_1 \wh F^2 (x) f(a)\equiv - \frac{1}{2}\int_0^L dx_1
\frac{\delta^2}{\delta A_1^2 (x)}f(a) =Ef(a)
\edi
with the solution
\bdi
f(a)=e^{-iFa}\quad ,\quad E=\int_0^L dx_1 \frac{F^2}{2}=L\frac{F^2}{2},
\edi
where $F$ is the eigenvalue of $\wh F$,
\bdi
\wh F f(a) =\langle \wh F\rangle f(a) =: Ff(a)
\edi
So $F$ is just a constant field with energy density $\frac{F^2}{2}$.
Finally, when we perform a large gauge transformation $eA_1 \ra eA_1
-\partial_1 \lambda$, $\lambda (L)=\lambda (0) +2\pi n$, $\Psi [A_1]$
changes according to
\bdi
\Psi [A_1 -\frac{1}{e}\partial_1 \lambda]=\Psi [A_1]e^{\frac{2\pi in}{e}F}
\edi
Therefore, the state vector $\Psi [A_1]$ changes under large gauge 
transformations precisely like the $\theta$ vacuum (18), provided we 
make the identification $F=-\frac{e\theta}{2\pi}$, equ. (26), which
again shows that $\theta$ is a constant background field.]
 
When matter is included, it was shown in \cite{Co1} that whenever $\theta >
\pi$ ($\theta <-\pi$), it is energetically favourable to create a real
particle-antiparticle pair (with fundamental charge $\pm e$) that partially
screens the background field (26). This explains the angular character of 
$\theta$. 

Quantum effects further change the picture. For the massless Schwinger model
($m=0$) it is obvious from (24) that the background field is 
completely screened.
The anomaly (23) stems from the vacuum polarization graph in QED$_2$
(for a lengthy discussion and explicit computation of the QED$_2$-anomaly
see e.g. \cite{ABH,Diss}),
therefore the screening is due to vacuum polarization. In the massive case this
screening is not complete and we will find (see Section 6)
\beq
\langle F\rangle_m \sim -m\sin\theta +o(m^2).
\eeq

\section{Instanton vacuum and zero modes}

The following two sections are devoted to a short review of some properties
of the massless Schwinger model which we need in the sequel.
First we remember that the exact fermion propagator in an external gauge
field may be given explicitly,
\beq
G^\beta (x,y)=e^{i\gaf (\beta (x)-\beta (y))}G_0 (x-y)
\eeq
(more precisely, (28) is correct for gauge fields with vanishing instanton
number, $k=0$. For $k\ne 0$ $G^\beta$ acquires an additional term. This term, 
however, does not contribute to the path integral computations which we want 
to perform and may therefore be omitted, see the remark after (32)).

Further we should recapitulate some wellknown properties of the Dirac operator
in an instanton field. For sufficiently simple space-time manifolds the Dirac
operator
\beq
\Dsla =\dsla +ie\Asla
\eeq
in a gauge field with instanton number $k$ has precisely $|k|$ zero modes,
which have positive chirality for $k>0$ and negative chirality otherwise.
For the Schwinger model these zero modes may be computed explicitly
(see \cite{Adam,Diss}),
\bdi
\Psi_{i_0}^{\beta_k}=\frac{1}{\sqrt{2\pi}}(x^{-})^{i_0}e^{i\beta_k}
{1 \choose 0}\quad ,\quad
k>0 \quad ,\quad i_0 =0\ldots k-1 ,
\edi
\beq
\Psi_{i_0}^{\beta_k}=\frac{1}{\sqrt{2\pi}}(x^{+})^{i_0}e^{-i\beta_k}
{0 \choose 1}\quad ,\quad
k<0 \quad ,\quad i_0 =0\ldots \vert k\vert -1 ,
\eeq
\bdi
x^{+}=x_1 +ix_0 \quad ,\quad x^{-}=x_1 -ix_0 .
\edi
Next let us investigate the pure fermionic part of the path integral in an 
external gauge field (for a proper treatment of the zero modes we keep a small
fermion mass $m$ that we set equal to zero at the end of the computation).

Because of the identity
\beqa
Z[A_\mu^k]&=&\lim_{m\to 0}\int D\bar \Psi D\Psi e^{\int dx\bar\Psi (i\DDsla +m)
\Psi} \no \\
&=&\lim_{m\to 0}\det (i\Dsla +m)=\lim_{m\to 0} m^{|k|}{\det}' (i\Dsla +m)
\eeqa
the vacuum functional obviously vanishes for $k\ne 0$ (the prime indicates
omission of the zero eigenvalues).

This remains true for VEVs of operators containing only gauge fields, because 
they do not influence the fermionic integration. On the other hand, fermionic
VEVs may get contributions from nontrivial instanton sectors.

By the use of Grassmann integration rules a general fermionic VEV may be 
written like
\bdi
\langle \Psi^{\alpha_1}(x_1)\bar\Psi^{\beta_1}(y_1)\ldots \Psi^{\alpha_n}(x_n)
\bar\Psi^{\beta_n}(y_n)\rangle_0 [A_\mu^k] =
\edi
\beq
\lim_{m\to 0}\sum_{\pi (1\ldots n)}(-)^{\sigma (\pi)}\prod_{i=1}^n
\sum_{l_i}\frac{\psi^{\alpha_i}_{l_i}(x_i)\bar\psi^{\beta_{\pi (i)}}_{l_i}
(y_{\pi (i)})}{i\lambda_{l_i} +m}\, m^{|k|}{\prod_{l}}'(i\lambda_l +m).
\eeq
Here $\psi_l (x)$ and $\lambda_l$ denote eigenfunctions and eigenvalues
of the Dirac operator, and the sum just means summation of all possible Wick 
contractions (for a more detailed discussion see e.g. \cite{Jay,Adam,Diss}).

This expression may give a nonvanishing result when in precisely $|k|$ 
Green functions only zero modes contribute, because then the $m^{|k|}$
factor is cancelled by a factor $m^{-|k|}$. Of course, every zero mode has
to occur exactly once because of the Pauli principle (the Pauli principle
is not explicitly written down in (32); however, because of the permutation
sign factor $(-)^{\sigma (\pi )}$ there is a pair-wise cancellation of all
terms with identical eigenfunctions). Therefore, a VEV of $n$ fermion bilinears,
like (32), may get contributions from instanton sectors $k=-n,\ldots ,n$.

[{\em Remark}: for a nontrivial instanton sector $k\ne 0$, the remaining
Green functions in (32) (the exact fermion propagators (28)) in
principle should be constructed on the subspace that is orthogonal to the
zero modes, i.e. they should satisfy
\bdi
\Dsla_x^{\beta_k} G^{\beta_k} (x,y)={\bf 1}\delta (x-y) - \sum_{i_0 =0}^{|k|-1}
c_{i_0}\Psi_{i_0}^{\beta_k}(x) \bar\Psi_{i_0}^{\beta_k}(y)
\edi
and, consequently, read
\beqa
G^{\beta_k}(x,y)& =& e^{i\gaf \beta_k (x)}G_0 (x-y)e^{i\gaf \beta_k (y)}
-e^{i\gaf \beta_k (x)}\sum_{i_0 =0}^{|k|-1} c_{i_0}\chi_{i_0}^{\beta_k}(x)
\bar\Psi_{i_0}^{\beta_k} (y) \no \\
&=&  e^{i\gaf \beta_k (x)}G_0 (x-y)e^{i\gaf \beta_k (y)}
-\sum_{i_0 =0}^{|k|-1} c_{i_0} \Psi_{i_0}^{\beta_k}
\bar\chi_{i_0}^{\beta_k}(y)e^{i\gaf \beta_k (y)} \no
\eeqa
where
\bdi
\dsla_x \chi_{i_0}^{\beta_k}(x) =e^{i\gaf\beta_k (x)}\Psi_{i_0}^{\beta_k}(x)
\quad ,\quad \partial^\mu \bar\chi_{i_0}^{\beta_k}(y)\gam =
\bar\Psi_{i_0}^{\beta_k}(y)e^{i\gaf\beta_k (y)}
\edi
(the $c_{i_0}$ are related to the zero mode normalizations and are unimportant 
for our argument). The important point is, of course, that the additional
term for $G^{\beta_k}$ contains the zero modes and, therefore, does not 
contribute to the fermionic path integral (32) due to the Pauli principle.
As a consequence, in our computations we may use the simple form (28) of
the exact fermion propagator for all instanton sectors (this argument can
be found e.g. in \cite{Zah}).] 

For further conclusions we have to specify the types of fermionic bilinears. 
E.g. pure vector-like VEVs get contributions only from the trivial sector
$k=0$ for the following reason: the zero modes contribute like 
$\Psi_{i_0}\bar\Psi_{i_0}\sim P_+$
(for $k>0$), the exact propagators are given in (28), and therefore all 
vector-like VEVs look like (for $k>0$)
\bdi
\tr P_+ \gamma_{\mu_1}G^\beta_2 \gamma_{\mu_2}G^\beta_3 \gamma_{\mu_3}\ldots
=0
\edi
\bdi
\tr P_+ \gamma_{\mu_1} P_+ \gamma_{\mu_2} G^\beta_3 \gamma_{\mu_3}\ldots
=\tr P_+P_-\gamma_{\mu_1}\gamma_{\mu_2}\ldots =0\quad ,\quad {\rm etc.}
\edi
and therefore vanish.

On the other hand, VEVs of scalar or chiral bilinears do get contributions 
from the $k\ne 0$ sectors.

For densities, like $S(x)=\bar\Psi (x) \Psi (x)$ 
(which are the only objects we need 
in the sequel), this may be inferred in a  very easy fashion from the various 
representations of the vacuum functional, (20) - (24). Suppose for the moment
that the fermion mass is space-time dependent. Then the scalar VEVs of the 
massless model are given by (see (20)) 
\bdi
\langle \prod_{i=1}^n \bar\Psi (x_i)\Psi (x_i)\rangle_0 =\frac{1}{Z(0,\theta)}
\prod_{i=1}^n \frac{\delta}{\delta m(x_i)}Z[m(x),\theta]|_{m=0} =
\langle\prod_{i=1}^n (e^{i\theta}S_+ (x_i) +e^{-i\theta}S_- (x_i))\rangle_0^{
\theta=0}
\edi
\beq
= e^{in\theta}\langle \prod_{i=1}^n S_+ (x_i)\rangle_0^{\theta =0}  +
e^{i(n-2)\theta}\sum_{j=1}^n \langle S_- (x_j) \prod_{i=1 \atop i\ne j}^n
S_+ (x_i)\rangle_0^{\theta =0} +\cdots +
 e^{-in\theta}\langle \prod_{i=1}^n S_- (x_i)\rangle_0^{\theta =0} 
\eeq
where we used (24), (25) in the first line.

Further we know from (20) that a factor $e^{ik\theta}$ indicates that the 
term stems from the instanton sector $k$. Therefore, we may draw the conclusion
that for a $n$-scalar VEV the sectors $k=n,n-2,\ldots ,-n$ contribute.
In addition, we find that for a VEV of $n_+$ positive chirality densities
$S_+$ and $n_-$ negative chirality densities $S_-$ only the sector 
$k=n_+ -n_-$ contributes and that ``instanton number equals chirality''.

[{\em Remark}: we gave a quite explicit construction of the vacuum structure
of the massless Schwinger model, because we need it for our further 
calculations. However, if one is only interested in the vacuum structure itself,
it may be shown to be an almost trivial consequence of gauge invariance.
First, imposing the Lorentz gauge condition $\alpha =0$ in the representation
(6) of the gauge field does not uniquely fix the gauge. There remains a
residual gauge freedom to introduce functions $\beta$ that fulfill the
condition $\Box\beta =0$. Of course, a constant $\beta =c$ is a possible
choice. Requiring gauge invariance and using the anomaly result (23)
(for a constant chiral rotation $c$) we find for the vacuum functional
\bdi
\int D\bar\Psi D\Psi D\beta e^{S[\bar\Psi ,\Psi ,\beta]}\equiv
\int D\bar\Psi D\Psi D\beta e^{S[\bar\Psi e^{ic\gaf},e^{ic\gaf}\Psi ,\beta]}=
\int D\bar\Psi D\Psi D\beta e^{S[\bar\Psi ,\Psi ,\beta]-2ic\frac{-ie}{2\pi}
\int dxF} 
\edi
and conclude $\nu =\frac{-ie}{2\pi}\int dxF=0$. Therefore, for the vacuum
functional only the sector $k=0$ may contribute. The conclusion remains the
same for gauge field VEVs and for vector current VEVs. For scalar and chiral 
VEVs, however, things change. E.g. for the positive chiral VEV $\langle
S_+ (x)\rangle_0 $ we find
\bdi
\int D\bar\Psi D\Psi D\beta \bar\Psi (x)P_+\Psi (x)
e^{S[\bar\Psi ,\Psi ,\beta]}\equiv 
\int D\bar\Psi D\Psi D\beta \bar\Psi (x) e^{ic\gaf}P_+e^{ic\gaf}
\Psi (x)e^{S[\bar\Psi ,\Psi ,\beta] -2ic\nu}
\edi
and conclude ($\gaf P_+ =P_+\gaf =P_+$) that $\nu =1$. Therefore, here only
the sector $k=1$ may contribute. The generalization to higher VEVs is
straight forward, and we may indeed conclude that the vacuum structure of
the Schwinger model is a consequence of gauge invariance.
(For the massive model the same gauge invariance requirement leads to the
equivalence of the different representations (21), (24) for the 
vacuum functional $Z(m,\theta)$. A similar discussion can be found in
\cite{Fry}, and more about instantons and $\theta$ vacua e.g. in \cite{PQ1}.)]

\section{Solution of the massless model}

The key observation for the exact solution of the massless model is the fact
that the interaction term in the fermionic Lagrangian may be transformed away
by a chiral rotation ($A_\mu =\emn \partial^\nu \beta$),
\beq
L_\Psi =\bar\Psi (i\dsla -e\Asla )\Psi =\bar\Psi e^{-i\beta\gaf}i\dsla
e^{-i\beta\gaf}\Psi .
\eeq
In the fermionic path integral such a chiral rotation causes the chiral
anomaly. In the model at hand this anomaly may be computed for finite
chiral rotations, too, leading to the result for general VEVs (for $\theta =0$;
$N$ is the path integral normalization)
\bdi
N\int D\bar\Psi D\Psi D\beta \hat O(\Psi ,\bar\Psi ,\beta)e^S =
\edi
\beq
N\int D\bar\Psi_f D\Psi_f D\beta e^{\int dx\bar\Psi_f i\ddsla \Psi_f }
\hat O(e^{i\beta\gaf} \Psi_f ,\bar\Psi_f e^{i\beta\gaf},\beta)e^{S_{\rm eff}}
\eeq
where
\beq
S_{\rm eff}=\frac{1}{2e^2}\int dx[(\Box\beta)^2 -\frac{e^2}{\pi}\beta\Box
\beta ] =:\int dx\beta {\rm\bf D}\beta
\eeq
\beq
{\rm\bf D}=\frac{\Box}{e^2}(\Box -\mu_0^2 )\quad ,\quad {\rm\bf G}(x)=
\pi (D_{\mu_0}(x)-D_0 (x))\quad ,\quad {\rm\bf D}_x {\rm\bf G}(x-y)=
\delta (x-y).
\eeq
The first term in $S_{\rm eff}$ is the "photon" kinetic energy, the second
one stems from the anomaly; $\Psi_f$ is the free fermion spinor. 

For a further evaluation the presence of zero modes for $k\ne 0$ has to be 
taken into account. In $\hat O(\Psi ,\bar\Psi ,\beta)$ all Wick contractions
among the fermions $\Psi =e^{i\beta\gaf}\Psi_f$ have to be done and the
corresponding number of $k$ zero modes and remaining $n-k$ exact propagators
have to be inserted, according to our discussion in the last section.
All this may be written down in a compact way by introducing the
generating functional for fermions in the instanton sector $k$,
\beq
Z_k [\bar\eta ,\eta]=e^{ik\theta}N\int D\beta_k \prod_{i_0 =0}^{k-1}
(\bar\eta |\Psi_{i_0}^{\beta_k})(\bar\Psi_{i_0}^{\beta_k}|\eta)
 e^{-i\int dxdy\bar\eta (x)G^{\beta_k}(x,y)\eta (y)}e^{S_{\rm eff}},
\eeq
where $\eta ,\bar\eta $ are Grassmann-valued external spinor sources.

Both exact propagator (28) and zero modes (30) depend exponentially on
$\beta$, therefore they add linear $\beta$ terms to $S_{\rm eff}$,
rendering thereby the $\beta$ path integral Gaussian. As a consequence, 
all VEVs may be computed explicitly, as we now demonstrate briefly.

E.g. for the chiral density $\langle S_+ (x)\rangle_0$ only the $k=1$ sector 
contributes; we have to insert one zero mode (30) and find
\bdi
\langle S_+ (x)\rangle_0 =e^{i\theta}N\int D\beta \, \tr P_+ \frac{1}{2\pi}
e^{i(\beta (x)+\beta (x))}P_+ e^{\int dz\beta {\rm\bf D}\beta }=
\edi
\beq
\frac{e^{i\theta}}{2\pi}e^{2{\rm\bf G}(0)} = e^{i\theta}\frac{e^\gamma
\mu_0}{4\pi} =:e^{i\theta}\frac{\Sigma}{2}
\eeq
where we completed the square and performed the $\beta$ integration in the
first step; further we introduced the fermion condensate
\beq
\Sigma\equiv \langle\bar\Psi (x)\Psi (x)\rangle_0^{\theta =0} =\frac{e^\gamma
\mu_0}{2\pi}.
\eeq
Analogously we find for the two-point functions
\bdi
\langle S_+ (x)S_- (y)\rangle_0 =-N\int D\beta e^{\int dz\beta{\rm\bf D}\beta
}\cdot
\edi
\bdi
\cdot \tr e^{2i\gaf (\beta (x)-\beta (y))}P_+ G_0 (x-y)P_- G_0 (y-x) =
\edi
\beq 
\frac{e^{4({\rm\bf G}(0) -{\rm\bf G}(x-y))}}{4\pi^2 (x-y)^2}=
\bigl(\frac{e^\gamma \mu_0}{4\pi}\bigr)^2 e^{-4\pi D_{\mu_0}(x-y)}
\eeq
(here only one of the two possible Wick contractions contributed due to
$\tr P_+ G =0$) and
\beq
\langle S_+ (x)S_+ (y)\rangle_0 =\ldots =e^{2i\theta}\frac{(x-y)^2}{4\pi^2}
e^{4( {\rm\bf G}(0) +{\rm\bf G}(x-y))}=e^{2i\theta}\bigl( 
\frac{e^\gamma \mu_0}{4\pi}\bigr)^2 e^{4\pi D_{\mu_0}(x-y)}.
\eeq
(the details of all these computations can be found e.g. in \cite{Adam,Diss}).

Observe that in both cases the massless propagator part of ${\rm\bf G}(x-y)$
is cancelled by a contribution steming from the free fermion propagator or
from the zero modes, respectively. This feature remains true for all
physical VEVs and shows that the only physical particle in the Schwinger model
is the massive Schwinger boson.

Further it may be seen easily that all VEVs fulfill the vacuum clustering
property, e.g.
\beq
\lim_{x-y\to \infty} \langle S_\pm (x)S_\pm (y)\rangle_0 =\langle S_\pm (x)
\rangle_0 \langle S_\pm (y)\rangle_0 .
\eeq
In fact, vacuum clustering is another reason that makes it necessary to 
introduce an instanton vacuum.

The above calculations may be easily generalized to higher VEVs of chiral
densities. The general formula reads (see e.g. \cite{Zah})
\beq
\langle S_{H_1}(x_1)\cdots S_{H_n}(x_n)\rangle_0 = e^{ik\theta} 
\Bigl( \frac{\Sigma}{2}\Bigr)^n \exp
\Bigl[ \sum_{i<j}\sigma_i \sigma_j 4\pi D_{\mu_0} (x_i -x_j)\Bigr] 
\eeq
\bdi
k=\sum_{i=1}^n \sigma_i =n_+ -n_-
\edi
where $\sigma_i =\pm 1$ for $H_i =\pm $. This result we need for the mass 
perturbation theory.

Further VEVs that may be easily found are the field strength VEV
\beq
\langle F(x)F(y)\rangle_0 =\frac{1}{e^2}\Box_x \Box_y \langle \beta (x)
\beta (y)\rangle_0 =-\mu_0^2 D_{\mu_0} (x-y) -\delta (x-y)
\eeq
and the vector current correlator. For the latter it is most efficient to
introduce a vector current source into the path integral,
\beq
eB_\mu =\emn \partial^\nu b
\eeq
\beq
eB_\mu J^\mu =\frac{1}{\sqrt{\pi}}\emn \partial^\nu b \eML \partial_\lambda
\Phi = \frac{1}{\sqrt{\pi}} \Phi \Box b =:\Phi \lambda
\eeq
where $\lambda =\frac{1}{\sqrt{\pi}}\Box b$ is the source for the Schwinger
boson $\Phi$. Now the inclusion of the source consists in the substitution
$\beta \to \beta +b$ in the second (anomaly) term of the effective action
(36) and in the exact fermion propagators (28) and zero modes (30).
The remaining path integral computation is similar to the one which we
discussed previously and leads to the following VEVs (for $n$ chiral
densities, which are at the same time  generating functionals for the 
Schwinger boson)
\bdi
\langle S_{H_1}(x_1)\cdots S_{H_n}(x_n)\rangle_0 [\lambda]= e^{ik\theta} 
\Bigl( \frac{\Sigma}{2}\Bigr)^n \exp
\Bigl[ \sum_{i<j} \sigma_i \sigma_j 4\pi D_{\mu_0} (x_i -x_j)\Bigr] \cdot 
\edi
\beq
\cdot \exp\Bigl[-\int dy_1 dy_2 \lambda (y_1) D_{\mu_0}(y_1 -y_2)\lambda (y_2)
+2i\sqrt{\pi}\sum_{l=1}^n \sigma_l \int dy\lambda (y)D_{\mu_0}(y-x_l)
\Bigr] 
\eeq 
(for an explicit computation see \cite{Gatt2}).

Observe that, although $b$ was needed in the intermediate computations, the
final result can be expressed entirely in terms of $\lambda$, and only
massive propagators occur. This again shows that the massive Schwinger boson
is the only physical particle of the theory.

After this short discussion and formulae collection of the massless
Schwinger model we are prepared for the mass perturbation expansion of the 
massive model. This will be done in the forthcoming sections. 

\input psbox.tex
\let\fillinggrid=\relax

\section{Vacuum functional and vacuum energy density}

By simply expanding the mass term in (20), $e^{m\int dx\bar\Psi \Psi}$,
the Euclidean vacuum functional for the massive Schwinger model may be traced
back to VEVs of the massless model and some space-time integrations
(\cite{MSSM}), 
\beq
Z(m,\theta) =\sum_{n=0}^\infty \frac{m^n}{n!}\int dx_1 \ldots dx_n
\langle \prod_{i=1}^n \bar\Psi (x_i)\Psi (x_i)\rangle_0
\eeq
where $\langle \prod_{i=1}^n S(x_i)\rangle_0 =\langle \prod_{i=1}^n
(S_+ (x_i)+S_- (x_i) )\rangle_0 $ may be inferred from (44).
There, all contractions among $S_{H_i}(x_i)S_{H_j}(x_j)$ produce
exponentials of the massive propagator $D_{\mu_0}(x_i -x_j)$.
For a first, rough approximation we may use the fact that the massive scalar
propagator $D_{\mu_0} (x)$ vanishes exponentially for large argument. Therefore,
when integrating over space time and expanding the exponential, all
contributions from $D_{\mu_0}^l $ will be ignored for the moment, 
supposing that
the space time volume $V$ is sufficiently large. In this case the integrations
in (49) just produce factors of $V$. Further, when inserting (44) into (49) we
have to sum over all possible distributions of $n_+ =n-n_-$ pluses and $n_-$
minuses on $n$ scalar densities $S$. This results in a factor $n \choose n_-$.
Therefore we find for the $n$-th order term 
\beq
\langle\prod_{i=1}^n \int dx_i S(x_i )\rangle \sim \Bigl( \frac{\Sigma}{2}\Bigr) ^n V^n
\sum_{n_- =0}^n {n \choose n_-}e^{i(n-2n_-)\theta}=(\Sigma V\cos \theta )^n
\eeq
and for the normalized vacuum functional
\beq
\frac{Z(m,\theta)}{Z(0,0)}\sim \exp (m\Sigma V\cos\theta)
\eeq 
(we ignored terms like $m^n V^{n-1}$ in this approximation compared to
$m^{n-1}V^{n-1}$, therefore (51) is the first order result in $m$). This
result is wellknown, and its consequences for the vacuum structure and
spectrum of the Dirac operator are discussed in great detail in \cite{LSm}.

As we have seen, the exponentials $e^{\pm 4\pi D_{\mu_0}(x)}$ produce
volume factors $V$ upon integration, and, consequently, they will not lead
to an IR-finite perturbation expansion for $V\to \infty$. Therefore, we have to
expand the (products of) exponentials in (49) into the functions $E_\pm (x)$
\bdi
E_\pm (x):= e^{\pm 4\pi D_{\mu_0}(x)} - 1
\edi
\bdi
E^{(n)}_\pm (x) := e^{\pm 4\pi D_{\mu_0}(x)} - \sum_{l=0}^n \frac{1}{l!}
(\pm 4\pi D_{\mu_0}(x))^l
\edi
\beq
\widetilde E^{(n)}_\pm (p) = \int d^2 xe^{ipx}E^{(n)}_\pm (x) \quad ,\quad
E_\pm := \int dx E_\pm (x)
\eeq
(where we defined some related functions for later convenience). The 
$E_\pm (x)$ decay exponentially for large $|x|$.

Inserting the exact VEVs (44) into (49) and using the notation (52),
we obtain for the vacuum functional $Z(m,\theta)$, order by order
 \\ \\
n=1:
\beq
\frac{m}{1!}\frac{\Sigma}{2}\int dx(e^{i\theta}+e^{-i\theta})=
\frac{m}{1!}\frac{\Sigma}{2}V(e^{i\theta}+e^{-i\theta})
\eeq
n=2:
\bdi
\frac{m^2}{2!}\Bigl( \frac{\Sigma}{2}\Bigr)^2 \int dx_1 dx_2 
\Bigl[ e^{2i\theta}e^{4\pi D_\mu
(x_1 -x_2)}+2e^{-4\pi D_\mu (x_1 -x_2)}+e^{-2i\theta}e^{4\pi D_\mu (x_1
-x_2)}\Bigr] 
\edi
\beq
=\frac{m^2}{2!}\Bigl( \frac{\Sigma}{2}\Bigr)^2 \Bigl[ V^2 (e^{2i\theta}+2
+e^{-2i\theta})+
V(E_+ e^{2i\theta}+2E_- +E_+ e^{-2i\theta})\Bigr] 
\eeq
n=3:
\bdi
\ldots =\frac{m^3}{3!}\Bigl( \frac{\Sigma}{2}\Bigr)^3 \Bigl[ V^3 
(e^{3i\theta}+3e^{i\theta}+3e^{-i\theta}+e^{-3i\theta})+
\edi
\bdi
V^2 \Bigl( 3E_+ e^{3i\theta}+3(E_+ +2E_- )e^{i\theta}+
3(E_+ +2E_- )e^{-i\theta}+3E_+ e^{-3i\theta}\Bigr) +
\edi
\beq
V\Bigl( (3E_+^2 +E_+ \times E_+ \times E_+ )(e^{3i\theta}+e^{-3i\theta})
+3(2E_+ E_- +E_-^2 +E_+ \times E_- \times E_- )(e^{i\theta}+e^{-i\theta})
\Bigr) \Bigr] 
\eeq
n=4:
\bdi
\ldots =\frac{m^4}{4!}\Bigl( \frac{\Sigma}{2}\Bigr)^4 \Bigl[ V^4 
(e^{4i\theta}+4e^{2i\theta}
+6+4e^{-2i\theta}+e^{-4i\theta})+
\edi
\bdi
V^3 \Bigl( 6E_+ e^{4i\theta}+12(E_+ +E_- )e^{2i\theta}+12(E_+ +2E_- )
+12(E_+ +E_- )e^{-2i\theta} +6E_+ e^{-4i\theta}\Bigr) +
\edi
\bdi
V^2 \Bigl( (15E_+^2 +4E_+ \times E_+ \times E_+ )(e^{4i\theta}+e^{-4i\theta})+
\edi
\bdi
4(3E_+^2 +9E_+ E_- +3E_-^2 +E_+ \times E_+ \times
E_+ +3E_+ \times E_- \times E_- )(e^{2i\theta}+e^{-2i\theta})+
\edi
\beq
+6(E_+^2 +8E_+ E_- +E_-^2 +4E_+ \times E_- \times E_- ) \Bigr) +\ldots \Bigr] 
\eeq
\bdi
\ldots
\edi
where the cross indicates convolutions, e.g.
\beq
E_+ \times E_+ \times E_+ \equiv \int dy_1 dy_2 E_+ (y_1)
E_+ (y_1 +y_2)E_+ (y_2)
\eeq
and we displayed the result up to the accuracy we need.
Observe that the result is not obtained by just expanding polynomials like
$(1+E_+ (x_i))^l$, because e.g. a third power in $E_+ (x_i)$ may contribute to
$V^{n-3}E_+^3$ or to $V^{n-2}E_+ \times E_+ \times E_+$. 
Concerning the dimensions, observe that $E_\pm (x)$ is dimensionless and,
therefore, e.g. $[E_\pm]\sim [V]$, $[E_\pm \times E_\pm \times E_\pm ]\sim
[V^2 ]$, etc.

In a next step we rearrange the terms (53) -- (56) in rising powers of $V$:
\bdi
\frac{V}{1!}\Bigl[ m\frac{\Sigma}{2}(e^{i\theta}+e^{-i\theta})+\frac{m^2}{2}
\Bigl( \frac{\Sigma}{2}\Bigr)^2 (E_+ e^{2i\theta}+2E_- +E_+ e^{-2i\theta})+
\edi
\bdi
\frac{m^3}{6}\Bigl( \frac{\Sigma}{2}\Bigr)^3 \Bigl( (3E_+^2 +E_+ \times 
E_+ \times E_+ )(e^{3i\theta}+
e^{-3i\theta})+3(2E_+ E_- +E_-^2 +E_+ \times E_- \times E_- )
(e^{i\theta}+e^{-i\theta})\Bigr) +\ldots\Bigr] +
\edi
\bdi
\frac{V^2}{2!}\Bigl[ m^2 \Bigl( \frac{\Sigma}{2}\Bigr)^2 
(e^{2i\theta}+2+e^{-2i\theta})+
m^3 \Bigl( \frac{\Sigma}{2}\Bigr)^3 
\Bigl( E_+ (e^{3i\theta}+e^{-3i\theta})+(E_+ +2E_- )(e^{i\theta}
+e^{-i\theta})\Bigr) +
\edi
\bdi
\frac{m^4}{12}\Bigl( \frac{\Sigma}{2}\Bigr)^4 \Bigl( 
(15E_+^2 +4E_+ \times E_+ \times E_+ )(e^{4i\theta}+e^{-4i\theta})+
\edi
\bdi
4(3E_+^2 +9E_+ E_- +3E_-^2 +E_+ \times E_+ \times E_+ +3E_+ \times 
E_- \times E_- )(e^{2i\theta}+ e^{-2i\theta})+
\edi
\bdi
6(E_+^2 +8E_+ E_- +E_-^2 +4E_+ \times E_- \times E_- )\Bigr) +\ldots\Bigr] +
\edi
\bdi
\frac{V^3}{3!}\Bigl[ m^3 \Bigl( \frac{\Sigma}{2}\Bigr)^3 (e^{3i\theta}+
3e^{i\theta}+3e^{-i\theta}+e^{-3i\theta})+
\edi
\bdi
\frac{m^4}{2}\Bigl( \frac{\Sigma}{2}\Bigr)^4 \Bigl( 3E_+ (e^{4i\theta}+
e^{-4i\theta})+
6(E_+ +E_- )(e^{2i\theta}+e^{-2i\theta})+(E_+ +2E_- )\Bigr) +\ldots \Bigr] +
\edi
\bdi
\frac{V^4}{4!}\Bigl[ m^4 \Bigl( \frac{\Sigma}{2}\Bigr)^4 
(e^{4i\theta}+4e^{2i\theta}+6
+4e^{-2i\theta}+e^{-4i\theta})+\ldots \Bigr] +\ldots
\edi
\beq
=:\frac{V}{1!}\epsilon +\frac{V^2}{2!}\epsilon^2 +\frac{V^3}{3!}\epsilon^3 
+\ldots
\eeq
This result indicates an exponentiation  of the exact vacuum functional, too,
\beq
\frac{Z(m,\theta)}{Z(0,0)}=e^{V\epsilon (m,\theta)}
\eeq
where
\bdi
\epsilon (m,\theta )=m\frac{\Sigma}{2}2\cos \theta
+\frac{m^2}{2!}\Bigl( \frac{\Sigma}{2}\Bigr)^2 (2E_+ \cos 2 \theta +2E_- )+
\edi
\beq
\frac{m^3}{3!}\Bigl( \frac{\Sigma}{2}\Bigr) ^3 
\Bigl( (3E_+^2 +E_+ \times E_+ \times E_+ )2\cos 3\theta 
+3(2E_+ E_- +E_-^2 +E_+ \times E_- \times E_- )2\cos \theta \Bigr) +\ldots
\eeq
This exponentiation, equ. (59), is, of course, very important, because it
guarantees that physical VEVs do not depend on the space-time volume $V$
and are therefore IR-finite. Therefore, it would be nice if equ. (59)
could be proven generally. This is indeed possible and shall be discussed
next. 

For this purpose, let us write $Z(m,\theta)$ as a double sum,
\beq
Z(m,\theta) =\sum_{l=0}^\infty \frac{m^l}{l!} \sum_{n=0}^l V^n c_{l,n} ,
\eeq
where the factorial has been introduced as in the perturbation expansion (49).

Because of the multinomial formula
\bdi
(\epsilon (m,\theta))^n =(mc_{1,1} +\frac{m^2}{2!}c_{2,1} +\ldots )^n =
\edi
\beq
\sum_{l=0}^\infty m^{n+l} \sum_{k_i =0 \atop \sum_i k_i =n,\sum_i ik_i =n+l}^n
\frac{n!}{k_1 !\ldots k_{n+l}!}\Bigl( \frac{c_{1,1}}{1!}\Bigr)^{k_1} \cdots
\Bigl(\frac{c_{n+l,1}}{(n+l)!}\Bigr)^{k_{n+l}}
\eeq
the exponentiation condition reads (the $\frac{1}{n!}$ stems from the factor 
$\frac{V^n}{n!}$)
\bdi
\frac{c_{n+l,n}}{(n+l)!} =\frac{1}{n!}
 \sum_{k_i =0 \atop \sum_i k_i =n,\sum_i ik_i =n+l}^n
\frac{n!}{k_1 !\ldots k_{n+l}!}\Bigl( \frac{c_{1,1}}{1!}\Bigr)^{k_1} \cdots
\Bigl(\frac{c_{n+l,1}}{(n+l)!}\Bigr)^{k_{n+l}}
\edi
or
\beq
c_{n+l,n}= \sum_{k_i =0 \atop \sum_i k_i =n,\sum_i ik_i =n+l}^n
\frac{(n+l)!}{k_1 !\ldots k_{n+l}!}\Bigl( \frac{c_{1,1}}{1!}\Bigr)^{k_1} \cdots
\Bigl(\frac{c_{n+l,1}}{(n+l)!}\Bigr)^{k_{n+l}} .
\eeq
Next we have to answer the question where the volume factors $V$ come from.
The first power $V$ stems from the fact that all ``propagators'' $E_\pm (x_i
-x_j)$ depend on coordinate {\em differences}. Higher powers occur when
the corresponding ``Feynman graph'' is disconnected. Let us show a graphical 
example in Fig. 1.

$$\psannotate{\psboxscaled{700}{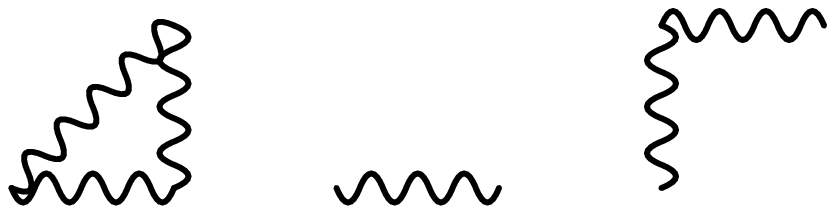}}{\fillinggrid
\at(3.9\pscm;-1\pscm){Fig. 1}}$$

\vspace{1.5cm}

Here each wavy line represents a $E_\pm (x)$ and their endpoints represent
``vertices'' $\frac{m\Sigma}{2} e^{\pm i\theta}$. These Feynman rules will
be discussed in detail in Section 7. 

Obviously, Fig. 1
is of third power in $V$ and of 8th power in $m$ (8 vertices). In general, 
$c_{l,n}$ is just the sum of all graphs of $l$-th order that consist of
$n$ connected pieces.

Therefore, we just have to prove that (63) is the correct combinatorial
formula for the distribution of $n$ connected graphs of total order $n+l$
on $n+l$ vertices.

But this is easy to understand. Consider e.g. a graph $c_{l_1 +l_2,
2}$ consisting of two connected pieces $c_{l_1 ,1}$, $c_{l_2 ,1}$.
There are $(l_1 +l_2)!$ possibilities to distribute the two connected
graphs on $l_1 +l_2$ vertices. However, $l_1 !$ ($l_2 !$) ways exist to
rearrange $c_{l_1 ,1}$ ($c_{l_2 ,1}$) on a given subset of $l_1$ ($l_2$)
vertices, therefore these factors must be divided out. This leads to
\bdi
c_{l_1 +l_2 ,2}=(l_1 +l_2 )!\frac{c_{l_1 ,1}}{l_1 !}\frac{c_{l_2 ,1}}{l_2 !}
\edi
and easily may be generalized to explain (63) up to the $k_i$ factors.
When some connected subgraphs $c_{i,1}$ occur more than once, $k_i >1$,
there is an additional symmetry factor $\frac{1}{k_i !}$ that counts for
the fact that an exchange of identical subgraphs $c_{i,1}$ leads to the 
same contribution to $c_{n+l,n}$. This explains formula (63).

So we have proven the exponentiation of the vacuum functional, (59) and,
as a consequence, the IR-finiteness of the mass perturbation theory.
In fact, the expansion of $e^{\pm 4\pi D_{\mu_0}(x)}=(1+E_\pm (x))$ into
$E_\pm (x)$ is analogous to the cluster expansion of statistical physics
models. There, our result correspond to the fact that the free energy is
an extensive quantity. A more rigorous discussion of these features can
be found in \cite{Froe1,FS1}.

Before ending this section we want to give an explicit (numerical) formula
for the lowest orders of the vacuum energy density, $\epsilon (m,\theta)$,
and, as a consequence, we will meet the problem of UV regularization.
For a numerical evaluation of order $m^2$ we have to compute the coefficients
$E_+$ and $E_-$. First, both $E_+$ and $E_-$ are proportional to 
$\frac{1}{\mu_0^2}$. Scaling $\mu_0$ out, we find
\beqa
\mu_0^2 E_+ =\mu_0^2 \int d^2 xE_+ (x) &=& \int d^2 x(e^{-2K_0 (|x|)}-1) \no \\
&=& 2\pi \int_0^\infty drr(e^{-2K_0 (r)}-1) = -8.9139
\eeqa
$E_+ (x)$ is well behaving ($E_+ (0)=-1$), 
so the numerical integration is straight
forward. $E_- (x)$ is singular at $x=0$, $E_- (x)\sim \frac{1}{x^2}$ 
for $x\to 0$,
but this singularity can easily be understood and removed in a unique way.
Indeed, this singularity is just the free fermion singularity, as can be seen
by rewriting $E_- (x)$ like (see e.g. equ. (41))
\beq
\Bigl( \frac{\Sigma}{2}\Bigr)^2 (E_- (x)+1)= \langle S_+ (x)S_- (0)
\rangle_0 = G_0^2 (x)e^{4\pi ({\rm\bf G}(0)-{\rm\bf G}(x))}
\stackrel{|x| \ra 0}{\longrightarrow} G_0^2 (x)=\frac{1}{4\pi^2 x^2}.
\eeq
This singularity may be isolated by a partial integration:
\beqa
\mu_0^2 E_- &=& \int d^2 x(e^{2K_0 (|x|)}-1)=2\pi \int_0^\infty \frac{dr}{r}
(e^{2K_0 (r)+2\ln r}-r^2) \no \\
&=& 2\pi\Bigl[ (\ln r)(e^{2K_0 (r)+2\ln r}-r^2)\Bigr]_{\epsilon\to 0}^\infty +
\no \\
&& 2\pi\int_0^\infty dr2(\ln r)((K_1 (r)-\frac{1}{r})e^{2K_0 (r)+2\ln r} + r)
\eeqa
($K_0^{'} =-K_1$). Observe that the first term precisely leads to the free 
fermionic
singularity at the lower boundary (and vanishes at the upper boundary). So the
second term is the unique and finite result we are looking for. The numerical
integration gives
\beq
\mu_0^2 E_- =9.7384
\eeq

In the literature there exist other UV regularizations as, e.g. 
the introduction
of an additional Thirring-type interaction (\cite{FS1,Gatt2}). Of course,
when this Thirring-type coupling constant is set equal to zero at the end
of the computations, the result agrees with our regularization.

With the help of (64), (67) the vacuum energy density reads
\beq
\epsilon (m,\theta) =m\Sigma\cos\theta +\frac{m^2}{\mu_0^2}\Bigl(\frac{\Sigma
}{2}\Bigr)^2 (-8.9139\cos 2\theta +9.7384) +o(m^3).
\eeq
%Here, of course, the question arises if this type of regularization may be
%generalized to higher orders of mass perturbation theory. The answer is
%that it is not even necessary to regularize higher order contributions,
%because they are UV finite. More precisely, individual terms contributing
%to a given order (in $m$) of $\epsilon (m,\theta)$ may contain a logarithmic
%singularity similar to (66). However, it can be proven that all terms that
%contribute to $\epsilon (m,\theta)$ in a given order sum up in such a way
%that all divergencies cancel completely (\cite{DIV}). Therefore, the
%super-renormalizability of the massive Schwinger model, that is obvious
%within electrical charge perturbation theory, is present in the framework of
%mass perturbation theory as well, although in a somewhat hidden
%fashion.
Higher order contributions to $\epsilon (m,\theta)$ (and to bosonic
$n$-point functions to be discussed in later sections) may contain
logarithmic singularities like in (66) and may be regulated in a
similar manner. However, one may perhaps hope for an even better UV
behaviour in higher order computations. After all, the model is known
to be super-renormalizable (in ordinary electrical coupling
perturbation theory this feature is obvious, as the vacuum
polarization graph (= a loop of two fermions) is the only divergent
diagram). So one might expect cancellations of divergencies in higher
order computations. And, indeed, for the $o(m^2)$ contribution to the
Schwinger mass precisely this happens (see Section 8, (99)).
A more detailed discussion of this problem will be given elsewhere.

\section{Condensates, susceptibilities and confinement}

In this section we will try to extract some physical information from our 
results obtained so far. The simplest task is the computation of the
fermion condensate,
\bdi
\langle S(x)\rangle =\langle \bar\Psi (x)\Psi (x)\rangle =\frac{1}{V}
\frac{1}{Z(m,\theta)}\frac{\partial}{\partial m}Z(m,\theta) =
\frac{\partial}{\partial m}\epsilon (m,\theta) 
\edi
\beq
\langle \bar\Psi (x)\Psi (x)\rangle =\Sigma\cos\theta +\frac{m}{2}\Sigma^2
(E_+ \cos 2\theta +E_- ) +o(m^2).
\eeq
Of course, the order zero result is the condensate of the massless Schwinger
model. From this result the pseudoscalar VEV $\langle P(x)\rangle $ may be
obtained, e.g. by the substitution $\cos n\theta\ra i\sin n\theta $
($n=0,1,\ldots$),
\beq
\langle P(x)\rangle =\langle \bar\Psi (x)\gaf \Psi (x)\rangle 
=i\Sigma\sin\theta
+i\frac{m}{2}\Sigma^2 E_+ \sin 2\theta +o(m^2).
\eeq
Another quantity that may be easily obtained is the field strength
condensate (see equ. (21))
\beq
\langle F(x)\rangle =\frac{2\pi}{e}\frac{\partial}{\partial\theta}
\epsilon (m,\theta) =-\frac{2\pi}{e}m\Sigma\sin\theta -\frac{\pi}{e}
m^2 \Sigma^2 E_+ \sin 2\theta +o(m^3).
\eeq 
Therefore, as discussed in Section 2, whenever there is a classical background
field, $F\sim \theta$, there remains some effect on the quantum level and
the screening is not complete in the massive model.

Observe that there is a relation between $\langle P\rangle$ and $\langle
F\rangle$. This is due to the anomaly equation (11),
\beq
0=\partial_\mu \langle J_5^\mu \rangle =\frac{e}{\pi}\langle F\rangle -
2im\langle P\rangle
\eeq
(one-point functions are always constants because of translational invariance).

By performing a second derivative on $\epsilon (m,\theta)$ one may infer the
susceptibilities. For the scalar susceptibility we get
\beq
\chi_{\rm s}=\int dx\langle S(x)S(0)\rangle_c =\frac{\partial^2}{
\partial m^2}\epsilon (m,\theta)=\frac{1}{2}\Sigma^2 (E_+ \cos 2\theta +E_- )
+o(m)
\eeq
(the subscript $c$ indicates the connected component; our discussion of the
last section explains why only connected components may contribute to
$\epsilon (m,\theta)$). For the pseudo-scalar susceptibility the mixed
contractions $S_+$, $S_-$ get a minus sign,
\beq
\chi_{\rm p}=\int dx\langle P(x)P(0)\rangle_c =\frac{1}{2} \Sigma^2
(E_+ \cos 2\theta -E_-) +o(m).
\eeq
The topological susceptibility is
\bdi
\chi_{\rm top}=\Bigl( \frac{-ie}{2\pi}\Bigr)^2 \int dx\langle F(x)F(0)
\rangle_c =-\frac{\partial^2}{\partial \theta^2} \epsilon (m,\theta)  =
\edi
\beq
m\Sigma\cos\theta +m^2 \Sigma^2 E_+ \cos 2\theta +o(m^3).
\eeq
There is a relation between $\chi_{\rm p}$ and $\chi_{\rm top}$; however,
this and related questions will be discussed in a later section, where the
Dyson-Schwinger equations corresponding to the equations of motion
(10), (11) are derived. More on the physical meaning of susceptibilities
in two-dimensional models may be found e.g. in \cite{SV1}.

A further physical feature that we are able to discuss by using our results 
obtained so far is the confinement behaviour of the massive Schwinger model.
This was recently discussed in \cite{GKMS} in first order mass perturbation, 
and we follow their approach and generalize their result to arbitrary order.

A usual way to derive the confinement property is the computation of the string
tension from the Wilson loop. The Wilson loop for a test particle of
arbitrary charge $g=qe$ is defined as (the additional factor $i$ in
Stokes' law is due to our Euclidean conventions (imaginary $F$), see e.g.
\cite{Diss,ABH})
\beq
W_D =\langle e^{ig\int_{\partial D} A_\mu dx^\mu}\rangle = \langle e^{g
\int_D F(x) d^2 x}\rangle =\langle e^{2\pi iq\int_D \nu (x) d^2 x}\rangle
\eeq
where $\nu (x)$ is the Pontryagin index density. Further $\partial D$ is
the contour of a closed loop and $D$ the enclosed region of space-time.
We are interested in the string tension for very large distances; further
we are able to explicitly separate the area dependence, therefore we may set
$D\ra V$ in the sequel.

For the VEV of an exponential the following formula holds,
\beq
\langle e^{2\pi iq\nu }\rangle =\exp 
\Bigl[ \sum_{n=1}^\infty \frac{(2\pi iq)^n}{n!}
\langle \nu^n \rangle_c \Bigr]
\eeq
where $\langle \rangle_c$ denotes the connected part of the $n$-point function.
These VEVs are given by
\beq
\langle \nu^n \rangle_c =V\int dx_2 \ldots dx_n \langle \nu (0) \nu (x_2)
\ldots \nu (x_n)\rangle_c =V(-i)^n \frac{\partial^n}{\partial\theta^n}
\epsilon (m,\theta)
\eeq
as is obvious from the vacuum functional (21) (as usual, disconnected VEVs
are of higher order in $V$). The vacuum energy density of arbitrary order
may be written like
\beq
\epsilon (m,\theta)=\sum_{l=0}^\infty \epsilon_l \cos l\theta
\eeq
where we used the symmetry of $\epsilon (m,\theta)$ with respect to
positive and negative instanton number, and $\epsilon_l$ acquires contributions
from instanton sectors $k=\pm l$. Performing the derivatives (78) we have
to separate even ($\sim \cos l\theta$) and odd ($\sim \sin l\theta$)
powers of derivatives. We find for the Wilson loop
\beqa
W&=& \exp \Bigl[V\sum_{n=1}^\infty\frac{(2\pi q)^{2n}}{(2n)!}
\sum_{l=0}^\infty (-1)^n l^{2n}\epsilon_l \cos l\theta 
+V\sum_{n=1}^\infty \frac{(2\pi q)^{2n-1}}{(2n-1)!}
\sum_{l=0}^\infty (-1)^n l^{2n-1} \epsilon_l \sin l\theta \Bigr] \no \\
&=& \exp \Bigl[ V\sum_{l=0}^\infty \epsilon_l \cos l\theta
\sum_{n=1}^\infty (-1)^n \frac{(2\pi ql)^{2n}}{(2n)!} 
+V\sum_{l=l}^\infty \epsilon_l \sin l\theta
\sum_{n=1}^\infty (-1)^n \frac{(2\pi ql)^{2n-1}}{(2n-1)!} \Bigr] \no \\
&=& \exp \Bigl[ V\sum_{l=0}^\infty \epsilon_l \cos l\theta
(\cos 2\pi ql -1) 
 -V\sum_{l=0}^\infty \epsilon_l \sin l\theta \sin 2\pi ql \Bigr] .
\eeqa
The string tension is defined as
\beq
\sigma := -\frac{1}{V} \ln W 
= \sum_{l=0}^\infty \epsilon_l \Bigl( \cos l\theta (1-\cos 2\pi ql) +
\sin l\theta \sin 2\pi ql \Bigr)
\eeq
and may be interpreted as the force between two widely separated probe
charges $g=qe$, where all quantum effects are included.

We find that, whenever the probe charge is an {\em integer} multiple of
the fundamental charge, $q \in {\rm\bf N}$, the charges are screened, and
the Wilson loop does not obey an area law. Observe that this result is
{\em exact} !

For noninteger probe charges there is no complete screening and the string
tension may be computed perturbatively,
\beq
\sigma =m\Sigma \Bigl( \cos\theta (1-\cos 2\pi q)+\sin\theta \sin 2\pi q
\Bigr) +o(m^2)
\eeq
showing that in the massive Schwinger model and for noninteger probe charges
there remains a constant force (linearly rising potential) for very
large distances. So in the massive Schwinger model true confinement is
realized instead of charge screening in the general case.
Observe that this formation of a long range force is a strictly nonperturbative
phenomenon in the sense of conventional perturbation theory, because only
nontrivial instanton sectors ($k\ne 0$) contribute to the string tension,
as is obvious from (78).

On the other hand, in the massless model arbitrary probe charges are
screened. (Discussions of screening and confinement in the massive Schwinger 
model within other approaches may be found e.g. in \cite{AAR,CJS,RRS1,RS1,AMZ},
and the behaviour found there agrees with our result.)

\section{Feynman rules for mass perturbation theory}

The interaction Lagrangian for the mass perturbation expansion reads
$L_{\rm I}= m\bar\Psi \Psi$, see (49). On the other hand, the formulae
(44), (48) for VEVs of the massless model, that we need for the perturbation 
expansion, contain chiral currents $S_\pm$ instead of the scalar one in
$L_{\rm I}$. As a consequence, each vertex corresponding to 
$L_{\rm I}$ contains in fact two vertices,
\beq
m\langle S(x)\rangle_0 =m\langle S_+ (x)\rangle_0 +m\langle S_- (x)\rangle_0
=\frac{m\Sigma}{2}e^{i\theta}+\frac{m\Sigma}{2}e^{-i\theta}
\eeq
corresponding to the two chiralities.

Further, 
these two types of vertices are connected by two types of propagators, namely
$S_+ (x) S_+ (y)$ and $S_- (x) S_- (y)$ by $E_+ (x-y)$, and $S_+ (x)S_- (y)$
and $S_- (x) S_+ (y)$ by $E_- (x-y)$. Further, because all vertices may be
connected to each other (see (44)), 
up to $n-1$ lines $E_\pm (x-y_i)$ may run from one
vertex $x$ to the other vertices $y_i$ for a $n$-th order mass 
perturbation contribution.

As a consequence, the Feynman rules acquire a matrix structure. More precisely,
the propagator corresponding to the $E_\pm (x)$ is a matrix, which in momentum
space reads
\beq
{\cal E}(p) =\left( \begin{array}{cc}\wt E_+ (p) & \wt E_- (p) \\ \wt E_- (p)
 & \wt E_+ (p) \end{array} \right)
\eeq
where the individual entries correspond to the individual $\langle S_i (x)S_j
(y)\rangle_0$, $i,j=\pm$, propagators. 

Each vertex, where $n$ propagator lines ${\cal E}(p_i)$
meet, is a $n$-th rank tensor ${\cal
G}_0$. Only two components of this tensor are nonzero, namely
\beq
{\cal G}_{0++\cdots +} =\frac{m\Sigma}{2}e^{i\theta} \quad ,\quad 
{\cal G}_{0--\cdots -} =\frac{m\Sigma}{2}e^{-i\theta}
\eeq
(corresponding to $S=S_+ +S_-$). E.g. the vertex where two propagators meet is
a matrix
\beq
{\cal G}_0 =\left( \begin{array}{cc} \frac{m\Sigma}{2}e^{i\theta} & 0 \\ 0
 & \frac{m\Sigma}{2}e^{-i\theta} \end{array} \right) .
\eeq
Internal lines must be ${\cal E}(p)$ propagators; external lines, however, may
be boson lines, too, when we treat bosonic $n$-point functions 
$\langle i\Phi (x_1)\ldots i\Phi (x_n)\rangle_m$. In (48) we see that each
boson that is connected to a $S_-$ vertex acquires a minus sign. Therefore,
the rule for a boson line is that each vertex $S=S_+ +S_-$, where a boson
line meets, is multiplied by $2\sqrt{\pi}$ times the boson propagator
$\frac{-1}{p^2 +\mu_0^2}$ times the pseudoscalar vector $P$, where
\beq
P = {1 \choose -1} \quad ,\quad S={1 \choose 1} .
\eeq
When $n$ bosons meet at one vertex, one may, instead of contracting that
vertex with $n$ vectors $P$, contract it with one $P$ ($S$) if $n$ is odd
(even). Of course, the number of indices of the vertex must be reduced 
accordingly.

These Feynman rules may be given by the graphs of Fig. 2 (we display them 
in momentum space).

$$\psannotate{\psboxscaled{650}{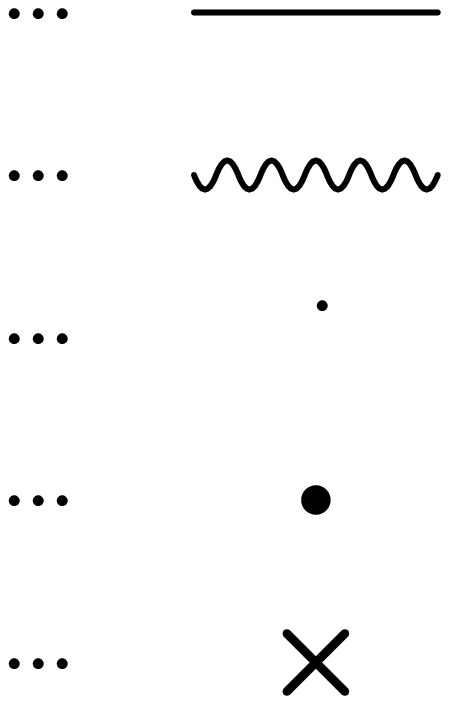}}{\fillinggrid
\at(0.7\pscm;-0.5\pscm){Fig. 2} \at(-1.9\pscm;2.8\pscm){${\cal G}$}
\at(-2.2\pscm;4.5\pscm){${\cal G}_0$} \at(-2\pscm;1.1\pscm){1}
\at(-2.6\pscm;6.2\pscm){${\cal E}(p)$} \at(-3\pscm;7.9\pscm){$\wt
D_{\mu_0}(p)$}}$$

\vspace{0.3cm}

Here ${\cal G}$ denotes the renormalized coupling that may be found like 
follows. The bare couplings are $m\langle S_\pm \rangle_0 =\frac{m\Sigma}{2}
e^{\pm i\theta}$, therefore the renormalized couplings are defined as
\beq
g_\theta =m\langle S_+ (x)\rangle_m \quad ,\quad g_\theta^* =m\langle
S_- (x)\rangle_m
\eeq
and ${\cal G}$ is constructed out of $g_\theta$, $g_\theta ^*$ like
${\cal G}_0$ out of $\frac{m\Sigma}{2}e^{\pm i\theta}$, see (85).
Graphically, ${\cal G}$ is just the sum of all graphs that may be attached
to the bare vertex ${\cal G}_0$, see Fig. 3,

$$\psannotate{\psboxscaled{800}{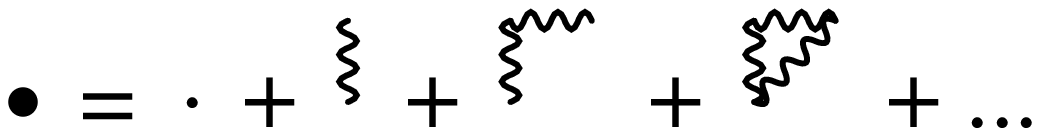}}{\fillinggrid
\at(4\pscm;-0.2\pscm){Fig. 3}}$$

\vspace{0.3cm}

i.e. the sum of all graphs where a line of propagators either starts and ends
at the same point, or it ends in the vacuum with zero momentum.
$g_\theta$ may be easily computed to be
\beqa
g_\theta &=& \frac{m\Sigma}{2}e^{i\theta}+\Bigl(\frac{m\Sigma}{2}\Bigr)^2
(e^{2i\theta}E_+ +E_- )\, +\, o(m^3) \no \\
&=:& g_1 +g_2 +\, o(m^3)
\eeqa
In fact, because of $g_\theta =m\langle S_+\rangle_m$, it is related to the 
vacuum energy density, see (69),
\beq
g_\theta +g_\theta^* =m\frac{\partial}{\partial m}\epsilon (m,\theta).
\eeq
Further, because each vertex may be connected to all the other vertices in
our theory, the contributions to the renormalized coupling, Fig. 3, may
be attached to each vertex of an arbitrary graph and, therefore, all the
vertices of the theory may be renormalized from ${\cal G}_0$ to ${\cal G}$.

As a further example of these Feynman rules we will investigate the
bosonic two-point function in the next section (we will compute it
explicitly there). In addition, we will use these Feynman rules extensively
in Sections 9 -- 11.

\section{The Schwinger mass}

As usual, in order to compute VEVs for the massive Schwinger 
model one has to insert the
corresponding operators into the path integral (20) and divide by the
vacuum functional $Z(m,\theta)$ (see (59), (60)),
\beq
\langle \hat O \rangle_m =\frac{1}{Z(m,\theta)} \langle \hat O
\sum_{n=0}^\infty \frac{m^n}{n!}\prod_{i=1}^n \int dx_i \bar\Psi (x_i) \Psi
(x_i) \rangle_0
\eeq
We will find that via the normalization all volume factors 
cancel completely, as it certainly has to be.

The Schwinger mass may be inferred from the Schwinger-boson two-point function.
Therefore, as a starting point for the mass perturbation computation we need
the following $n$-point functions of the massless model,
\bdi
\langle i\Phi (y_2)i\Phi (y_1)\prod_{i=1}^n S_{H_i}(x_i)\rangle_0 =e^{ik\theta}
\Bigl(\frac{\Sigma}{2}\Bigr)^n e^{4\pi \sum_{k<l} \sigma_k \sigma_l
D_{\mu_0}(x_k -x_l)}\cdot
\edi
\beq
\cdot \Bigl[ D_{\mu_0}(y_1 -y_2) +4\pi \Bigl( \sum_{i=1}^n (-)^{\sigma_i}
D_{\mu_0}(x_i -y_2)\Bigr) \Bigl( \sum_{j=1}^n (-)^{\sigma_j}
D_{\mu_0}(x_j -y_1)\Bigr) \Bigr]
\eeq
\bdi
k=\sum_{i=1}^n \sigma_i
\edi
which may be easily computed from the generating functionals (48).

For a perturbative computation of the Schwinger boson propagator we simply 
have to insert successive orders of (92) into the perturbation formula 
(91). Doing so, we find up to second order
\bdi
\langle i\Phi (y_1)i\Phi (y_2)\rangle_m = 
\frac{1}{Z(m,\theta )}\Bigl[ D_{\mu_0}(y_1 -y_2) +
m\frac{\Sigma}{2}(e^{i\theta}+e^{-i\theta})V D_{\mu_0}(y_1 -y_2) + 
\edi
\bdi
4\pi m\frac{\Sigma}{2}(e^{i\theta}+e^{-i\theta})\int
dxD_{\mu_0} (x-y_1)D_{\mu_0}(x-y_2) +
\edi
\bdi
\frac{m^2}{2!}\Bigl( \frac{\Sigma}{2}\Bigr)^2 (e^{2i\theta}+e^{-2i\theta})
\int dx_1 dx_2 [D_{\mu_0}(y_1
-y_2)+ 4\pi (D_{\mu_0}(x_1 -y_1)+D_{\mu_0}(x_2 -y_1))\cdot
\edi
\bdi
\cdot (D_{\mu_0}(x_1 -y_2) +
D_{\mu_0}(x_2 -y_2))]e^{4\pi D_{\mu_0}(x_1 -x_2)} \, + 
\edi
\bdi
\frac{m^2}{2!}\Bigl( \frac{\Sigma}{2}\Bigr)^2 2
\int dx_1 dx_2 [D_{\mu_0}(y_1
-y_2)+ 4\pi (D_{\mu_0}(x_1 -y_1)-D_{\mu_0}(x_2 -y_1))\cdot
\edi
\beq
\cdot (D_{\mu_0}(x_1 -y_2) -
D_{\mu_0}(x_2 -y_2))]e^{-4\pi D_{\mu_0}(x_1 -x_2)}\Bigr] .
\eeq
Inserting $Z(m,\theta)$ up to second order (see (59), (60)) and expanding
the denominator in the usual perturbative fashion, we arrive at
\bdi
\langle i\Phi (y_1)i\Phi (y_2)\rangle_m 
=D_{\mu_0}(y_1 -y_2)+4\pi m\Sigma \cos\theta \int dx D_{\mu_0}(x-y_1)
D_{\mu_0}(x-y_2) +
\edi
\bdi
4\pi m^2 \Bigl(\frac{\Sigma}{2}\Bigr)^2 \cos 2\theta \int dx_1 dx_2 [2D_{\mu_0}
(x_1 -y_1)D_{\mu_0}(x_1 -y_2)+2D_{\mu_0}(x_1 -y_1)D_{\mu_0}(x_2 -y_2)]\cdot
\edi
\bdi
\cdot (E_+ (x_1 -x_2)+1) - 8\pi m^2 \Bigl(\frac{\Sigma}{2}\Bigr)^2
\cos 2\theta \, V\int dx
D_{\mu_0} (x-y_1)D_{\mu_0}(x-y_2)\, +
\edi
\bdi
4\pi m^2 \Bigl(\frac{\Sigma}{2}\Bigr)^2  \int dx_1 dx_2 [2D_{\mu_0}
(x_1 -y_1)D_{\mu_0}(x_1 -y_2)-2D_{\mu_0}(x_1 -y_1)D_{\mu_0}(x_2 -y_2)]\cdot
\edi
\bdi
\cdot (E_- (x_1 -x_2)+1) - 8\pi m^2 \Bigl(\frac{\Sigma}{2}\Bigr)^2
V\int dx
D_{\mu_0} (x-y_1)D_{\mu_0}(x-y_2)
\edi
\bdi
=D_{\mu_0}(y_1 -y_2)+4\pi m\frac{\Sigma}{2}\cos\theta \int dx D_{\mu_0}(x-y_1)
D_{\mu_0}(x-y_2) +
\edi
\bdi
8\pi m^2 \Bigl(\frac{\Sigma}{2}\Bigr)^2   \int dx_1 dx_2 D_{\mu_0}
(x_1 -y_1)D_{\mu_0}(x_2 -y_2)[\cos 2\theta E_+ (x_1 -x_2) -E_- (x_1 -x_2)] +
\edi
\bdi
8\pi m^2 \Bigl(\frac{\Sigma}{2}\Bigr)^2 (\cos 2\theta E_+ +E_- )
\int dx D_{\mu_0}(x-y_1)D_{\mu_0}(x-y_2) +
\edi
\beq
8\pi m^2 \Bigl(\frac{\Sigma}{2}\Bigr)^2 (\cos 2\theta -1)
\int dx_1 dx_2   D_{\mu_0}(x_1 -y_1)D_{\mu_0}(x_2 -y_2)
\eeq
where $E_\pm ,E_\pm (x)$ are given in (52) and we used 
the $x\ra -x$ symmetry of all
occurring functions. The last term stems from a disconnected part of (93) and
must be subtracted. Observe that, as claimed, all volume factors $V$ have
dropped out.

We may easily check that (94) is the right expression by depicting
the corresponding Feynman diagrams in Fig. 4

$$\psannotate{\psboxscaled{650}{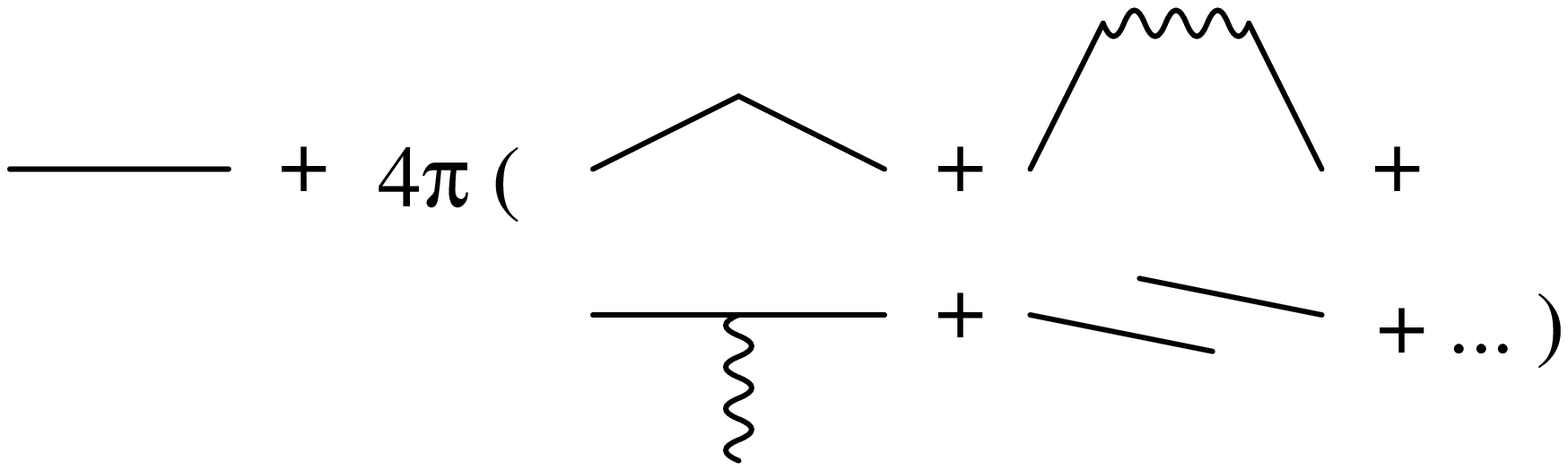}}{\fillinggrid
\at(8\pscm;-0.5\pscm){Fig. 4}}$$

\vspace{0.5cm}

and find that (94) and Fig. 4 coincide. Again, the last graph is 
disconnected and must be subtracted. 

In order to obtain the second order result for the Schwinger mass, 
we rewrite expression (94)
in momentum space and substitute all functions by their
Fourier transforms (thereby the convolutions turn into products; the
disconnected term in (94) is omitted),
\bdi
\wt{\langle i\Phi i\Phi \rangle}_m^c (p)=
\frac{-1}{p^2 +\mu_0^2}+4\pi m\Sigma\cos\theta \frac{1}{(p^2
+\mu_0^2)^2} +
\edi
\bdi
2\pi m^2 \Sigma^2 
\frac{1}{(p^2 +\mu_0^2)^2}[\cos 2\theta (E_+ +\widetilde E_+ (p))
+E_- -\wt E_- (p)]=
\edi
\bdi
\frac{-1}{p^2 +\mu_0^2}\Bigl( 1-4\pi m\Sigma\cos\theta \frac{1}{p^2 
+\mu_0^2}- 2\pi
m^2 \Sigma^2 [\cos 2\theta (E_+ +\widetilde E_+ (p))+E_- -\wt E_- (p)]
\frac{1}{p^2 +\mu_0^2}\Bigr) =
\edi
\bdi
\frac{-1}{p^2 +\mu_0^2 +4\pi m\Sigma\cos\theta +2\pi m^2 \Sigma^2 [\cos 2\theta
(E_+ +\widetilde E_+ (p))+E_- -\wt E_- (p)]+
(4\pi m \Sigma\cos\theta)^2 \frac{1}{p^2 +\mu_0^2}}
\edi
\beq
+o(m^3) 
\eeq
Therefore, for finding the mass pole, $p^2$ has to obey the 
equation (after a rescaling ${p'}^2 =\frac{p^2}{\mu_0^2}$, $E_\pm' =E_\pm 
(\mu_0^2 \equiv 1)=\mu_0^2 E_\pm $ etc.)
\bdi
{p'}^2 =-1-4\pi\frac{m\Sigma}{\mu_0^2}\cos\theta -2\pi \frac{m^2
\Sigma^2}{\mu_0^4}\cos 2\theta [E_+' +\widetilde E_+'(p') +
\frac{4\pi}{{p'}^2 +1}]-
\edi
\beq
2\pi \frac{m^2 \Sigma^2}{\mu_0^4}[E_-' -\wt E_- (p')
+\frac{4\pi}{{p'}^2 +1}].
\eeq
The second order part (the term in square brackets) may be further evaluated 
like (for the $\cos 2\theta$ term)
\bdi
[\cdots ]=\int d^2 x[e^{-2K_0 (|x|)}-1+e^{ip' x}(e^{-2K_0 (|x|)}-1+ 2K_0
(|x|)]=
\edi
\bdi
\int_0^\infty dr r[2\pi (e^{-2K_0 (r)}-1) + \int_0^{2\pi} d\theta
e^{i|p'|r\cos\theta} (e^{-2K_0 (r)}-1+2K_0 (r))]=
\edi
\beq
2\pi \int_0^\infty dr r[e^{-2K_0 (r)}-1 + J_0 (|p'|r)(e^{-2K_0 (r)}-1+ 2K_0
(r))] 
\eeq
where $J_0$ is the Bessel function of the first kind. This expression behaves
well around $|p'|=i$ and therefore we may set $|p'|=i$ because deviations
from this value are of higher order in $m$. Using $I_0 (r)=J_0 (ir)$ we find
\beq
\frac{1}{2\pi}[\ldots ]=:A= \int_0^\infty dr r[e^{-2K_0 (r)}-1 + 
I_0 (r)(e^{-2K_0 (r)}-1+2K_0 (r))] =-0.6599
\eeq
Analogously we find for the other second order term (containing the $E_-$)
\beq
\frac{1}{2\pi}[\ldots ]=: B=\int_0^\infty drr[e^{+2K_0 (r)}-1 + 
I_0 (r)(-e^{+2K_0 (r)}+1+2K_0 (r))]=1.7277
\eeq
In this expression (99) 
the nice feature of cancellation of UV divergencies occurs.
Indeed, both $e^{2K_0 (r)}$ and $I_0 (r)e^{2K_0 (r)}$ diverge like
$\frac{1}{r^2}$ for small $r$ (this divergency corresponds to the free fermion
field divergency of the underlying theory that we discussed in
Section 5, see (65)), 
but obviously the divergencies
cancel each other. In fact, this cancellation was already observed twenty
years ago in \cite{KS1} within a bosonization approach.

Collecting all results we find
for the Schwinger mass in second order
\beq
-{p'}^2 \equiv \frac{\mu_2^2}{\mu_0^2}=1+4\pi
\frac{m}{\mu_0}\frac{\Sigma}{\mu_0} \cos\theta + 4\pi^2 \frac{m^2}{\mu_0^2}
\Bigl( \frac{\Sigma}{\mu_0}\Bigr)^2 (A\cos 2\theta + B)
\eeq
or, inserting all numbers (remember
$\frac{\Sigma}{\mu_0}=\frac{e^\gamma}{2\pi}$, equ. (40))
\beq
\mu_2^2 =\mu_0^2 (1 + 3.5621\cdot \frac{m}{\mu_0}\cos\theta + 5.4807 \cdot
\frac{m^2}{\mu_0^2} - 2.0933\cdot\frac{m^2}{\mu_0^2}\cos 2\theta ).
\eeq

For the special case $\theta =0$ our result (101) precisely coincides with
the result in \cite{Vary}, where the second order correction for $\theta =0$
was computed within bosonization and using near light cone coordinates. In the
same article this result was compared to a lattice calculation
(\cite{Crew}), and a good agreement is obtained within the range of the 
expansion
parameter $\frac{m}{\mu_0}$ where the lattice calculations were performed.

\section{Dyson-Schwinger equations and exact $n$-point functions}

The Schwinger boson -- which we discussed in the last section -- is an
interacting particle in the massive Schwinger model. As a consequence, we
will find that it forms $n$-boson bound states. In principle, their masses
could be computed analogously to the previous section, by a computation
of the corresponding Schwinger-boson $2n$-point functions and by the 
determination of their mass poles.

Here we will adapt a slightly different method. We will discuss the
Dyson-Schwinger equations that follow from the equations of motion (10), (11),
acquiring thereby a deeper insight into the structure of the model. 
With the help of these Dyson-Schwinger equations we will be able to re-express
the $n$-point functions of the model in a way that is more suitable for
our discussion. We will find in this way that the spectrum of the theory is
even richer than expected. There is a second stable particle in the theory
in addition to the Schwinger boson, namely the two-boson bound state, and
unstable higher bound states may be formed out of both these stable particles.
Further, we will find that both particles may occur in final states of 
decays and scattering processes (see also \cite{DECAY,THREE,SCAT}).

When we use the generating functional for Schwinger bosons (48) 
for a computation
of the Schwinger boson $2n$-point function in lowest order in $m$, we find a 
contribution that may be depicted graphically like in Fig. 5,

$$\psannotate{\psboxscaled{650}{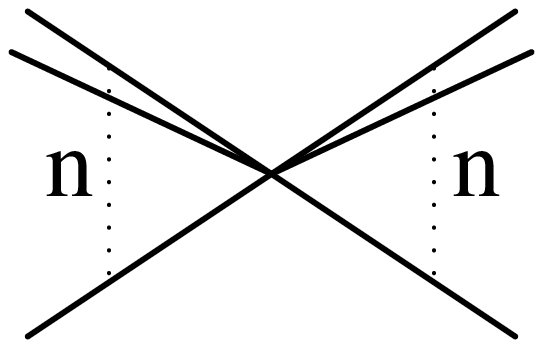}}{\fillinggrid
\at(2.1\pscm;-0.5\pscm){Fig. 5}}$$

\vspace{0.5cm}

and the vertex corresponds to a coupling constant $c$, where
\beq
c=(4\pi )^n m\Sigma\cos\theta .
\eeq
Therefore, this lowest order coupling mediates an attractive force for
$|\theta |<\frac{\pi}{2}$, and in this range of the vacuum angle $\theta$ 
the formation of bound states has to be expected, at least for sufficiently
small fermion mass $m$. The restriction $|\theta |<\frac{\pi}{2}$ shall
be assumed in the sequel.

In the introduction we wrote down the two equations of motion that relate the
field strength operator and the fermionic vector current operator, namely the
Maxwell equation (10) and the anomaly equation (11). By introducing the 
Schwinger boson (9) and by eliminating the field strength we may derive the
equation (expressed in real fields $P$, $i\Phi$)
\beq
M_x i\Phi (x):=
(\Box_x -\mu_0^2 )i\Phi (x)=2\sqrt{\pi}mP(x).
\eeq
(where we introduced the operator $M_x$ for convenience).

[{\em Remark}: there is a slight difference between the equations (10), (103) 
and equation (11). Whereas the anomaly equation 
(11) holds as an operator relation, and
consequently on all states, (10) and (103) are only true on physical states 
(i.e. on states that are invariant with respect to small {\em and large}
gauge transformations; for a deeper discussion of this feature see e.g.
\cite{LS1,AAR}). However, the introduction of the $\theta$ vacuum
within our path integral approach precisely corresponds to the
introduction of the physical vacuum (see Section 2). Therefore, for our
physical VEVs we may use equ. (103), as we will verify by explicit
perturbative computations.]

Next we define the amputated, connected bosonic $n$-point functions
(FT\ldots Fourier transform)
\beq
M^{(n)}(p_1 ,\ldots ,p_n)={\rm FT}(M_{x_1}\ldots M_{x_n}\langle i\Phi (x_1)
\ldots i\Phi (x_n)\rangle_m^c ).
\eeq

\subsection{Two-point function}

So let us study e.g. the connected two-point function $\langle i\Phi (x_1)
i\Phi (x_2)\rangle_m^c$. Graphically it may be depicted like in Fig. 6.

$$\psannotate{\psboxscaled{650}{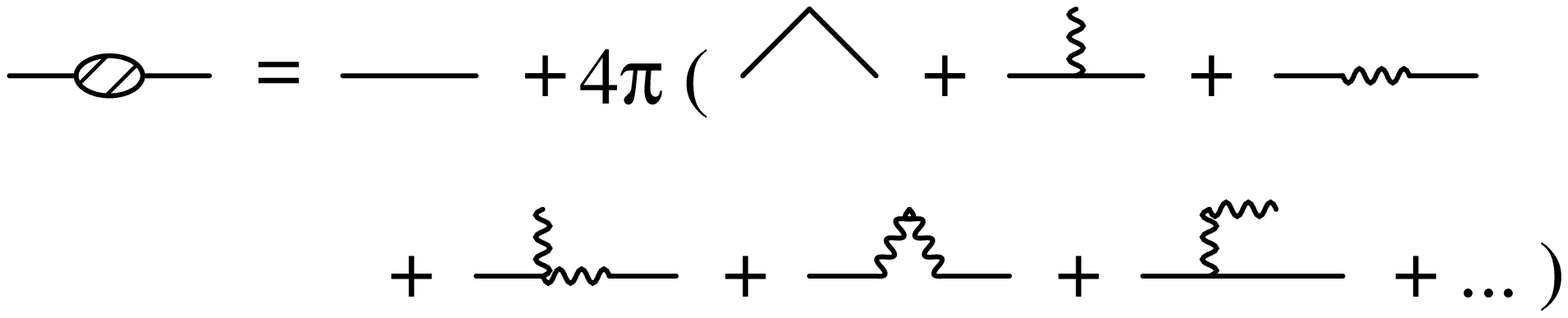}}{\fillinggrid
\at(9\pscm;-0.5\pscm){Fig. 6}}$$

\vspace{0.5cm}

We find the following behaviour: all graphs where both boson lines meet at
one and the same vertex contain just the corrections that change this vertex 
from the bare one ${\cal G}_0$ to the exact (renormalized) vertex ${\cal G}$,
see Fig. 3. In all the other graphs, each individual vertex is renormalized in
the same way, too.

So let us re-express Fig. 6 in terms of the renormalized coupling, and, in
addition, amputate the two external boson lines (and the pseudoscalar 
vectors $P$, see (87)). We obtain Fig. 7, where we introduced the exact 
propagator that is defined in Fig. 8.

$$\psannotate{\psboxscaled{600}{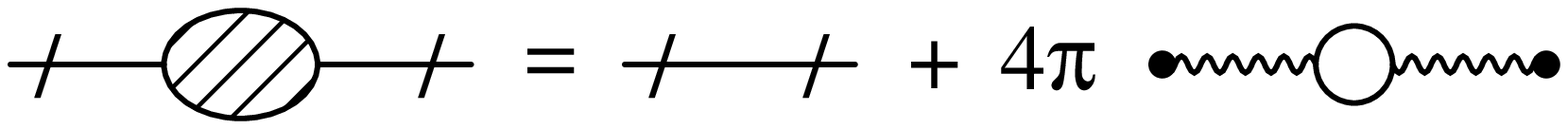}}{\fillinggrid
\at(6.9\pscm;-0.5\pscm){Fig. 7}}$$

$$\psannotate{\psboxscaled{600}{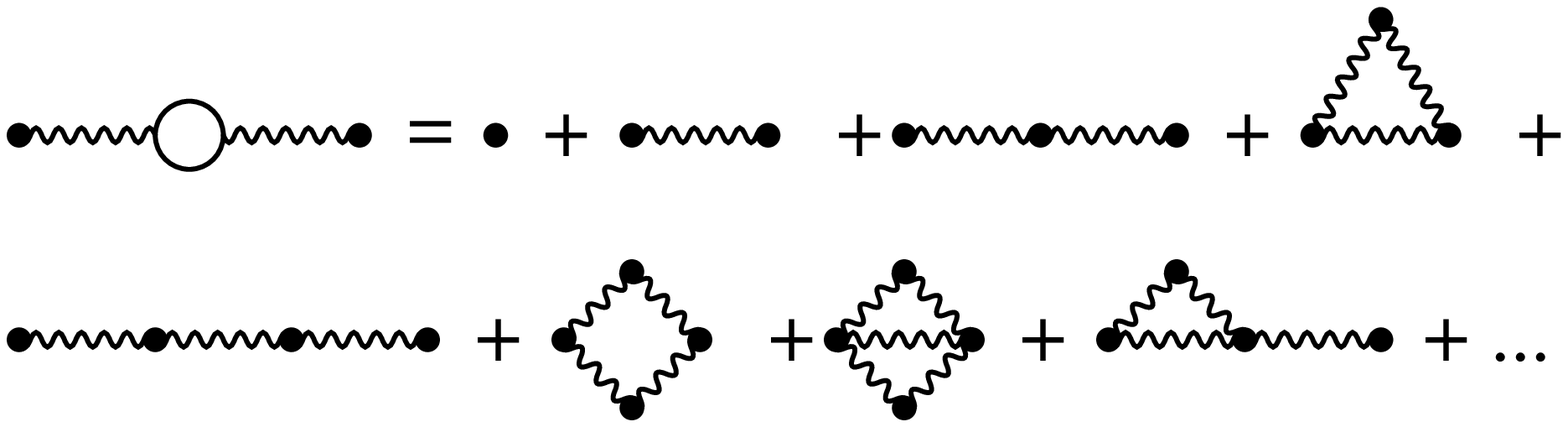}}{\fillinggrid
\at(8.9\pscm;-0.5\pscm){Fig. 8}}$$

\vspace{0.1cm}

Here it is understood that the left and right vertices of each graph in Fig. 8
are the initial and final ones where we amputated the bosons.
We introduce for the above exact propagator of Fig. 8 the name ${\cal G}
\Pi (p)$ in momentum space (matrix multiplication is understood, see the
Feynman rules of Fig. 2), because we will need it frequently
\beq
{\cal G}\Pi (p) := {\cal G} +{\cal G}{\cal E}(p){\cal G}+{\cal G}{\cal E}(p)
{\cal G}{\cal E}(p){\cal G} +\ldots 
\eeq
This exact propagator $\Pi (p)$ consists of a constant part (the sole vertex
in Fig. 8 without ${\cal E}(p)$ lines) that is a scalar (because two boson
lines meet on this one vertex), and of a true two-point function (depending
on $p$), where the initial and final vertices are pseudoscalars. Inserting 
all factors properly, Fig. 7 may be written like ($M^{(2)}(p,p)\equiv
M^{(2)}(p)$)
\beq
M^{(2)}(p)=-(p^2 +\mu_0^2 )+4\pi m\langle S\rangle_m +4\pi m^2 \wt{\langle
PP\rangle}_m^c (p)
\eeq
where
\beq
m\langle S\rangle_m \equiv P^T {\cal G}P=S_i{\cal G}_i =g_\theta +g^*_\theta
\eeq
\beq
m^2 \wt{\langle PP\rangle}_m^c (p)\equiv P^T {\cal G}(\Pi (p) -{\bf 1})P
\eeq
where $P^T =(1,-1)$ is the transpose of the vector $P$, (87), and matrix
multiplication is understood (the single vertex ${\cal G}$ we may interpret
either as a two-component object that is contracted by two vectors $P$ or as a
one-component object that is contracted by one vector $S$).

This is just the momentum space version of the Dyson-Schwinger equation for
the two-point function,
\bdi
M_{y_1}M_{y_2}\langle i\Phi (y_1)i\Phi (y_2)\rangle = M_{y_1}\delta (y_1 -y_2)+
\edi
\beq
4\pi m\langle S(y_1)\rangle_m \delta (y_1 -y_2) +4\pi m^2 
\langle P(y_1)P(y_2)\rangle_m^c .
\eeq 
Observe that the $\langle P(y_1)P(y_2)\rangle$ propagator includes an
arbitrary number of bosons propagating from $y_1$ to $y_2$, even in least
order. This will be important in the sequel.

The key observation for the computation of bound states is the fact that the 
exact propagator $\Pi (p)$ may be resummed. This resummation
 relies on the following
observation. All diagrams that fall into two pieces when they are cut at a
vertex, factorize in momentum space, i.e. they are a product of two functions
of $p$, see Fig. 8. 
The opposite type graphs are called non-factorizable (n.f.).

Here we should be more precise about the cutting. We stated in Section 7 
that the
vertices are tensors, so how to cut such a vertex? Suppose e.g. we have a
vertex where three lines meet and we want to cut it in a way that two wavy
lines belong to the left hand side, and one line to the right hand side. Then
we rewrite the vertex like
\beq
{\cal G}_{ijk}=\delta_{ijl}{\cal G}_{ll'}\delta_{l' k}\quad ,\quad
i,j,k,l,l' =\pm
\eeq
where ${\cal G}_{ll'}$ is the vertex matrix (85,88) and the $\delta_{i_1 \cdots
i_n}$ are generalizations of the Kronecker delta $\delta_{ij}$, i.e.
\beq
\delta_{++\cdots +}=\delta_{--\cdots -}=1\quad ,\quad \delta_{i_1 \cdots i_n}=0
\quad \mbox{otherwise}
\eeq
Therefore, we may write for the sum of non-factorizable graphs, which we
call ${\cal A}$ (see Fig. 9)

$$\psannotate{\psboxscaled{600}{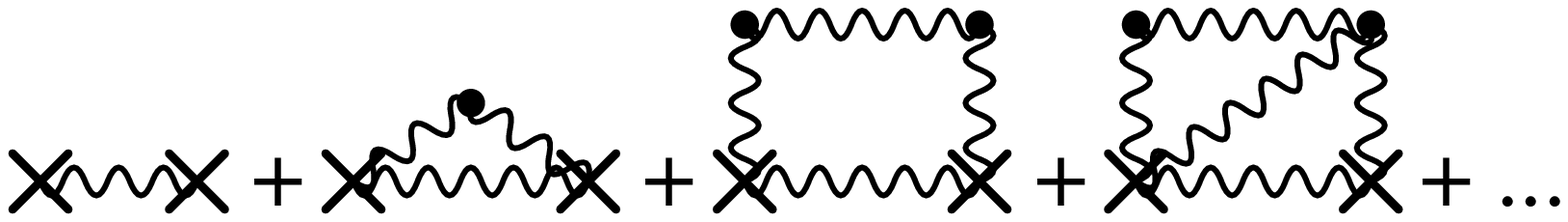}}{\fillinggrid
\at(6.9\pscm;-0.5\pscm){Fig. 9}}$$ 

\beq
{\cal A}_{ij}(p)={\cal E}_{ij}(p)+\int\frac{d^2 q}{(2\pi)^2}\delta_{ikk'}
{\cal E}_{kl}(q){\cal G}_{ll'}{\cal E}_{l' m}(q){\cal E}_{k' m'}(q-p)
\delta_{jmm'} +\ldots 
\eeq
The matrix ${\cal A}(p)$ may be rewritten like
 \beq
{\cal A}(p) =\left( \begin{array}{cc} \wt{\langle S_+ S_+ \rangle}_{\rm n.f.} 
(p)
 & \wt{\langle S_+ S_- \rangle}_{\rm n.f.} (p) \\ \wt{\langle S_- S_+
 \rangle}_{\rm n.f.} (p)
 & \wt{\langle S_- S_- \rangle}_{\rm n.f.} (p) \end{array} \right)
\eeq
The entries of this matrix are, however, related (e.g. $\wt{\langle S_- S_-
\rangle}_{\rm n.f.}(g_\theta ,p)=\wt{\langle S_+ S_+ \rangle}_{\rm n.f.}
(g^*_\theta ,p)$, as may be checked from the perturbative expansion) and,
therefore, we find for the product ${\cal G}{\cal A}(p)$ (which we need in the
sequel)
\beq
{\cal G}{\cal A}(p)=:\left( \begin{array}{cc} \alpha (g_\theta ,p) & \beta
(g_\theta ,p) \\ \beta (g^*_\theta ,p) & \alpha (g^*_\theta ,p) \end{array}
\right)
\eeq
\beq
\alpha (g_\theta ,p)=g_\theta \wt{\langle S_+ S_+ \rangle}_{\rm n.f.}(g_\theta
,p) \quad ,\quad \beta (g_\theta ,p)=g_\theta \wt{\langle S_+ S_- \rangle}_{\rm
n.f.}(g_\theta ,p).
\eeq
Now we may collect all n.f. graphs in (105), 
Fig. 8, e.g. on the left hand side,
and find that they are again multiplied by {\em all} graphs that occur in Fig.
8. Therefore we may write for ${\cal G}\Pi (p)$ of equ. (105)
\beq
{\cal G}\Pi (p)={\cal G}({\bf 1}+{\cal A}(p){\cal G}\Pi (p)).
\eeq
Equation (116) may be solved for $\Pi (p)$ by a matrix inversion and has the
solution
\beq
\Pi (p)=\frac{1}{N(p)} \left( \begin{array}{cc} 1-\alpha (g^*_\theta ,p) & 
\beta (g^*_\theta ,p) \\ \beta (g_\theta ,p) &
1-\alpha (g_\theta ,p) \end{array} \right)
\eeq
where $N(p)$ is the determinant of the matrix that had to be inverted,
\beq
N(p)=\det ({\bf 1}-{\cal G}{\cal A}(p))=1-\alpha (g_\theta ,p)-\alpha
(g^*_\theta ,p)+\alpha (g_\theta ,p)\alpha (g^*_\theta ,p) -\beta (g_\theta
,p)\beta (g^*_\theta ,p).
\eeq
We will find that the zeros of the real part of the denominator $N(p)$ will
give us all bound-state masses, whereas its imaginary parts at the bound-state
masses are related to the decay widths.

\subsection{Higher $n$-point functions}

Dyson-Schwinger equations for higher $n$-point functions may be derived in a 
way that is similar to the case of the two-point function. 
 Before showing them
we need some more graphical rules (see Fig. 10),

$$\psannotate{\psboxscaled{600}{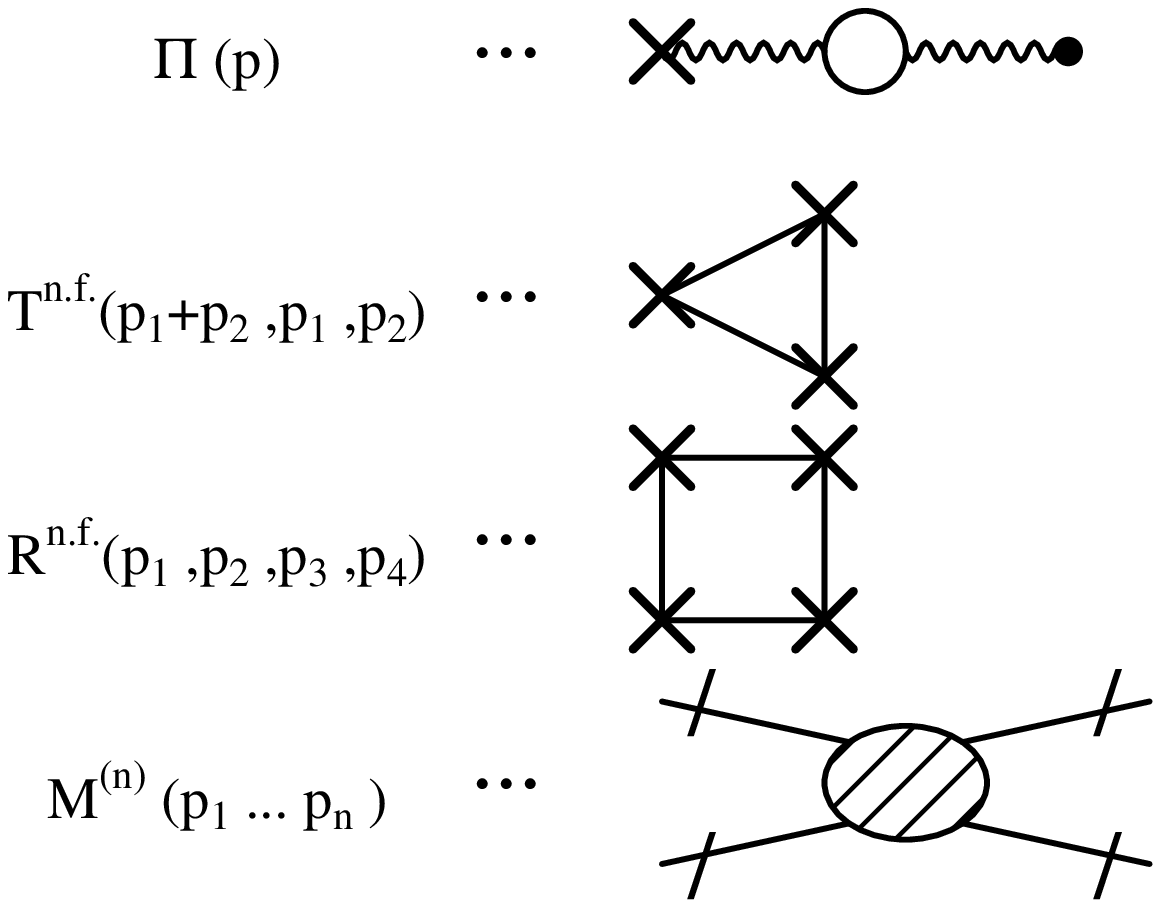}}{\fillinggrid
\at(5.9\pscm;-0.5\pscm){Fig. 10}}$$ 

\vspace{0.1cm}

where $M^{(n)}$, of course, should have $n$ external (amputated) boson lines.

For the three-point function e.g. we find the Dyson-Schwinger equation
(in momentum space)
\bdi
M^{(3)}(p_1 +p_2 ,p_1 ,p_2 )=(2\sqrt{\pi})^3 [m\langle P\rangle_m +m^2
\wt{\langle SP\rangle}_m^c (p_1 +p_2 )
\edi
\beq
+m^2 \wt{\langle SP\rangle}_m^c (p_1)+m^2 \wt{\langle SP\rangle}_m^c (p_2) + 
m^3 \wt{\langle PPP\rangle}_m^c (p_1 +p_2 ,p_1 ,p_2) ]
\eeq
where $m\langle P\rangle_m$ and $m^2 \wt{\langle SP\rangle}_m^c (p)$ are
analogous to (107), (108) whereas the last term is given by
\beq
m^3 \wt{\langle PPP\rangle}_m^c (p_1 +p_2 ,p_1 ,p_2)=P_iP_jP_k {\cal G}_{ii'}
{\cal G}_{jj'}{\cal G}_{kk'}T_{i' j' k'}(p_1 +p_2 ,p_1 ,p_2 )
\eeq
and $T_{ijk}$ is just the exact three-point function for general chiral
indices and external couplings equal to 1, see Fig. 11.

$$\psannotate{\psboxscaled{600}{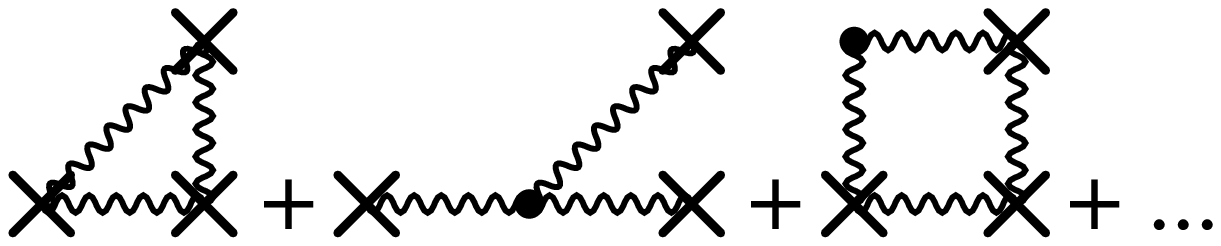}}{\fillinggrid
\at(5.9\pscm;-0.5\pscm){Fig. 11}}$$ 

\vspace{0.1cm}

The essential point is that $M^{(3)}$, again, may be reexpressed entirely in
terms of non-factorizable $n$-point functions, namely
\bdi
M^{(3)}(p_1 +p_2 ,p_1 ,p_2)=(2\sqrt{\pi})^3 P_iP_jP_k \Pi_{ii'}(p_1 +p_2 )
\Pi_{jj'}(p_1 )\Pi_{kk'}(p_2) \cdot
\edi
\beq
\cdot \Bigl( {\cal G}_{i' j' k'}+ {\cal G}_{i' l}
{\cal G}_{j' m}{\cal G}_{k' n}T^{\rm n.f.}_{lmn}(p_1 +p_2 ,p_1 ,p_2)\Bigr)
\eeq
or, graphically (see Fig. 12; we ignore an overall factor $(2\sqrt{\pi})^3$
on the r.h.s. of Fig. 12),

$$\psannotate{\psboxscaled{450}{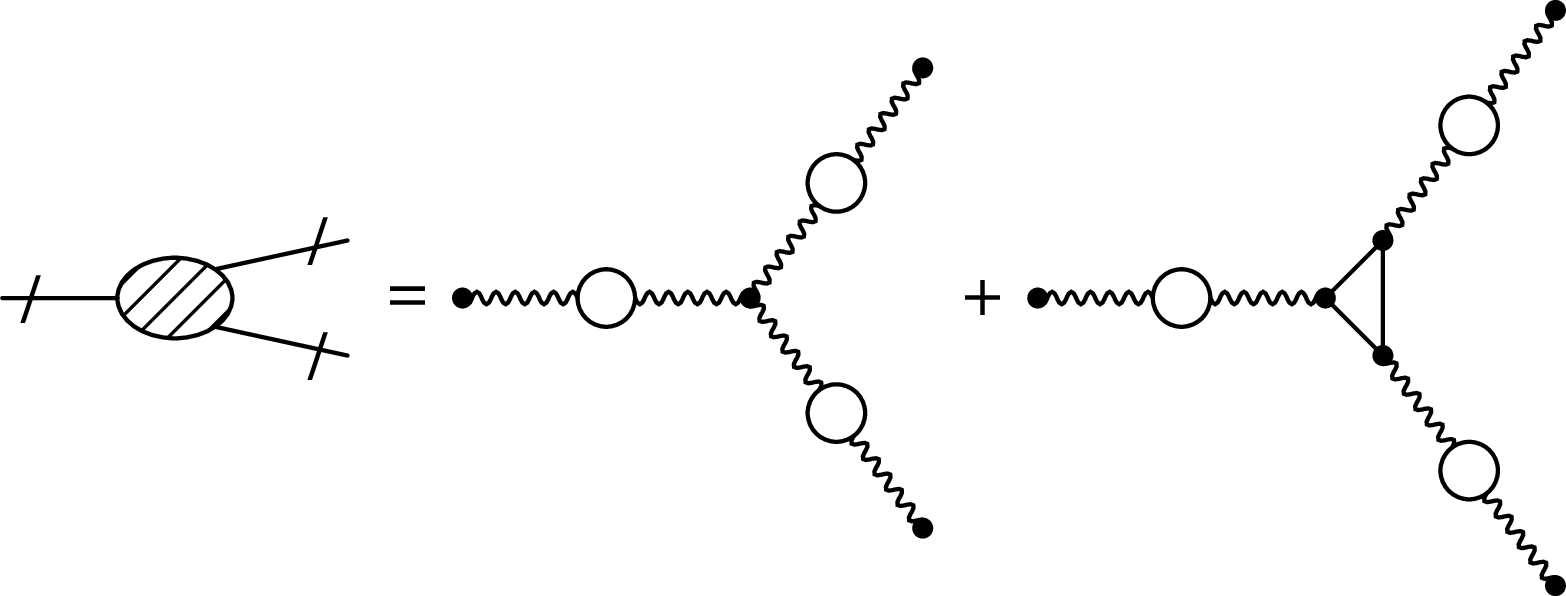}}{\fillinggrid
\at(10.9\pscm;-0.5\pscm){Fig. 12}}$$ 

where the non-factorizable three-point function $T_{\rm n.f.}$ is given by
Fig. 13.

$$\psannotate{\psboxscaled{600}{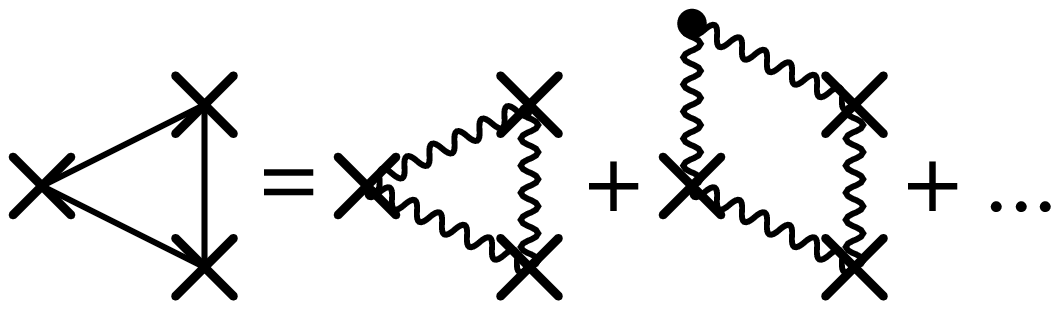}}{\fillinggrid
\at(5.9\pscm;-0.5\pscm){Fig. 13}}$$ 

\vspace{0.1cm}

The actual validity of (121), Fig. 12, 
has to be checked by a closer inspection of
the Feynman graphs (it is just tedious combinatorics).

We find that the non-factorizable $n$-point functions in our theory play a role
analogous to the 1PI Green functions in other theories.

The four-point function $M^{(4)}$ may be treated along similar lines. Again,
the Dyson-Schwinger equation allows to express $M^{(4)}$ in terms of $\langle
P\ldots \rangle_m^c$ and $\langle S\ldots \rangle_m^c$ 
$n$-point functions (which we show in coordinate space this time),
\bdi
M_{y_1} M_{y_2} M_{y_3} M_{y_4} \langle \Phi (y_1)\Phi (y_2)\Phi (y_3)
\Phi (y_4)\rangle^c =
\edi
\bdi
16\pi^2 \Bigl[ m\langle S(y_1)\rangle 
\delta (y_1 -y_2)\delta (y_1 -y_3)\delta (y_1 -y_4) +
\edi
\bdi
m^2 \delta (y_1 -y_2) \delta (y_3 -y_4) \langle S(y_1)S(y_3)
\rangle^c + \mbox{ perm. } +
\edi
\bdi
m^2 \delta (y_1 -y_2)\delta (y_1 -y_3) \langle P(y_1)P(y_4)\rangle^c
+\mbox{ perm. }+
\edi
\bdi
m^3 \delta (y_1 -y_2)\langle S(y_1)P(y_3)P(y_4)\rangle^c +
\mbox{ perm. } +
\edi
\beq
m^4 \langle P(y_1)P(y_2)P(y_3)P(y_4)\rangle^4 \Bigr]
\eeq

Further, $M^{(4)}$ may be reexpressed in terms of non-factorizable $n$-point
functions and reads
\bdi
M^{(4)}(p_1 ,\ldots ,p_4)=(4\pi )^2 P_iP_jP_kP_l
\Pi_{ii'}(p_1)\Pi_{jj'}(p_2) \Pi_{kk'}(p_3) \Pi_{ll'}(p_4)\Bigl[ {\cal
G}_{i' j' k' l'}
\edi
\bdi 
+ {\cal G}_{i' m}{\cal G}_{j' m'}{\cal G}_{k' n}{\cal G}_{l' n'} R^{\rm
n.f.}_{mm' nn'}(p_1 ,p_2 ,p_3 ,p_4)
\edi
\bdi
+\Bigl( {\cal G}_{i' j' m}(\Pi_{mm'}(p_1 +p_2 )-\delta_{mm'})\delta_{m'
k' l'} +\mbox{ perm. }\Bigr)
\edi
\bdi
+\Bigl( {\cal G}_{i' j' m}\Pi_{mm'}(p_1 +p_2)T^{\rm n.f.}_{m' nn'}
(p_1 +p_2 ,p_3,
p_4){\cal G}_{nk'}{\cal G}_{n' l'} + \mbox{ perm. }\Bigr)
\edi
\beq
+\Bigl( {\cal G}_{i' m}{\cal G}_{j' m'}T^{\rm n.f.}_{mm' n}(p_1 +p_2 ,p_1
,p_2){\cal G}_{nn'}\Pi_{n' r}(p_1 +p_2)T^{\rm n.f.}_{rr' s}(p_3 +p_4 ,p_3
,p_4){\cal G}_{r' k'}{\cal G}_{sl'} + \mbox{ perm. }\Bigr) \Bigr]
\eeq
where momentum conservation requires $p_1 +p_2 =p_3 +p_4$. Graphically, this
identity may be depicted like in Fig. 14 (again we suppress an overall factor
$(2\sqrt{\pi})^4$ in Fig. 14).

$$\psannotate{\psboxscaled{400}{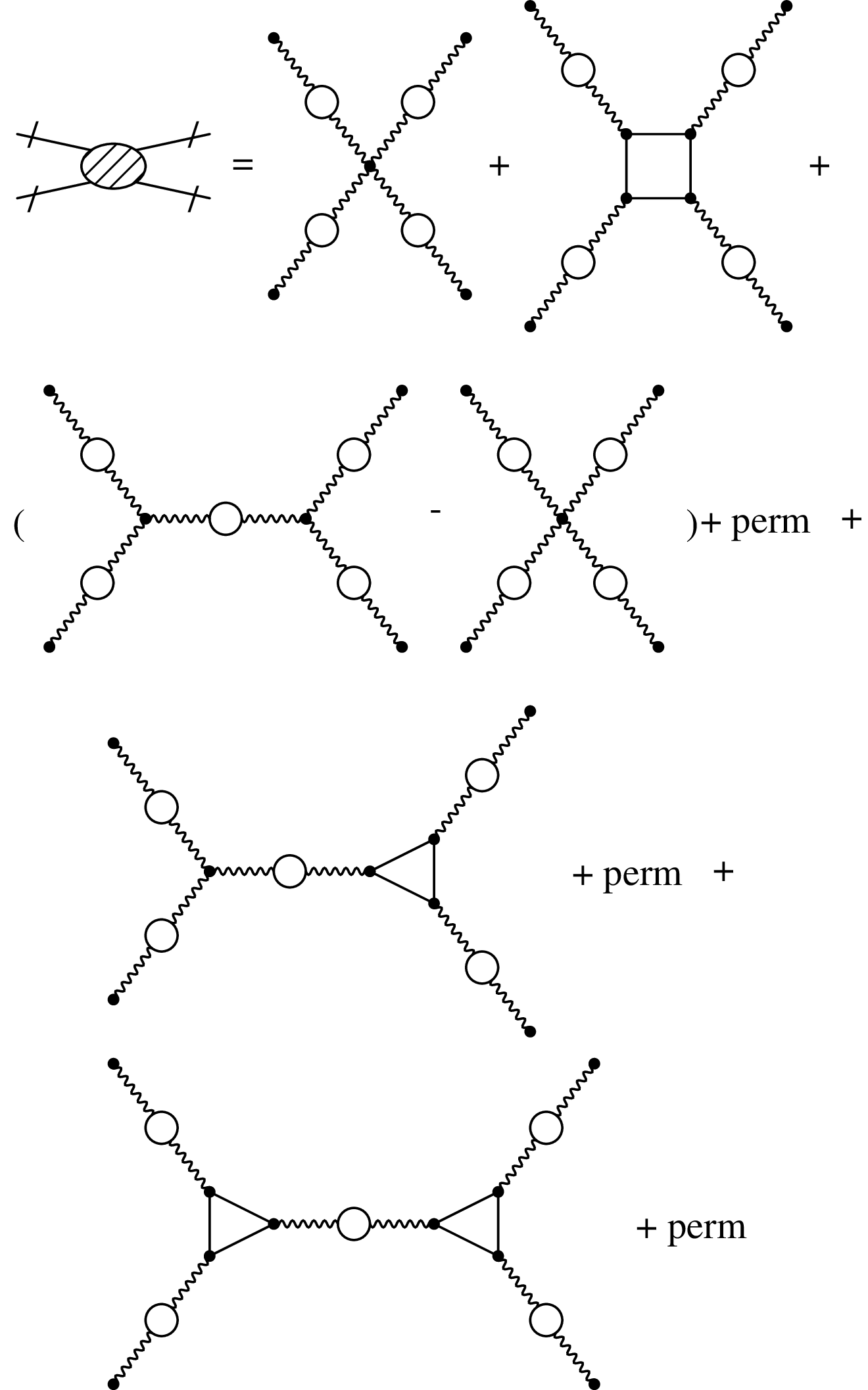}}{\fillinggrid
\at(10.9\pscm;-0.5\pscm){Fig. 14}}$$ 

The permutations in Fig. 14 contain all attachments of the external
$\Pi (p_i)$ lines that are topologically distinct (i.e. 3, 6 and 3
permutations).

Observe that in each of the third type diagrams of Fig. 14 
the lowest order diagram has to
be subtracted in order to avoid an overcounting (this is so because $\Pi (p)$
contains the lowest order, ${\cal G}\Pi (p)={\cal G} +o(g_\theta^2)$).

\section{Bound-state masses and decay widths}

\subsection{General bound-state structure}

We claimed at the end of Subsection 9.1 that we could infer all the boson 
states and decay widths from the two-point function $\Pi (p)$, (117), which we 
want to discuss now.
 First, observe that the $\Pi (p)$ propagator also occurs in higher bosonic
$n$-point functions. E.g. for $M^{(4)}$ (Fig. 14), when one takes the third
type of diagrams and inserts the lowest order ($\Pi \sim {\bf 1}$) for the four
{\em external} $\Pi (p_i)$ lines, there remains precisely an internal $\Pi (p_1
+p_2)$ propagator (times ${\cal G}$). Therefore it is not a surprize that we
can provide information on higher bosonic states, too, from $\Pi (p)$. In fact,
most of the information may be inferred from the denominator $N(p)$, (118), of
$\Pi (p)$. The zeros of the real part of $N(p)$ will give all the bound-state
masses of the theory -- at least the leading order contribution -- 
and the imaginary parts
will give the corresponding decay widths
(\cite{GBOUND,DECAY,THREE}).

This denominator $N(p)$ reads, in lowest order
\beq
N(p)=1-\alpha (g_\theta ,p)-\alpha (g_\theta^* ,p)
\eeq
where, again in lowest order
\beq
\alpha (g_\theta ,p)=g_\theta \wt E_+ (p) \quad ,\quad \beta (g_\theta ,p)=
g_\theta \wt E_- (p) \quad ,\quad g_\theta =m\frac{\Sigma}{2}e^{i\theta}
\eeq
and the $\wt E_\pm (p)$ are the exponentials of bosonic propagators,
\beq
\wt E_\pm (p)=\sum_{n=1}^\infty (\pm 1)^n d_n (p) \quad ,\quad d_n (p):=
\frac{(4\pi)^n}{n!}\wt{D^n_{\mu_0}}(p) .
\eeq
The $d_n (p)$ are just $n$-boson blobs (see Fig. 15 for $d_2$, $d_3$)

$$\psannotate{\psboxscaled{700}{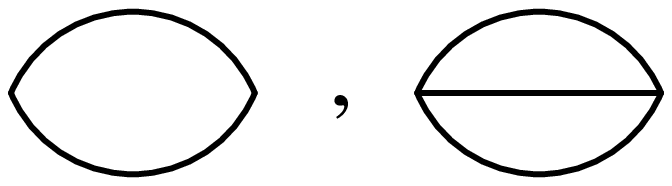}}{\fillinggrid
\at(2.9\pscm;-0.5\pscm){Fig. 15}}$$ 

\vspace{0.5cm}

and have the following properties: at $s=-p^2 =(n\mu)^2$, $d_n (p)$ has a
singularity (real particle production threshold), and above this threshold it
has an imaginary part. Therefore, slightly below the threshold $(n\mu)^2$, $d_n
(p)$ is large enough to balance the coupling constant and make the real part of
$N(p)$, (124), vanish,
\beq
m\Sigma \cos\theta d_n (p)\sim 1+o(m)
\eeq
and, therefore, causes an $n$-boson bound state. At the position of the
two-boson bound state, $s=M_2^2 =4\mu^2 -\Delta_2$, $N(p)$ has no imaginary
part and, therefore, the two-boson bound state is stable. At the three-boson
bound-state mass $M_3$, $d_2 (s=M_3^2)$ has an imaginary part and, therefore, a
decay into two Schwinger bosons (with mass $\mu$) is possible. For higher
$n$-boson bound states the functions $d_2
,\ldots ,d_{n-1}$ have imaginary parts at $M_n^2$,
therefore decays into $2,\ldots ,n-1$ Schwinger bosons are
possible.

So far this is a lowest order reasoning, but we will find that we have to
take into account some higher order effects, too, in order to obtain the
physical spectrum of the theory.

One higher-order effect may be understood easily. Remember that the 
$\alpha (g_\theta ,p)$ of (124) stems from the non-factorizable two-point 
function ${\cal A}$, (114), (in fact, $\alpha$ is the $++$ component of 
${\cal GA}$). To get more insight, let us rewrite ${\cal A}$ in terms of
internal bosons, see Fig. 16.

$$\psannotate{\psboxscaled{700}{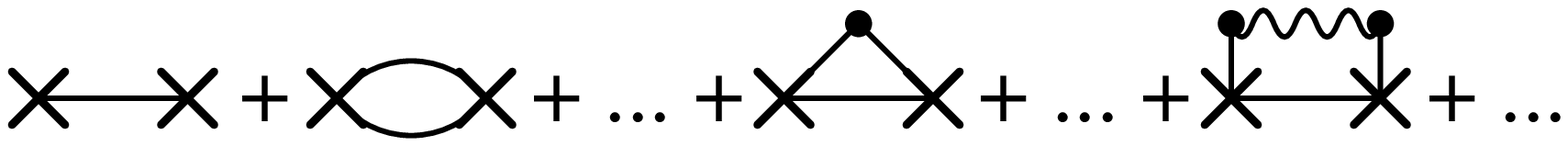}}{\fillinggrid
\at(8\pscm;-0.5\pscm){Fig. 16}}$$

\vspace{0.4cm}

We find that the one-boson line propagating from the initial to the final
external vertex acquires {\em no} correction, because such corrections
would be factorizable and are, therefore, excluded from $\alpha$. 
On the other hand, all the higher $d_n (p)$, $n\ge 2$, do get corrections.

This means that for the computation of the lowest pole mass (the Schwinger
mass) one needs the bare Schwinger mass as an input, and the renormalized 
Schwinger mass is provided by the computation. For the higher bound states, 
on the other hand, one needs the renormalized Schwinger mass as an input
in order to compute the bound-state mass poles.
 The reason is that the mass corrections
for the bosons just shift the position of the threshold singularity and
are therefore important in lowest order. There are other corrections present,
too (internal boson interactions), however, they are unimportant in lowest
order. 

This result is very plausible physically: the higher bound states should
consist of {\em physical} Schwinger bosons with their physical masses $\mu$
(not the bare masses $\mu_0$).

 Therefore we redefine the $d_n (p)$ ($n\ge 2$) for the rest
of the paper to be
\beq
d_n (p):=\frac{(4\pi)^n}{n!}\wt{D^n_\mu}(p) .
\eeq

So we found, up to now, bound states composed of an arbitrary number of 
Schwinger bosons, where $\mu$ and $M_2$ are stable, and the higher bound states
are unstable.

However, this can not yet be the whole story. To understand why, look at the
lowest order contribution to the three-point function, Fig. 12, with one
incoming $\Pi (p_1)$, one vertex and two outgoing $\Pi (p_2)$, $\Pi (p_3)$.
Suppose the incoming $\Pi (p_1)$ is at the mass $-p_1^2 =M_n^2$ of a
sufficiently heavy unstable bound state. For a decay into stable final
particles all the stable mass poles of $\Pi (p_2)$, $\Pi (p_3)$ are possible.
But by our above arguments the mass pole of the stable $M_2$ particle is
present in $\Pi (p_i)$ as well as the mass pole of the Schwinger boson $\mu$.
Therefore, Fig. 12 describes decays into $M_2$ particles as well as $\mu$
particles. On the other hand, we did not find imaginary parts (up to now) in
$N(p)$ that describe decays into some $M_2$, so obviously something is missing.

The $M_2$ bound state itself was found by a resummation, therefore it is a
reasonable idea to use the higher order contributions to $\alpha$, $\beta$ for
a further resummation. $\alpha$ and $\beta$ are just components of the
non-factorizable propagator ${\cal A}(p)$, (114), so let us investigate it more
closely.

By a partial resummation we may find the following contribution to ${\cal
A}(p)$,
\beq
H_{ii'}(p):=\int\frac{d^2 q}{(2\pi)^2} \delta_{ijk}{\cal A}_{jj'}(q){\cal
G}_{j' l}\Pi_{ll'}(q){\cal A}_{l' k'}(q){\cal A}_{km}(q-p) \delta_{i' k'
m}.
\eeq
This is just a blob where ${\cal A}(q-p)$ runs along one line, the other terms
run along the other line. We want to discuss the $\mu$-$M_2$ contribution,
therefore we substitute ${\cal A}(q-p)$ by its lowest order, one-boson part,
\beq
{\cal A}(q-p)\sim 4\pi \wt D_\mu (q-p)
\left( \begin{array}{cc} 1 & -1 \\ -1
& 1 \end{array} \right) 
\eeq
and the two further ${\cal A}(q)$ by their lowest order contribution
\beq
 {\cal A}(q)\sim 
\left( \begin{array}{cc} \wt E_+ (q) & \wt E_- (q) \\ \wt E_- (q)
& \wt E_+ (q) \end{array} \right) 
\eeq
The resummation that we need in (129) is taken into account by $\Pi (q)$. 
With these restrictions $H(p)$ corresponds to the graph of Fig. 17.

$$\psannotate{\psboxscaled{600}{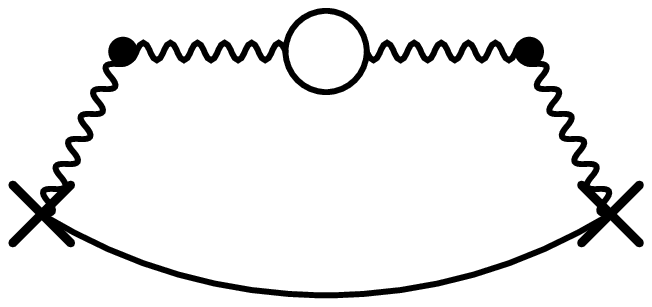}}{\fillinggrid
\at(2.9\pscm;-0.7\pscm){Fig. 17}}$$ 

\vspace{0.6cm}

Observe that all internal bosons may be renormalized to their physical masses
$\mu$, because this does not spoil non-factorizability in Fig. 17. The two
factors ${\cal A}(q)$ in (129) are necessary in order to avoid an overcounting,
but they cannot influence the presence of higher poles in Fig. 17.

Now suppose that $H(p)$ is at the $M_2 +\mu$-threshold, $s=-p^2=(M_2 +\mu)^2$.
Then $\wt D_\mu (q-p)$ is at its $\mu$-singularity and $\Pi (q)$ at its
$M_2$-singularity, and Fig. 17 corresponds (up to a normalization) to a
$\mu$-$M_2$ two-boson loop, i.e. Fig. 17 may effectively be substituted by 
Fig. 18,

$$\psannotate{\psboxscaled{700}{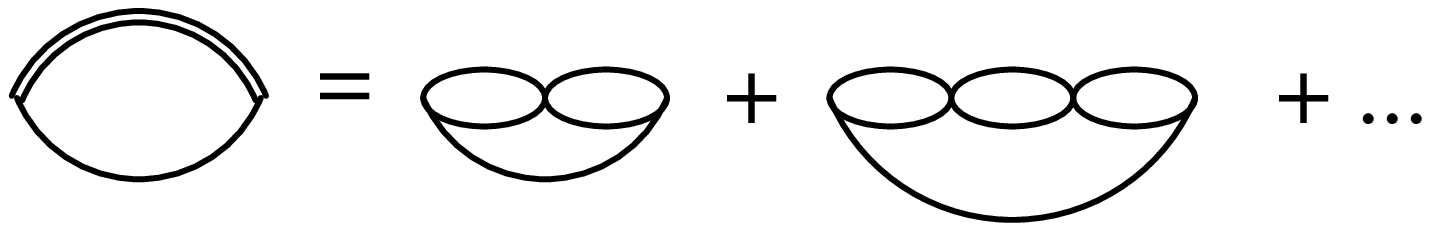}}{\fillinggrid
\at(6.9\pscm;-0.5\pscm){Fig. 18}}$$ 

\vspace{0.4cm}

where the double line represents the two-boson bound-state propagator.

Therefore, $H(p)$ is singular at $-p^2=(M_2 +\mu)^2$, and has a large real part
slightly below and a large imaginary part slightly above this threshold. As a
consequence, when the contribution of $H(p)$ to $\alpha (p)$ is
taken into account in the denominator $N(p)$, (118,124), it will give rise to a
further $\mu $-$M_2$ bound state slightly below $s=(M_2 +\mu)^2$. Further it
will open the $\mu$-$M_2$ decay channel at $s=(M_2 +\mu)^2$.

Now suppose we put $\Pi (q)$ in (129) on a higher (unstable) bound-state mass
$M_n$, $n>2$.
Then in the denominator $N(q)$ of $\Pi (q)$ the real part again
vanishes, but there remains an imaginary part. Therefore, $H(p)$ is finite and
imaginary at $s=-p^2 =(M_n +\mu)^2$ and {\em cannot} give rise to a $\mu$-$M_n$
bound-state formation.

Further, because there is no threshold singularity at $s=(M_n +\mu)^2$, this
means that {\em no} new decay channel opens at that point (i.e. the imaginary
part of $H(p)$ varies smoothly around $s\sim (M_n +\mu^2$)), which simply means
that the unstable higher $n$-boson bound states are no possible final states
(of course, they are possible as intermediate resonances).

We could substitute the one-boson line in Fig. 17 by another ${\cal A}{\cal
G}\Pi {\cal A}$ line and would find that this graph behaves like a $M_2$-$M_2$
blob near $s=(M_2 +M_2)^2$, and we could allow for even more ${\cal A}{\cal G}
\Pi {\cal A}$ lines. The physical picture that evolves from these
considerations is like follows: in addition to the unstable $n$-boson bound
states there exist further (unstable) bound states that are composed of
Schwinger bosons $\mu$ and (stable) two-boson bound states $M_2$. 
Further, the
unstable bound states may decay into all combinations of $\mu$ and $M_2$
particles that are possible kinematically. The imaginary parts of the
corresponding $n$-particle blobs (where particle means $\mu$ or $M_2$) are
large near their thresholds, therefore there is a kinematical tendency to
rise the decay probabilities for decays with {\em small} kinetic energy. 
This is not so
surprizing, because in $1+1$ dimensions the phase space "volume" does not grow
with kinetic energy.

Further we want to emphasize the following point for later convenience. We
found, by a further resummation, contributions to $N(p)$ that may be 
substituted by a $n$-particle blob that contains $\mu$ and $M_2$ bosons
(or only $M_2$), {\em near their respective $n$-particle thresholds}.
This is true for the {\em real} parts. The {\em imaginary} parts are given 
precisely by the threshold singularities which stem solely from the
$n$-particle blobs. Therefore, the imaginary parts of these resummed 
contributions may be substituted by the corresponding imaginary parts
of the $n$-particle blobs for arbitrary values of $p$.

Finally let us briefly comment on the special case $\theta =0$. Here parity is
conserved and we have to investigate scalar ($S{\cal G}\Pi (p)S$) and 
pseudoscalar
($P{\cal G}\Pi (p)P$) propagators separatly. 
We find a partial cancellation between
numerator and denominator in (117) ($\alpha (g_{\theta =0},p)\equiv \alpha
(g^*_{\theta =0},p)$, etc.)
\bdi
S{\cal G}\Pi (p)S=
\frac{m\Sigma}{1-\alpha (g_{\theta =0},p)-\beta (g_{\theta =0},p)}
\edi
\beq
 P{\cal G}\Pi (p)P=
\frac{m\Sigma}{1-\alpha (g_{\theta =0},p)+\beta (g_{\theta =0},p)}
\eeq
and, therefore, using the lowest order expression (125), we find that the odd
(even) $n$-boson blobs $d_n (p)$ are cancelled in $S{\cal G}\Pi (p)S$ 
($P{\cal G}\Pi (p)P$),
so that only even (odd) mass poles remain. Of course, these parity
considerations may be generalized easily to the mixed bound states (where
each $\mu$ is odd and each $M_2$ is even).

\subsection{$n$-boson bound-state masses}

The mass pole equation of the $n$-boson bound state in lowest order is given 
by (see (127))
\beq
f_n (p):=1-m\Sigma\cos\theta d_n (p) =0
\eeq
For the Schwinger boson ($n=1$) this equation reads
\beq 
1=m\Sigma\cos\theta\frac{-4\pi}{p^2 +\mu_0^2}
\eeq
with the solution
\beq
-p^2 =:\mu_1^2 =\mu_0^2 +4\pi m\Sigma\cos\theta =:\mu_0^2 +\Delta_1
\eeq
(we will denote the lowest order corrections to all the pole masses $M_n^2$
by $\Delta_n$). 

When we want to recover the second order result for the Schwinger mass
of Section 8 we have to 
include all terms of $N(p)$, (118). The mass pole equation up to second order
reads
\bdi
1=\frac{-4\pi}{p^2 +\mu_0^2}(g_1 +g_1^* +g_2 +g_2^*) + (g_1 +g_1^*)\wt
E_+^{(1)}(p)-
\edi
\beq
g_1 g_1^* \Bigl[ (\wt E_+^{(1)}(p)+\frac{-4\pi}{p^2 -\mu_0^2})^2 -
(\wt E_-^{(1)}(p)-\frac{-4\pi}{p^2 +\mu_0^2})^2 \Bigr]
\eeq
where $g_1$ ($g_2$)  are the first (second) order contributions to $g_\theta$ 
(see (89)) and $\wt E_\pm^{(1)}(p)$ are the exponentials $\wt E_\pm (p)$
without the one-boson term, see (52). In (136), the $g_1 g_1^* 
(\wt E_\pm^{(1)}(p))^2$ parts are, in fact, $o(m^3)$, and may be omitted. 
The solution to (136) is 
\beq
\mu_2^2 := -p^2 =
\mu_0^2 \Bigl[ 1+4\pi \frac{\Sigma m}{\mu_0^2}\cos\theta +2\pi \frac{
m^2 \Sigma^2}{\mu_0^4}\Bigl( ( E_+ + \widetilde E_+^{(1)} (1))\cos 2\theta +
E_- -\widetilde E_-^{(1)} (1)\Bigr) \Bigr]
\eeq
and, indeed, coincides with (96) of Section 8 that was obtained by a direct 
perturbative calculation. 

For the two-boson bound state mass in lowest order we have to solve
\beq
1=\frac{1}{2!}(g_1 +g_1^* )16\pi^2 \widetilde{(D_\mu^2 )} (p)
\eeq
where now $\mu$ is the {\em physical} Schwinger mass (137) including 
fermion mass corrections. $(\widetilde D_\mu^2 ) (p)$ is just the two-boson
blob of Fig. 15
and may be evaluated by standard methods:
\bdi
(\widetilde D_\mu^2 ) (p)=\int\frac{d^2 q}{(2\pi)^2}\frac{-1}{q^2 +\mu^2}
\frac{-1}{(q-p)^2
+\mu^2}=
\edi
\bdi
\int\frac{d^2 q}{(2\pi)^2}\int_0^1 \frac{dx}{[q^2 +2pq(x-1)
+p^2 (1-x) +\mu^2 ]^2}
\edi
\bdi
=\frac{1}{4\pi}\int_0^1 \frac{dx}{p^2 x(1-x)+\mu^2} =\frac{1}{\pi(-p^2)}
\int_0^1\frac{dy}{y^2 +(\frac{4\mu^2}{-p^2} -1)}
\edi
\beq
= \frac{1}{\pi (-p^2)}\frac{1}{R(p^2 )}\arctan\frac{1}{R(p^2 )} ,
\eeq
\beq
R(p^2 ):=\sqrt{\frac{4\mu^2}{-p^2} -1}
\eeq
where we used the fact that, for the bound state, $-p^2$ has to be beyond 
the threshold, $-p^2 <4\mu^2$.

Now we simply have to insert this result into (138) in order to get the mass 
pole:
\beq
1=\frac{8\pi m\Sigma\cos\theta}{(-p^2)}
\frac{1}{R(p^2)}\arctan\frac{1}{R(p^2)} .
\eeq
For small fermion mass $m$ $R(p^2)$ is very small, too. Therefore,
the leading order result will stem from a matching between these two factors,
where we may set $\frac{1}{-p^2}=\frac{1}{4\mu^2}$ and $\arctan\frac{1}{R(p^2)}
=\frac{\pi}{2}$. Doing so we get
\beq
R(p^2)=\frac{\pi^2 m\Sigma\cos\theta}{\mu^2}
\eeq
or
\beq
M_2^2 :=4\mu^2 \frac{1}{1+(\frac{\pi^2 m\Sigma\cos\theta}{\mu^2})^2} 
\simeq 4\mu^2 (1-\frac{\pi^4 m^2 \Sigma^2 \cos^2 \theta}{\mu^4})=:
4\mu^2 -\Delta_2 ,
\eeq
\beq
M_2^2 =4\mu^2 (1-7.83 \frac{m^2}{\mu_0^2}\cos^2 
\theta +o(\frac{m^3}{\mu_0^3}))
\eeq
which is of second order in $m$. Again, this result coincides with the one
from a direct perturbative calculation (\cite{BOUND}).

For the computation of the 
three-boson bound-state mass $M_3$ we need the three-boson propagator $d_3$ 
and find, in lowest order (see (133))
\beq
1=\frac{1}{3!}m\Sigma \cos \theta \cdot 64\pi^3 \widetilde{D_\mu^3 } (p)
\eeq
or, after a rescaling $p\ra \frac{p}{\mu}$ to dimensionless momenta
\beq
1=\frac{64\pi^3}{6}\frac{m\Sigma}{\mu^2}\cos\theta
\, \widetilde{D_\mu^3 } (p).
\eeq
$\widetilde{D_\mu^3 } (p)$ is given by the three-boson loop of Fig. 15
 (in the sequel we introduce positive squared momentum $s =-p^2 > 0$)

\bdi
\widetilde{D_\mu^3 } (p) =
-\int \frac{d^2 q_1 d^2 q_2}{(2\pi)^4}\frac{1}{(p+q_1 +q_2)^2 +1}
\frac{1}{q_1^2 +1}\frac{1}{q_2^2 +1}=
\edi
\bdi
-2\int_0^1 dx\int_0^x dy\int  \frac{d^2 q_1 d^2 q_2}{(2\pi)^4}
\frac{1}{\Big[ q_1^2 +1+(q_2^2 -q_1^2)x + ((p+q_1 +q_2 )^2 -q_2^2)y\Bigr]^3}=
\edi
\bdi
\int \frac{dx}{(4\pi)^2}\int_0^x \frac{dy}{s (xy-x^2 y-y^2 +xy^2 )-x+x^2 -xy 
+y^2} =
\edi
\bdi
\int \frac{dx}{8\pi^2 (1-s (1-x))}\int_0^{\frac{x}{2}}
\frac{dz}{z^2 +T^2 (s ,x)} =
\edi
\beq
\int_0^1 \frac{dx}{8\pi^2 (1-s (1-x))}\frac{1}{T(s ,x)}
\arctan \frac{x}{2T(s ,x)} ,
\eeq
\beq
T^2 (s ,x)=\frac{x^2 -s x^2 (1-x)+4x(1-x)}{4(s (1-x)-1)} ,
\eeq
where, as usual, we introduced Feynman parameter integrals and performed the
momentum integrations. Further, the first Feynman parameter integral could be
done analytically. 
The numerator of $T^2$ has a double zero at $s =9$:
\beq
9x(x-\frac{2}{3})^2 .
\eeq
This double zero is in the integration range of $x$ and is precisely the 
threshold singularity. Setting
\beq
s=9-\Delta_3
\eeq
in the numerator of $T^2$ in the factor $\frac{1}{T}$, and $s =9$ everywhere
else, where it is safe, one arrives at:
\beq
\frac{1}{12\pi^2}\int_0^1 \frac{dx}{\sqrt{\vert 9x-8\vert }}
\frac{\arctan \frac{\sqrt{x\vert 9x-8\vert }}{3(x-\frac{2}{3})}}{
\sqrt{(x-\frac{2}{3})^2 x+\frac{\Delta_3}{9} x^2 (1-x)}}=:I(\Delta_3 ).
\eeq
The mass-pole equation reads 
\beq
1=\frac{64\pi^3}{6}m\Sigma\cos\theta I(\Delta_3 )
\eeq
and must be evaluated numerically. It gives rise to an extremely tiny mass 
correction $\Delta_3$. For sufficiently small $m$ it is very well 
saturated by
\beq
\Delta_3 (m\Sigma\cos\theta )
\simeq 6.993 \exp (-\frac{0.263}{m\Sigma\cos\theta})
\eeq
and is therefore smaller than polynomial in $m$. (I checked the 
numerical formula (153) for $30 <\frac{1}{m\Sigma\cos\theta}<1000$, 
corresponding to
$10^{-2}<\Delta_3 <10^{-100}$, but I am convinced that it remains true for
even larger $\frac{1}{m\Sigma\cos\theta}$; 
however, there the numerical integration
is quite difficult because of the pole in (151).) 

We conclude that the three-boson bound state mass is nearly entirely given by 
three times the Schwinger boson mass 
(we change back to dimensionfull quantities now),
\beq
M_3^2 = 9\mu^2 -\Delta_3 \quad ,\quad \Delta_3 =6.993\mu^2 \exp
(-0.263\frac{\mu^2}{m\Sigma\cos\theta})
\eeq
or, differently stated, that the binding of three bosons is extremely weak.

Therefore it holds that $M_3 >\mu +M_2$, and, 
consequently, a decay of
$M_3$ into $\mu +M_2$ is possible. This has the consequence that the
three-boson bound state is unstable even for $\theta =0$.

\subsection{A mixed bound-state mass}

For our further discussion we will need the residues of the propagator
$\Pi (p)$ at its various mass poles. The $n$-boson mass poles were, in
leading order, the zeros of the functions $f_n (p)$, see (153). Therefore,
we need the first Taylor coefficient $c_n$ of each $f_n (p)$ at its mass pole
$M_n^2 =(n\mu)^2 -\Delta_n$ ($n\ge 2$) or $M_1^2 =\mu_0^2 +\Delta_1$,
\beq
f_n (s)\simeq c_n (s-M_n^2 )\quad ,\quad c_n =\frac{d}{ds}f_n (s)|_{s=M_n^2}
\eeq
where $s=-p^2$. The $c_n$ may be easily obtained from our mass computations. 
From (134) and (141) we find for $c_1$ and $c_2$
\beq
c_1 =\frac{1}{4\pi m\Sigma\cos\theta}=\frac{1}{\Delta_1}
\eeq
\beq
c_2 =\frac{\mu^2}{8\pi^4 (m\Sigma\cos\theta)^2}=\frac{1}{2\Delta_2}
\eeq
For the computation of $c_3$ we observe
 that because of formulae (150), (153)
$m\Sigma\cos\theta d_3 (s)$ may be written, in the vicinity of $s=M_3^2$, like
\beq
m\Sigma\cos\theta d_3 (s)\sim \frac{m\Sigma\cos\theta}{0.263}\ln\frac{6.993
\mu^2}{9\mu^2 -s}.
\eeq
Therefore, we find the Taylor coefficient
\beq
c_3 =\frac{m\Sigma\cos\theta}{0.263 \Delta_3}.
\eeq
Further we need, for the computation of the lowest mixed bound-state mass
$M_{1,1}$ (which is composed of one $M_2$ and one $\mu \equiv M_1$),
the residues of the propagator $\Pi (p)$ at the two lowest mass poles.
The denominator $N(p)$, (118), is given by (155), and for the numerator of
$\Pi (p)$, (117), we use  $\alpha (g_\theta ,s\sim M_n^2)\sim g_\theta
\wt E_+ (s\sim M_n^2)\sim\frac{g_\theta}{g_\theta +g^*_\theta}$, etc.,
which holds near the pole, see (124) -- (127), and find (here $n=1,2$)
\beq
\Pi (s\sim M_n^2 ) \sim \frac{1}{(g_\theta +g^*_\theta )c_n
(s-M_n^2 )} \left( \begin{array}{cc} g_\theta & (-1)^n g_\theta^* 
\\ (-1)^n g_\theta & g_\theta^* \end{array} \right) 
\eeq
For the computation of the $\mu$-$M_2$ bound state we need, in addition, the
matrix ${\cal A}$, (113), at the $n$-boson mass poles (here $n\ne 1$, because
there ${\cal A}$ itself has a pole like (160)),
\beq
{\cal A}(s=M_n^2 )=\frac{1}{g_\theta +g^*_\theta } \left( \begin{array}{cc} 1 &
(-1)^n \\ (-1)^n & 1 \end{array} \right) 
\eeq
Now we are prepared for the computation of the $M_{1,1}$ mixed bound-state
mass. In Subsection 10.1 we claimed that we could substitute the resummed
contribution $H(p)$ to $N(p)$, see (129) and Fig. 17, near its threshold by a
two-particle blob consisting of one $\mu$ and one $M_2$ (times a not yet
specified factor), see Fig. 18. This we achieve by inserting expression (160) 
for $\Pi (p)$ near its two poles $\mu$, $M_2$, into (129), and by using (161) 
for the remaining ${\cal A}(-q^2 \sim M_2^2)$.
Althogether we find for $H(p)$ near its threshold
\bdi
H_{ii'}(-p^2 \sim (M_2 +\mu)^2)\sim \int\frac{d^2 q}{(2\pi)^2} \delta_{ijk}
\frac{1}{g_\theta +g^*_\theta } \left( \begin{array}{cc} 1 & 1 \\ 1 & 1
\end{array} \right)_{jj'} \left( \begin{array}{cc} g_\theta & 0 \\ 0 &
g^*_\theta \end{array} \right)_{j' l} \frac{1}{(g_\theta +g^*_\theta)c_2 (-q^2
-M_2^2 )} \cdot
\edi
\bdi
\cdot \left( \begin{array}{cc} g_\theta & g_\theta^* \\ 
g_\theta & g_\theta^* \end{array} \right)_{ll'}
\frac{1}{g_\theta +g^*_\theta } \left( \begin{array}{cc} 1 & 1 \\ 1 & 1
\end{array} \right)_{l' k'} \frac{1}{(g_\theta +g^*_\theta )c_1 (-(q-p)^2
-\mu^2 )} \left( \begin{array}{cc} 1 & -1 \\ -1 & 1 \end{array} \right)_{km}
\delta_{i' k' m} =
\edi
\beq
\delta_{ijk} \left( \begin{array}{cc} 1 & 1 \\ 1 & 1 \end{array} \right)_{jj'}
\left( \begin{array}{cc} 1 & -1 \\ -1 & 1 \end{array} \right)_{kk'} \delta_{i'
j' k'} \int\frac{d^2 q}{(2\pi)^2} \frac{1}{(g_\theta +g^*_\theta )^2 c_1 c_2
(-q^2 -M_2^2 )(-(q-p)^2 -\mu^2 )}
\eeq
The contribution of $H_{ii'}$ to $\alpha (g_\theta) +\alpha (g^*_\theta )$ in
the denominator $N(p)$, (124), is
\bdi
g_\theta H_{++} (p) +g^*_\theta H_{--} (p) =: (g_\theta +g^*_\theta )d_{1,1}
(p)=
\edi
\bdi
(g_\theta +g^*_\theta )\int\frac{d^2 q}{(2\pi)^2}\frac{8\pi^4
m\Sigma\cos\theta}{\mu^2 (q^2 +M_2^2 )} \frac{4\pi}{(p-q)^2 +\mu^2 }=
\edi
\bdi
 \frac{32\pi^5 m^2 \Sigma^2 \cos^2 \theta}{2\pi \mu^2\bar w(s ,M_2^2
,\mu^2 )}\Bigl( \pi +
\edi
\beq
\arctan \frac{2s}{\bar w(s ,M_2^2 ,\mu^2 ) 
-\frac{1}{\bar w(s ,M_2^2 ,\mu^2 )}(s +\mu^2 -M_2^2)(s -\mu^2 +M_2^2 )}
\Bigr)
\eeq
\beq
\bar w(x,y,z):=(-x^2 -y^2 -z^2 +2xy+2xz+2yz)^{\frac{1}{2}}
\eeq
where $s=-p^2 \sim (\mu +M_2)^2 $. The $\mu$-$M_2$ bound-state mass fulfills
the equation 
\beq
1=(g_\theta +g^*_\theta )d_{1,1}(p)
\eeq
with the solution in leading order (here $M_{1,1}$ denotes the $\mu$-$M_2$
bound-state mass)
 \beq
M_{1,1}^2 =(\mu +M_2)^2 -\Delta_{1,1}\quad ,\quad \Delta_{1,1}=
\frac{32\pi^{10}(m\Sigma\cos\theta)^4}{\mu^6}
\eeq
which is valid for sufficiently small $\Delta_{1,1}$. $M_{1,1}$ was computed
from a two-boson blob (Fig. 18), like $M_2$, 
therefore the Taylor coefficient of
$(s-M_{1,1}^2 )$ is analogous to $c_2$, equ. (157),
\beq
c_{1,1} =\frac{1}{2\Delta_{1,1}}=\frac{\mu^6}{64\pi^{10}(m\Sigma\cos\theta)^4}.
\eeq
Further, the above equ.
(162) shows that the $\mu$-$M_2$ blob $d_{1,1}(p)$ enters into the functions
$\alpha$, $\beta$ of ${\cal A}$, (113), like any other odd $n$-boson blob $d_n
(p)$.

In principle, even higher mixed bound-state masses could be computed along
similar lines, but we want to change now to the computation of the decay
widths of the lowest unstable bound states.

\subsection{Decay width computations}

In order to find the decay widths of some bound states, we have to examine
the imaginary parts of the denominator $N(p)$ in the vicinity of the
corresponding mass poles.
 Using the first order approximation (125) 
for $\alpha$,
$\beta$ we find for $N(p)$ ( e.g. in the vicinity of $s\sim M_3^2$ for
definiteness)
\beqa
N(p) & \simeq & 1-m\Sigma \cos\theta\wt E_+ (p) +\frac{m^2 \Sigma^2}{4} (\wt
E_+^2 (p) -\wt E_-^2 (p)) \no \\
& = & 1-m\Sigma\cos\theta (d_1 (p) +d_2 (p) +d_{1,1} (p) +d_3 (p) +\ldots )+ 
\no \\
&& m^2 \Sigma^2 \Bigl( d_1 (p)(d_2 (p) +d_4 (p) +\ldots ) +d_{1,1}(p) (d_2 (p)
+ d_4 (p) +\ldots ) + \no \\ 
&& d_3 (p) (d_2 (p) +d_4 (p) +\ldots )+\ldots \Bigr)
\eeqa
where we included the $\mu$-$M_2$ blob $d_{1,1}$, as discussed above, because
we need it for the subsequent discussion (we ignore, for the moment, 
higher $M_2$
blobs that are, in principle, present). Near $s=M_3^2$ the real part of (168) 
is given by $c_3 (s-M_3^2 )$ and we find
\beqa
N(s\sim M_3^2) & \sim & c_3 (s-M_3^2 ) 
-im\Sigma\cos\theta ({\rm Im} d_2 (s\sim M_3^2 )
+{\rm Im} d_{1,1}(s\sim M_3^2 )) + \no \\
&& im^2 \Sigma^2 d_3 (s\sim M_3^2 ){\rm Im} d_2
(s\sim M_3^2 ) +o(m^2) \no \\
&=& c_3 (s-M_3^2 )-im\Sigma (\cos\theta -\frac{1}{\cos\theta}){\rm Im} d_2
(M_3^2 ) -im\Sigma\cos\theta \, {\rm Im} d_{1,1}(M_3^2 ) +o(m^2) \no \\
&&
\eeqa
where we used $d_3 (M_3^2)\sim \frac{1}{m\Sigma\cos\theta}$, see (133).

This computation may be generalized and tells us that parity forbidden
imaginary parts (decay channels) acquire a factor $(\cos\theta
-\frac{1}{\cos\theta})$, whereas parity allowed imaginary parts have the usual
$\cos\theta$ factor.

[{\em Remark}:
There seems to be something wrong with the sign of the parity forbidden 
imaginary part (the $d_2$ term). Actually the sign is o.k.
and the problem is a remnant of the Euclidean conventions that are implicit in
the whole computation (see Section 1). In these conventions $\theta$
is imaginary and therefore $(\cos\theta -\frac{1}{\cos\theta}) \ge 0$. 
Of course,
this is not a reasonable convention for a final result. When performing the
whole computation in Minkowski space and for real $\theta$, roughly speaking,
the roles of $E_+$ and $E_-$ are exchanged in (168). This gives an additional
relative sign between parity even and odd $n$-boson propagators and, therefore,
changes the factor of $d_2$ to $(\frac{1}{\cos\theta} -\cos\theta )$, which is
$\ge 0$ for real $\theta$.
We will keep this remark in mind and express the final results in Minkowski
space and for real $\theta$.] 

Now we may find the $M_3$ decay widths by comparing the inverse of (169) to
the general formula
\beq
G(p)\sim \frac{{\rm const.}}{s-M^2 -iM\Gamma} ,
\eeq
where $\Gamma$ is the decay width.
We find (for real $\theta$)
\beq
\frac{1}{N(s\sim M_3^2)}
\simeq \frac{{\rm const.}}{s-M_3^2 -i\frac{m\Sigma}{c_3}\Bigl[ (
\frac{1}{\cos\theta}-\cos\theta ){\rm Im \,}d_2 (M_3^2) + \cos\theta {\rm Im
\,}d_{1,1}(M_3^2)\Bigr] } .
\eeq
Next we need the imaginary parts ${\rm Im \,}d_2$, ${\rm Im \,}d_{1,1}$. Both
of them stem from a two-boson blob, so let us write down the general result
(which is standard) ($s=-p^2$)
\beqa
{\rm Im \,}\wt{(D_{M_1}D_{M_2})}(s)&=& {\rm Im \,}\int \frac{d^2 q}{(2\pi)^2}
\frac{-1}{q^2 +M_1^2}\frac{-1}{(p-q)^2 +M_2^2} \no \\
&=& \frac{1}{2w(s,M_1^2 ,M_2^2 )},
\eeqa
\beq
w(x,y,z)=(x^2 +y^2 +z^2 -2xy-2xz-2yz)^\frac{1}{2} .
\eeq
Therefore we can write for (171) (where the normalization factors
of $d_2$ and $d_{1,1}$ are $\frac{c_1^2}{2}$ and $c_1 c_2$, respectively)
\beq
\frac{{\rm const.}}{s-M_3^2 -i\frac{m\Sigma}{c_3}(
\frac{1}{\cos\theta}-\cos\theta )\frac{4\pi^2}{w(M_3^2 ,\mu^2 ,\mu^2 )}
-i\frac{(m\Sigma\cos\theta )^2}{c_3}\frac{16\pi^5}{\mu^2 w(M_3^2 ,\mu^2 ,M_2^2
)}}
\eeq
and therefore, by using the approximations
\beq
w(M_3^2 ,\mu^2 ,\mu^2 )\simeq w(9\mu^2 ,\mu^2 ,\mu^2 )=3\sqrt{5}\mu^2
\eeq
\beq
w(M_3^2 ,M_2^2 ,\mu^2 )\simeq w(9\mu^2 ,M_2^2 ,\mu^2 )=2\sqrt{3}\mu
\sqrt{\Delta_3} +o(m^2),
\eeq
we find the following results:
\beqa
\Gamma_{M_3 \ra 2\mu}&=&
0.263\frac{4\pi^2\Delta_3}{9\sqrt{5}\mu}(\frac{1}{\cos^2 \theta}-1) \no \\
&\simeq & 3.608 \mu (\frac{1}{\cos^2 \theta}-1)\exp
(-0.929\frac{\mu}{m\cos\theta})
\eeqa
and
\beqa
\Gamma_{M_3 \ra \mu +M_2}&=& 0.263\frac{4\pi^3 \Delta_3}{3\sqrt{3}\mu} \no \\
&\simeq & 43.9 \mu \exp (-0.929\frac{\mu}{m\cos\theta})
\eeqa
where we inserted the numerical value $\Sigma =\frac{e^\gamma \mu}{2\pi} =0.283
\mu$. 

The ratio of the two partial decay widths does not depend on the approximations
that were used for the $M_3$ computation,
\beq
\frac{\Gamma_{M_3 \ra 2\mu}}{\Gamma_{M_3 \ra \mu +M_2}}= \frac{\frac{1}{\cos^2
\theta}-1}{\sqrt{15}\pi}.
\eeq
Analogously we may compute the decay width of the mixed bound state $M_{1,1}$,
starting from
\beq
N(s\sim M_{1,1}^2)\simeq c_{1,1}(s-M_{1,1}^2)-im\Sigma (\frac{1}{\cos\theta} 
-\cos\theta ){\rm Im} d_2 (M_{1,1}^2)
\eeq
which leads to the decay width 
\beq
\Gamma_{M_{1,1}\to 2\mu}=
\frac{2^8 \pi^{12} (m\Sigma\cos\theta)^5}{9\sqrt{5}\mu^9}
(\frac{1}{\cos^2 \theta} -1) \simeq 21340 \mu(\frac{m\cos\theta}{\mu})^5
(\frac{1}{\cos^2 \theta} -1)
\eeq
for the decay $M_{1,1}\ra 2\mu$. This decay is parity forbidden, and therefore
$M_{1,1}$ is stable for $\theta =0$.

In principle, we could have computed the above decay widths by another
method, too, namely by the use of the resummed three-point function of
Fig. 12. Choosing the first graph on the r.h.s. of Fig. 12 (consisting
of three exact propagators and one pure vertex), we could precisely
rederive our results (177), (178), and (181).

\section{Scattering}

\subsection{Two-dimensional kinematics}

For a discussion of scattering processes 
we need some basic facts about two-dimensional kinematics. We will
restrict our discussion to elastic scattering. Suppose we have two incoming
particles with masses $M_1$, $M_2$ and momenta $p_1$, $p_2$, and two outgoing
particles, again with masses $M_1$,  $M_2$, and with momenta $p_3$, $p_4$.
Momentum conservation requires 
\beq
p:=p_1 +p_2 =p_3 +p_4
\eeq
and all momenta are {\em Minkowskian} in the sequel. In the center of mass
system we may write 
\bdi
p_1 ={\sqrt{k^2 +M_1^2} \choose k} \quad ,\quad p_2 ={\sqrt{k^2 +M_2^2} \choose
-k}
\edi
\beq
p_3 ={\sqrt{k^2 +M_1^2} \choose \pm k} \quad ,\quad p_4 ={\sqrt{k^2 +M_2^2} 
\choose \mp k}
\eeq
where in $p_3$, $p_4$ the first sign is for transmission, the second sign is
for reflexion. For the kinematical variables we find for transmission
\beqa
s=(p_1 +p_2 )^2 &=& 2k^2 +M_1^2 +M_2^2 +2\sqrt{(k^2 +M_1^2)(k^2 +M_2^2)} \no \\
t_T =(p_1 -p_4 )^2 &=& -2k^2 +M_1^2 +M_2^2 -
2\sqrt{(k^2 +M_1^2)(k^2 +M_2^2)} \no \\
u_T =(p_1 -p_3)^2 &=&0
\eeqa
and for reflexion
\beqa
s=(p_1 +p_2 )^2 &=& 2k^2 +M_1^2 +M_2^2 +2\sqrt{(k^2 +M_1^2)(k^2 +M_2^2)} \no \\
t_R =(p_1 -p_4 )^2 &=& 2k^2 +M_1^2 +M_2^2 -
2\sqrt{(k^2 +M_1^2)(k^2 +M_2^2)} \no \\
u_R =(p_1 -p_3)^2 &=& -4k^2
\eeqa
When the two masses are equal, the two particles are identical in our theory
and the discrimination between transmission and reflexion does not make sense.
The kinematical variables turn into
\beqa
s=(p_1 +p_2 )^2 &=& 4(k^2 +M^2) \no \\
t =(p_1 -p_4 )^2 &=& -4k^2 \no \\
u =(p_1 -p_3)^2 &=& 0
\eeqa
The elastic scattering cross section of two particles is given by
\beq
\sigma_{M_a M_b \ra M_a M_b}(s)=\frac{C_{\rm sym}|{\cal M}(s)|^2}{2w^2 (s,M_a^2
,M_b^2)}
\eeq
where $w$ is defined in (173), 
${\cal M}$ is the transition matrix element and $C_{\rm sym}$ is a
symmetry factor that takes into account identical particles in the final state
($C_{\rm sym} =\frac{1}{n_1 !n_2 !}$ for $n_1$ particles $M_1$ and $n_2$
particles $M_2$ in the final state). As it stands, expression (187) holds
provided that the initial and final particle propagators are normalized in the
usual fashion ($\sim \frac{1}{s-M_i^2}$). Otherwise, (187) is multiplied by the
normalization factors (the residues of the propagators).

\subsection{Scattering processes}

Finally we are prepared for a discussion of scattering. Let us focus for the
moment on the lowest order graph of Fig. 14 for the four-point function (123).
It consists of
four external exact propagators $\Pi (p_i)$ and a simple vertex as the lowest
order transition matrix element. The $\Pi (p_i)$ contain two stable-particle
mass poles, $\mu$ and $M_2$, therefore this graph describes $\mu$ and $M_2$
scattering (this remains true for higher order contributions; as a consequence,
the same transition matrix elements contribute to $\mu$ and $M_2$ scattering
processes, and they may only differ by some kinematical and normalization
factors).

Let us consider elastic scattering of two Schwinger bosons for definiteness.
Then each external $\Pi (p_i)$ propagator is odd and contributes to the graph
like ($s_i =-p_i^2$) 
\beq
\Pi_{jk}(s_i =\mu^2)P_k =
\frac{4\pi (g_\theta +g^*_\theta)}{s_i -\mu^2}{1
\choose -1}_j
\eeq
Here we face the problem that the first graph of Fig. 14 is already of fifth 
order, because each propagator $\Pi (s_i)$ has an external vertex. We just
omit these external vertices (i.e. we omit the factor $(g_\theta +g^*_\theta)$
in (188) for each propagator), because we want to discuss first order 
scattering. Doing so, we find for this graph 
\beq
P_{j_1}P_{j_2}P_{j_3}P_{j_4}
\delta_{j_1 j_2 k_1}{\cal G}_{k_1 k_2}\delta_{j_3 j_4 k_2}
\prod_{i=1}^4 \frac{4\pi}{s_i -\mu^2}
= (g_\theta +g^*_\theta )\prod_{i=1}^4 \frac{4\pi}{s_i -\mu^2}
\eeq
i.e. each $\mu$ propagator has a residue $4\pi$. In order to obtain the
transition matrix element one has to amputate the external boson propagators in
the usual LSZ fashion. When the propagators are normalized by $r_1 =4\pi$, the
bosons themselves are normalized by $\sqrt{4\pi}$, which has to be divided out
for each amputation. This leaves a factor $\sqrt{4\pi}$ for each external boson
in the transition matrix element. However, the squared transition matrix 
element
enters the scattering cross section, therefore the net effect on the cross
section is a multiplication by the corresponding propagator residue $r_i$ for
each external line.

Therefore, we find for the lowest order boson-boson elastic scattering
\beq
\sigma_{\mu +\mu \ra\mu +\mu}(s)=r_1^4 \frac{\frac{1}{2}(m\Sigma\cos\theta)^2
}{2w^2 (s,\mu^2 ,\mu^2 )} 
\eeq
where we have for the propagator residues (they may be inferred from (160),
$r_i =\frac{1}{c_i (g_\theta +g^*_\theta)}$)
\beq
 r_1 =4\pi \quad ,\quad r_2 =\frac{8\pi^4 m\Sigma\cos\theta}{\mu^2}.
\eeq
(190), of course, coincides with a naive computation using the first order
bosonic four-point function $\langle \Phi (x_1)\ldots \Phi (x_4)\rangle_m^c $
(the latter may be inferred immediately from (48)). Observe that (190) 
is singular
at the real particle production threshold $s=4\mu^2$ ($w(4\mu^2 ,\mu^2
,\mu^2)=0$).

In a next step we want to consider the second order contribution of Fig. 14 
(the
third type graphs). There are three graphs of this type, namely $s$, $t$ and
$u$ channel, but we will consider only the $s$ channel (annihilation channel)
for the moment. In this diagram the lowest order graph must be subtracted in
order to avoid overcounting (see Fig. 14), therefore the graph of Fig. 19

$$\psannotate{\psboxscaled{500}{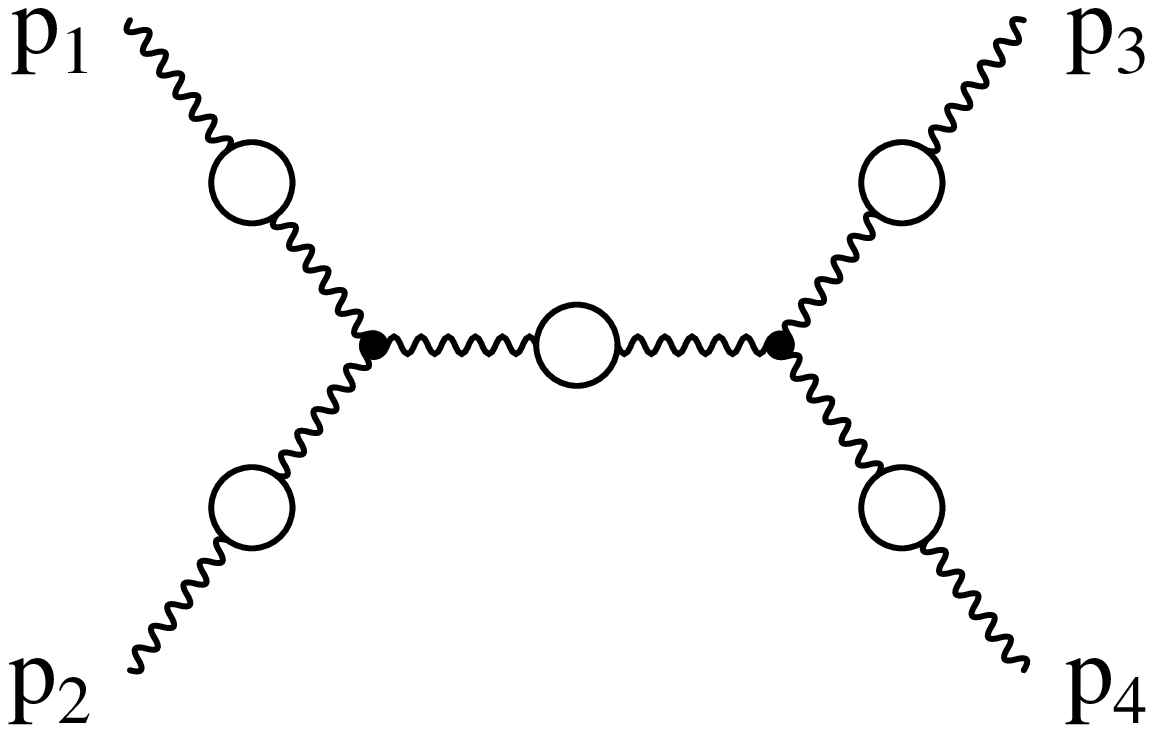}}{\fillinggrid
\at(5.7\pscm;0\pscm){Fig. 19}}$$ 

\vspace{0.1cm}

contains the lowest order graph and the second order $s$ channel contribution.

Actually we will allow for arbitrary final states in the sequel, $\mu +\mu \ra
f$, because this enables us to use the optical theorem, which may be written
for the current problem like
\beq
\sigma^{\rm tot}_{ab\ra f}(s)=\frac{r_a r_b}{w(s,M_a^2 ,M_b^2)}{\rm Im}{\cal
M}_{ab\ra ab}(s)
\eeq
where ${\cal M}_{ab\ra ab}$ is
the forward elastic scattering amplitude. $w(s,M_a^2 ,M_b^2)$ is an initial
state velocity factor; the final state factors must be produced by ${\cal
M}_{ab\ra ab}$, as we will find in the sequel. 

Specifically we choose $a=b=\mu$, and, therefore, both vertices of ${\cal M}$
are contracted by scalars $S$ (we use matrix notation)
\beq
{\cal M}_{2\mu \ra 2\mu}(s)=S^T {\cal G}\Pi (s)S
\eeq
Before starting the computations, we want to make some comments. First, as is
obvious from Fig. 19 and our discussion, in (192) all combinations of $n_1 \mu$
and $n_2 M_2$ are allowed as final states. Consequently, they must exist as
intermediate states in ${\cal M}_{2\mu \ra 2\mu}$, too, in order to saturate
the optical theorem (192). 
Therefore, we are forced to include the $M_2$ particle
into the two-point function $\Pi (p)$, as we did in the previous section, in
order to maintain unitarity.

Secondly, in finite order perturbation theory the optical theorem relates
graphs of different order. However, we use a resummed perturbation series in
(192) and, therefore, we will find a relation that holds for the whole, 
resummed two-point function $\Pi (s)$.

In a first step we want to discuss the special case $\theta =0$, because it is
much easier and shows the relevant features without technical complications.
For $\theta =0$ the amplitude (193) reads (see (132))
\beqa
{\cal M}^{\theta =0}_{2\mu\ra 2\mu}(s) &=&
\frac{m\Sigma}{1-\frac{m\Sigma}{2}(\wt E_+ (s) +\wt E_- (s))} \no \\
&=& \frac{m\Sigma}{1-m\Sigma (d_2 (s) +d_{2,0}(s) +d_4 (s)+\ldots )}
\eeqa
where we inserted the lowest order (125) 
and expanded the exponentials $\wt E_\pm
(s)$ like in (126). Again, we include the $M_2$ particle (which is found by a
further resummation) into $\wt E_\pm$, because this is absolutely necessary, as
we have just argued. Actually $d_{2,0}$ describes the $M_2$-$M_2$ blob, 
and in (194)
only parity even contributions may occur. For the optical theorem (192) we need
the imaginary part
\beq
{\rm Im}{\cal M}^{\theta =0}_{2\mu\ra 2\mu}(s)=\frac{m^2 \Sigma^2 ({\rm Im}d_2
(s) +{\rm Im}d_{2,0}(s)+{\rm Im}d_4 (s)+\ldots )}{[1-m\Sigma ({\rm Re}d_2 (s)+
\ldots )]^2 +m^2 \Sigma^2 ({\rm Im}d_2 (s)+\ldots )^2}
\eeq
We find the following physical picture: at $s=4\mu^2$ the elastic scattering
threshold ($f=2\mu$) opens, at $s=4M_2^2$ the $2\mu \ra 2M_2$ threshold is
added, at $s=16\mu^2$ the $2\mu\ra 4\mu$ threshold, etc. The $d_n (s)$ were
defined as $d_n (s)=\frac{r_1^n}{n!}\wt{D^n_\mu}(s)$, therefore their imaginary
parts are precisely the final state factors for the corresponding cross
section, including the phase space integration (the cutting of the
$\wt{D_\mu^n}(s)$), the propagator normalizations $r_1 =4\pi$, and the final
state symmetry factors for $n$ identical particles, $C_{\rm sym}=\frac{1}{n!}$.
For the multi-$M_2$ propagators $d_{m,0}(s)$ (and, more generally, for
$d_{m,n}(s)$) the first two points (propagators with their residues) are
obvious, the third one (correct $\frac{1}{m!}$ final state symmetry factor) may
be checked by a closer inspection of the mass perturbation series. We show it
for the lowest order contribution to the $M_2$-$M_2$ propagator $d_{2,0}(s)$,
where we depict in Fig. 20 this lowest order contribution and the perturbation
expansion graph where it stems from.

$$\psannotate{\psboxscaled{700}{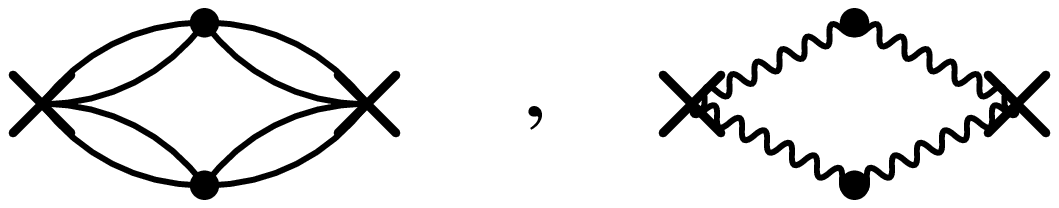}}{\fillinggrid
\at(4.9\pscm;-0.5\pscm){Fig. 20}}$$ 

\vspace{0.2cm}

The second graph in Fig. 20 is a second order mass perturbation, therefore it
contains a factor $\frac{m^2}{2!}$. Further there exists precisely one diagram
of this kind in the perturbation series, therefore the $\frac{1}{2!}$ factor
remains in $d_{2,0}(s)$ as the required final space symmetry factor. Via some
combinatorics this argument may be generalized to higher order contributions to
the $M_2$-$M_2$ loop $d_{2,0}(s)$ and to higher multi-$M_2$ loops.

For the total cross section (192) we get
\beq
\sigma^{{\rm tot},\theta =0}_{2\mu\ra f}(s)=\frac{r_1^2 m^2 \Sigma^2 ({\rm
Im}d_2 (s) +{\rm Im}d_{2,0}(s) +{\rm Im}d_4 (s) +\ldots )}{w(s,\mu^2 ,\mu^2)
\Bigl( [1-m\Sigma ({\rm Re}d_2 (s) +\ldots )]^2 +m^2 \Sigma^2 ({\rm Im}d_2 (s)
+\ldots )^2 \Bigr) }
\eeq
\beq
{\rm Im}d_2 (s)=\frac{r_1^2}{2!}\frac{1}{2w(s,\mu^2 ,\mu^2)} \qquad \mbox{etc.}
\eeq
which we want to evaluate for some specific values of $s$. At the elastic
scattering threshold $s=4\mu^2$, ${\rm Im}d_2 (s)$ is singular and we find
\beq
\sigma^{{\rm tot},\theta=0}_{2\mu\ra f}(4\mu^2)=4.
\eeq
Therefore, the singular behaviour of the lowest order cross section at
$s=4\mu^2$ is cancelled by higher order contributions. This behaviour is,
however, further changed by the $t$ and $u$ channel contributions.

In an intermediate range, far from all thresholds and bound state masses,
$4\mu^2 <s<4M_2^2$, $\sigma^{\rm tot}$ is well described by the lowest order
result (190), because there $m\Sigma d_n (s)$ is small compared to 1,
\beq
\sigma^{{\rm tot},\theta =0}_{2\mu\ra f}(s)\simeq \frac{r_1^2 m^2 \Sigma^2 {\rm
Im}d_2 (s)}{w(s,\mu^2 ,\mu^2)} =\frac{\frac{1}{2!}r_1^4 m^2 \Sigma^2}{2w^2
(s,\mu^2 ,\mu^2)}.
\eeq
At the first bound-state mass, $s=M_{2,0}^2 < 4M_2^2$, a resonance occurs.
There the real part contribution to the denominator of (196) vanishes by
definition and we find
\beq
\sigma^{{\rm tot},\theta=0}_{2\mu\ra f}(M_{2,0}^2)=\frac{r_1^2 m^2 \Sigma^2
{\rm Im}d_2 (M_{2,0}^2)}{w(M_{2,0}^2 ,\mu^2 ,\mu^2)m^2 \Sigma^2 ({\rm Im}d_2
(M_{2,0}^2))^2}=4
\eeq
and the resonance height does not depend on the coupling constant (of course,
the width does).

At the $2M_2$ production threshold $s=4M_2^2$ the scattering cross section goes
down to zero (here $d_{2,0}$ is singular)
\beq
\sigma^{{\rm tot},\theta =0}_{2\mu\ra f}(4M_2^2)\simeq \frac{r_1^2 m^2
\Sigma^2}{w(4M_2^2 ,\mu^2 ,\mu^2)}\frac{{\rm Im}d_{2,0}(4M_{2}^2)}{m^2
\Sigma^2 ({\rm Im}d_{2,0}(4M_2^2))^2}=0.
\eeq
In addition, at this point the $2\mu\ra 2M_2$ production channel opens. At the
four-boson bound-state mass $s=M_4^2$ we find the next resonance
\beq
\sigma^{{\rm tot},\theta =0}_{2\mu\ra f}(M_4^2)=\frac{r_1^2 ({\rm Im}d_2
(M_4^2) +{\rm Im}d_{2,0}(M_4^2))}{w(M_4^2 ,\mu^2 ,\mu^2)({\rm Im}d_2
(M_4^2)+{\rm Im}d_{2,0}(M_4^2))^2}
\eeq
Again, the resonance height does not depend on the coupling constant, and, in
addition, here already two decay channels are open for the $M_4$ resonance.

At the $2\mu\ra 4\mu$ real production threshold $s=16\mu^2$, $\sigma^{\rm tot}$
again vanishes, and for even higher $s$ the above pattern repeats.

Observe that, because $\sigma^{\rm tot}$ has a local maximum (resonance) at
the bound-state masses, whereas it is zero at the real particle production
thresholds, the resonance widths (decay widths) {\em must} be bounded by the
binding energies. For the $M_{1,1}$ and $M_3$ decay widths this may be seen
from the explicit results (177), (178) and (181).

The $t$ and $u$ channel contributions do not change this pattern (they have no
imaginary parts and are small for all $t$, $u$). 

Next let us turn to the $\theta\ne 0$ case. There parity forbidden transitions
are possible, and therefore we will find $M_{1,1}$ and $M_3$ resonances, too.
The forward scattering amplitude (193) reads
\bdi
{\cal M}_{2\mu\ra 2\mu}(s)= \frac{g_\theta +g^*_\theta -2g_\theta g^*_\theta
(\wt E_+ (s) -\wt E_- (s))}{1-(g_\theta +g^*_\theta )\wt E_+ (s) +g_\theta
g^*_\theta (\wt E_+^2 (s) +\wt E_-^2 (s))} = 
\edi
\beq
\frac{g_\theta +g^*_\theta -4g_\theta g^*_\theta (d_1 (s)+d_{1,1}(s)+d_3
(s) +\ldots )}{1-(g_\theta +g^*_\theta )(d_1 (s)+d_2 (s)+d_{1,1}(s)+\ldots )
+4g_\theta g^*_\theta [d_1 (s)(d_2 (s)+d_{2,0}(s)+\ldots )+\ldots ]}
\eeq
Please observe the presence of only odd $d_i$ in the numerator and of only
odd$\times$even $d_i \times d_j$ in the second term of the denominator.
Therefore, the $4g_\theta g^*_\theta$ terms in the numerator and denominator do
not contribute to parity allowed transitions, and the discussion
of such parity allowed transitions is analogous
to the $\theta =0$ case that we discussed above.

Again, we want to discuss the scattering cross section
\beq
\sigma^{\rm tot}_{2\mu\ra f}(s)=\frac{r_1^2}{w(s,\mu^2 ,\mu^2)}{\rm Im}{\cal
M}_{2\mu\ra 2\mu}(s)
\eeq
for some specific values of $s$. At $s=4\mu^2$ we find again
\beq
\sigma^{\rm tot}_{2\mu\ra f}(4\mu^2)=\frac{r_1^2}{w(4\mu^2 ,\mu^2
,\mu^2)}\frac{(g_\theta +g^*_\theta)^2 {\rm Im}d_2 (4\mu^2)}{1+(g_\theta
+g^*_\theta )^2 ({\rm Im}d_2 (4\mu^2))^2}=4
\eeq
At the first parity forbidden resonance $s=M_{1,1}^2$ we find (${\rm Re}d_{1,1}
=\frac{1}{g_\theta +g^*_\theta}$)
\beqa
\sigma^{\rm tot}_{2\mu\ra f}(M_{1,1}^2) &\simeq & \frac{r_1^2}{w(M_{1,1}^2
,\mu^2 ,\mu^2)}{\rm Im}\frac{g_\theta +g^*_\theta -4g_\theta g^*_\theta {\rm
Re}d_{1,1} (M_{1,1}^2)}{-i(g_\theta +g^*_\theta ){\rm Im}d_2 (M_{1,1}^2)
+4ig_\theta g^*_\theta {\rm Re}d_{1,1}(M_{1,1}^2) {\rm Im}d_2 (M_{1,1}^2)} \no
\\
&=& \frac{r_1^2}{w(M_{1,1}^2 ,\mu^2 ,\mu^2)}\frac{\Bigl( g_\theta +g^*_\theta
-\frac{4g_\theta g^*_\theta}{g_\theta +g^*_\theta}\Bigr)^2 {\rm Im}d_2
(M_{1,1}^2)}{\Bigl( g_\theta +g^*_\theta -\frac{4g_\theta g^*_\theta}{g_\theta
+g^*_\theta}\Bigr)^2 ({\rm Im}d_2 (M_{1,1}^2))^2} \no \\
&=&4
\eeqa
and, therefore, the same resonance height as for the first parity allowed
resonance in the $\theta =0$ case (200).

At the parity forbidden threshold $s=(M_2 +\mu)^2$, where ${\rm Im}d_{1,1}((M_2
+\mu)^2)$ is singular, we find
\bdi
\sigma^{\rm tot}_{2\mu\ra f}((M_2 +\mu)^2) \simeq \frac{r_1^2}{w((M_2 +\mu)^2
,\mu^2 ,\mu^2)}\cdot
\edi
\bdi
\cdot {\rm Im}\frac{g_\theta +g^*_\theta -4ig_\theta g^*_\theta {\rm
Im}d_{1,1}((M_2 +\mu)^2)}{1-i(g_\theta +g^*_\theta )({\rm Im}d_2 
+{\rm Im}d_{1,1}) - 4g_\theta g^*_\theta {\rm Im}d_2 
 {\rm Im}d_{1,1}} =
\edi
\bdi
\frac{r_1^2}{w((M_2 +\mu)^2 ,\mu^2 ,\mu^2)}\frac{(g_\theta +g^*_\theta )^2
({\rm Im}d_2 +{\rm Im}d_{1,1})-4g_\theta  g^*_\theta {\rm Im}d_{1,1}
(1-4g_\theta g^*_\theta {\rm Im}d_2 {\rm Im}d_{1,1})}{(1-4g_\theta g^*_\theta
{\rm Im}d_2 {\rm Im}d_{1,1})^2 +(g_\theta +g^*_\theta )^2 ({\rm Im}d_2 +{\rm
Im}d_{1,1})^2} 
\edi
\beq
\ra \quad  \frac{r_1^2}{w((M_2 +\mu)^2 ,\mu^2 ,\mu^2)}\frac{4g_\theta g^*_\theta
({\rm Im}d_{1,1})^2 {\rm Im}d_2 }{(g_\theta +g^*_\theta )^2 ({\rm Im}d_{1,1})^2
} \quad
\ra \quad \Bigl( \frac{4g_\theta g^*_\theta }{g_\theta +g^*_\theta }\Bigr)^2
\frac{r_1^2 {\rm Im}d_2 ((M_2 +\mu)^2)}{w((M_2 +\mu)^2 ,\mu^2 ,\mu^2)}
\eeq
where we performed the limit ${\rm Im}d_{1,1}\ra\infty$ and kept only the
lowest order contribution in $g_\theta$. Therefore, in contrast to the parity
allowed case, the parity forbidden thresholds do not give zero in $\sigma^{\rm
tot}$.

The reason for this behaviour may be easily understood. In the limit of
$\theta\ra 0$ there should not remain any effect of resonances or thresholds in
$\sigma^{\rm tot}$ for parity forbidden transitions, and $\sigma^{\rm tot}$
should be described by the lowest order result (190). 

Precisely this happens: Although the resonance height at $M_{1,1}^2$ remains
unchanged for $\theta\ra 0$, (206), its width tends to zero, (181). 
This means that
the resonance $M_{1,1}$ still exists but is stable against  $M_{1,1}\ra 2\mu$
decay for $\theta\ra 0$. Actually the $M_{1,1}$ bound state is a stable
particle at all for $\theta =0$. Further, at threshold $s=(M_2 +\mu)^2$,
$\sigma^{\rm tot}$ tends to the first order result (190) for $\theta\ra 0$,
\beq
\lim_{\theta\to 0} \Bigl(\frac{4g_\theta g^*_\theta }{g_\theta +g^*_\theta
}\Bigr)^2 =m^2 \Sigma^2 +o(m^3)
\eeq 
as it should hold.

For even higher $s$, when both parity allowed and parity forbidden final states
are possible, we again have the problem that the relative sign of the parity
forbidden process is "wrong" due to our conventions (see the remark after equ.
(169)). E.g. at the $M_3$ resonance we find from (203)
\beqa
\sigma^{\rm tot}_{2\mu\ra f}(s=M_3^2) &\simeq & \frac{r_1^2}{w(M_3^2 ,\mu^2
,\mu^2)}\cdot \no \\ 
&&{\rm Im}\frac{g_\theta +g^*_\theta -4g_\theta g^*_\theta {\rm Re}d_3
(M_3^2)}{-i (g_\theta +g^*_\theta )({\rm Im}d_2 (M_3^2)+{\rm Im}d_{1,1}(M_3^2))
+ 4ig_\theta g^*_\theta {\rm Re}d_3 (M_3^2){\rm Im}d_2 (M_3^2)} \no \\
&=& \frac{r_1^2}{w(M_3^2 ,\mu^2 ,\mu^2)}\frac{g_\theta +g^*_\theta
-\frac{4g_\theta g^*_\theta}{g_\theta +g^*_\theta}}{(g_\theta +g^*_\theta
-\frac{4g_\theta g^*_\theta}{g_\theta +g^*_\theta}){\rm Im}d_2 (M_3^2) +
(g_\theta +g^*_\theta ){\rm Im}d_{1,1}(M_3^2)} \no \\
&\simeq & \frac{r_1^2}{w(M_3^2 ,\mu^2 ,\mu^2)}\frac{m\Sigma (\cos\theta
-\frac{1}{\cos\theta})}{m\Sigma (\cos\theta -\frac{1}{\cos\theta}){\rm Im}d_2
(M_3^2) +m\Sigma\cos\theta {\rm Im}d_{1,1}(M_3^2)} \no \\
&\ra & \frac{r_1^2}{w(M_3^2 ,\mu^2
,\mu^2)}\frac{\frac{1}{\cos\theta}-\cos\theta}{(\frac{1}{\cos\theta}-\cos\theta){\rm
Im}d_2 (M_3^2) +\cos\theta {\rm Im}d_{1,1}(M_3^2)} \no \\
&=& \frac{r_1^2}{w(M_3^2 ,\mu^2 ,\mu^2 )}\frac{\sin^2 \theta (\sin^2 \theta
{\rm Im}d_2 (M_3^2) +\cos^2 \theta {\rm Im}d_{1,1}(M_3^2))}{
[\sin^2 \theta
{\rm Im}d_2 (M_3^2) +\cos^2 \theta {\rm Im}d_{1,1}(M_3^2)]^2}
\eeqa
where the last two lines are for real $\theta$ (the last line may be easily
checked by a low order reasoning, in case somebody does not trust our imaginary
$\theta$ convention). 

Again, the resonance height (containing two partial decay channels) does not
depend on the coupling constant.

For even higher $s$ the above pattern repeats.

The last thing to be discussed is the contribution of the $t$ and $u$ channel
diagrams. There the lowest order diagram must be subtracted (see Fig. 14),
\beq
{\cal M'}_{2\mu\ra 2\mu}(t)=S^T{\cal G}(\Pi (t) -{\bf 1})S
\eeq
where $t=4\mu^2 -s\le 0$, $u\equiv 0$.

It is a wellknown fact that the $t$ and $u$ channel amplitudes in the case at
hand have no singularities on the physical sheet of the complex $s$ plane, and,
therefore, no imaginary parts (see e.g. \cite{Eden}). They are, themselves,
imaginary parts of some higher order graphs (in fact, of the non-factorizable
four-point function of Fig. 14). As a consequence, the ${\cal M'}(t)$, ${\cal
M'}(u=0)$ contributions are small for all $t$ and cannot change the
above-discussed behaviour. The only point where ${\cal M'}(t)$, ${\cal M'}(u)$
cause a qualitative change is the elastic threshold $s=4\mu^2$, $t=0$. There
the lowest order singular behaviour, equ. (190), that was cancelled by the
$s$-channel contribution, equ. (205), is retained, but with a different
coefficient. We find indeed
\beqa
\sigma_{2\mu\ra 2\mu}(s\sim 4\mu^2) &=& \frac{r_1^4}{w^2 (s\sim 4\mu^2 ,\mu^2
,\mu^2)}|{\cal M}_{2\mu\ra 2\mu}(s\sim 4\mu^2) +2{\cal M'}_{2\mu\ra 2\mu}(0)|^2
\no \\
&\simeq & \frac{r_1^4}{w^2 (s\sim 4\mu^2 ,\mu^2
,\mu^2)}|2{\cal M'}_{2\mu\ra 2\mu}(0)|^2 \no \\
&\simeq & \frac{r_1^4}{w^2 (s\sim 4\mu^2 ,\mu^2
,\mu^2)} 4\Bigl(\frac{2g_\theta g^*_\theta \wt E_- (0) +(g_\theta^2
+g^{*2}_\theta )\wt E_+ (0)}{1-(g_\theta +g^*_\theta )\wt E_+ (0)}\Bigr)^2 \\
&\ra &\infty \qquad \mbox{for} \qquad s\ra 4\mu^2 \no
\eeqa
where ${\cal M}_{2\mu\ra 2\mu} (4\mu^2)$ is given by (193) 
(including the lowest
order), and ${\cal M'}_{2\mu\ra 2\mu}(0)$ is given by (210) (without lowest
order). The $\wt E_\pm (0)$ are just the $E_\pm$ of Section 5 (see (64),
(67)).

Further, whenever the $s$-channel cross section vanishes (at parity allowed
higher production thresholds), its value is changed from zero to a small
nonzero number (of order $(m\Sigma)^4)$).

All the other features of the $s$-channel scattering cross section remain
unchanged.

\section{Summary}

It is our hope that the discussion of the previous sections has convinced
the reader that the massive Schwinger model exhibits quite a rich
quantum-field theoretic structure and, therefore, is worth studying.

The mass perturbation theory -- which was the basic ingredient of our
approach -- is similar to the derivation of a low-energy effective
theory in realistic models like QCD. In our approach only bosonic states
(fermion-antifermion bound states) are present as physical states. The
(dimensionfull) coupling constant $e$ plays a role similar to
$\Lambda_{\rm QCD}$ within our model. In fact, because of the small number
of degrees of freedom ($N_f =1$) this model mimicks, up to a certain
extent, the $\eta'$ dynamics of QCD.

A further advantage of the mass perturbation expansion is the fact that --
in contrast to ordinary perturbation theory -- nontrivial phenomena, 
like instanton vacuum and fermion condensates, may be discussed quite
straight forwardly (actually, all the  computations throughout the article 
are for general vacuum angle $\theta$).
 
A first major point in the discussion of the model was the computation of
the vacuum functional and vacuum energy density. As a consequence, the mass 
perturbation theory could be shown to be IR-finite. 
%Concering the UV
%behaviour, the super-renormalizability of the model continues to hold
%within mass perturbation theory, although this feature is somehow hidden
%(\cite{DIV}).

After deriving the matrix-valued Feynman rules of the mass perturbation theory,
we could actually give quite a complete description of the physical 
properties of the model. We were able to compute the condensates and
susceptibilities and to give an exact description of the confinement
behaviour.

Further we succeeded in performing a resummation of the perturbation series
 for the $n$-point functions
with the help of the Dyson-Schwinger equations. These resummed $n$-point 
functions turned out to be very useful for a further discussion of the physical
properties of the model. We could infer the whole spectrum of (stable)
particles and (unstable) bound states from the two-point function. 
Further we were able to identify all the partial decay channels of all
bound states. As an illustration, we computed some bound-state masses and
decay widths. 

At last we discussed scattering processes and found that all the unstable 
bound states turn into resonances of the scattering cross section, as has
to be expected. Further, with the help of unitarity we were able to identify 
all the possible final states that may exist for the scattering of a given 
initial state.

As a result, the following physical picture emerged: there exist two
stable particles in the theory, namely the Schwinger boson $\mu$ and the
two-boson bound state $M_2$. Higher (unstable) bound states may be formed 
out of an arbitrary number of $\mu$ and $M_2$. Further, these unstable 
bound states may decay into all combinations of $\mu$ and $M_2$ that are
possible kinematically. (For the special case $\theta =0$ the lowest mixed
bound state, composed of one $\mu$ and one $M_2$, is stable, too, because of
parity conservation, and is therefore present both in final and
intermediate states, analogous to our discussion of the $M_2$ particle; 
however, we treated the generic $\theta \ne 0$ case throughout the 
article and, therefore, did not discuss this straight-forward generalization
in detail.)  

 For scattering processes we found that far from all resonances and particle 
production thresholds the scattering cross section is well described by
a lowest order computation. 
Whenever the squared momentum $s$ is near a bound-state mass, 
the scattering cross section has a local maximum, i.e. 
a resonance occurs. Moreover, for all values of $s$ where a new final state 
becomes possible kinematically, the corresponding real particle production 
threshold indeed occurs.

We want to emphasize that all these features result from our resummed 
mass perturbation theory, and that we did not have to impose further
assumptions or use further approximations in order to find this physical 
structure.

There are (at least) two directions of further study within this field that we 
believe to be important. On one hand, an increase of the number
of fermions ($N_f >1$) further enriches the complexity of the model and
makes it possible to mimick the light field dynamics (pions) of QCD, too.
There exists some work on the massive multi-flavour Schwinger model that
uses bosonization and/or semiclassical methods 
(\cite{Co1,Gatt2,Sm1,Para1,Hoso1,SaWe1}). It would be very interesting 
to discuss this model analogously to our discussion in the previous
sections; however, a direct application of the mass perturbation
expansion is not possible there, because it fails to be IR-convergent
\cite{Gatt2}. Therefore, more elaborate methods should be developed
for a further study.

On the other hand, the properties we found hold for sufficiently small
fermion mass $m$ (strong coupling $e$). Of course, it would be interesting
to understand whether and how the physical structure of the model
changes when the fermion mass is increased. Perhaps a discussion similar
to this article for the massive Schwinger model within ordinary (electrical
charge) perturbation theory and a comparison of the two approaches could be a 
first step towards getting more insight into this question.

In any case, QED$_2$ remains a fascinating subject of study that will offer
further insight into general concepts of quantum field theory as well as
into some deep problems of QCD.

\section*{Acknowledgement}

The author thanks R. Jackiw for the opportunity to join the Center of 
Theoretical Physics at MIT, where this work was finished, and the
CTP members for their hospitality.
Further, this work has greatly benefitted from discussions with various
scientists. The author owes special thanks to
C. Gattringer, H. Grosse, M. Hutter, R. Jackiw, 
H. Leutwyler, J. Pawlowski, A. V. Smilga and A. Wipf.

This work is supported by a Schr\"odinger stipendium of the Austrian FWF.


\begin{thebibliography}{999999}
\bibitem{Sc1}
J. Schwinger, {\em Phys. Rev.} {\bf 128} (1962), 2425.
\bibitem{LS1}
J. Lowenstein, J. Swieca, {\em Ann. Phys.} {\bf 68} (1971), 172.
\bibitem{Jay}
C. Jayewardena, {\em Helv. Phys. Acta} {\bf 61} (1988), 636.
\bibitem{SW1}
I. Sachs, A. Wipf, {\em Helv. Phys. Acta} {\bf 65} (1992), 653.
\bibitem{SW2}
I. Sachs, A. Wipf, 
{\em Ann. Phys.}  {\bf 249} (1996), 380. 
\bibitem{DSW}
A. Dettki, I. Sachs, A. Wipf, preprint ETH-TH/93-14.
\bibitem{DR1}
W. Dittrich, M. Reuter, "Selected topics \ldots ", {\em Lecture Notes in
Physics}  {\bf 244}, Springer Verlag, Berlin 1986.
\bibitem{Frish}
Y. Frishman, {\em Lecture Notes in Physics} {\bf 32}, Springer Verlag,
Berlin 1975.
\bibitem{Gro}
H. Grosse, "Models in statistical physics and quantum field theory",
Springer Verlag, Berlin 1988. 
\bibitem{GS1}
R. E. Gamboa Saravi, M. A. Muschietti, F. A. Schaposnik, J. E. Solomin,
{\em Ann. Phys.} {\bf 157} (1984), 360.
\bibitem{AAR}
E. Abdalla, M. Abdalla, K. D. Rothe, "2 dimensional Quantum Field Theory",
World Scientific, Singapore 1991.
\bibitem{Jac1}
R. Jackiw, "Topological investigations \ldots ", in: Treiman et al,
"Current algebras and anomalies", World Scientific, Singapore 1985.
\bibitem{Bert1}
R. A. Bertlmann, "Anomalies in quantum field theory", Clarendon Press,
Oxford 1996.
\bibitem{IP1}
N. P. Ilieva, V. N. Pervushin, {\em Sov. J. Part. Nucl.} {\bf 22} (1991), 275.
\bibitem{CKS}
A. Casher, J. Kogut, P. Susskind, {\em Phys. Rev.} {\bf D10} (1974), 732.
\bibitem{KS1}
J. Kogut, P. Susskind, {\em Phys. Rev.} {\bf D11} (1975), 3594.
\bibitem{ABH}
C. Adam, R. A. Bertlmann, P. Hofer, {\em Riv. Nuovo Cim.} {\bf 16} (1993), 1.
\bibitem{Adam}
C. Adam, {\em Z. Phys.} {\bf C63} (1994), 169.
\bibitem{Diss}
C. Adam, thesis, Universit\"at Wien 1993.
\bibitem{DSEQ}
C. Adam, {\em Czech. J. Phys.} {\bf 46} (1996), 893,
 HEP-PH 9501273.
\bibitem{Sm1}
A. V. Smilga, {\em Phys. Lett.} {\bf B278} (1992), 371.
\bibitem{Sm3}
A. V. Smilga, {\em Phys. Rev.} {\bf D46} (1992), 5598.
\bibitem{Sm2}
A. V. Smilga, {\em Phys. Rev.} {\bf D49} (1994), 5480.
\bibitem{CJS}
S. Coleman, R. Jackiw, L. Susskind, {\em Ann. Phys.} {\bf 93} (1975), 267.
\bibitem{Co1}
S. Coleman, {\em Ann. Phys.} {\bf 101} (1976), 239.
\bibitem{Fry}
M. P. Fry, {\em Phys. Rev.} {\bf D47} (1993), 2629.
\bibitem{MSSM}
C. Adam, {\em Phys. Lett.} {\bf B 363} (1995), 79, HEP-PH 9507279.
\bibitem{SMASS}
C. Adam,  {\em Phys. Lett.} {\bf B 382} (1996), 383,
HEP-PH 9507331.
\bibitem{Froe1}
J. Fr\"ohlich, {\em Comm. Math. Phys.} {\bf 47} (1976), 233.
\bibitem{FS1}
J. Fr\"ohlich, E. Seiler, {\em Helv. Phys. Acta} {\bf 49} (1976), 889.
\bibitem{Zah}
J. Steele, A. Subramanian, I. Zahed, {\em Nucl. Phys.} {\bf B452} (1995), 545,
HEP-TH 9503220.
\bibitem{Gatt1}
C. Gattringer, E. Seiler, {\em Ann. Phys.} {\bf 233} (1994), 97.
\bibitem{Gatt2}
C. Gattringer, HEP-TH 9503137; {\em Ann. Phys.} {\bf 250} (1996), 389, 
HEP-TH 9602027.
\bibitem{BOUND}
C. Adam, preprint PM 96/01, HEP-PH 9601227.
\bibitem{GBOUND}
C. Adam,  {\em Phys. Lett.} {\bf B 382} (1996), 111,
HEP-TH 9602175.
\bibitem{NS1}
N. K. Nielsen, B. Schroer, {\em Nucl. Phys.} {\bf B120} (1977), 62.
\bibitem{NS2}
N. K. Nielsen, B. Schroer, {\em Nucl. Phys.} {\bf B127} (1977), 493.
\bibitem{RRS1}
H. J. Rothe, K. D. Rothe, J. A. Swieca, {\em Phys. Rev.} {\bf D19} (1979), 
3020.
\bibitem{HRS1}
M. Hortacsu, K. D. Rothe, B. Schroer, {\em Phys. Rev.} {\bf D20} (1979), 3203.
\bibitem{RS1}
K. D. Rothe, B. Schroer, {\em Nucl. Phys.} {\bf B185} (1981), 429; 
{\bf B172} (1980), 383.
\bibitem{Bai}
R. Baier, E. Pilon, {\em Z. Phys.} {\bf C52} (1991), 339.
\bibitem{LSm}
H. Leutwyler, A. V. Smilga, {\em Phys. Rev.} {\bf D46} (1992), 5607.
\bibitem{SV1}
A. V. Smilga, J. J. M. Verbaarschot, {\em Phys. Rev.} {\bf D54} (1996), 1087,
HEP-PH 9511471.
\bibitem{GKMS}
D. J. Gross, I. R. Klebanov, A. V. Matytsin, A. V. Smilga, {\em Nucl. Phys.} 
{\bf B461} (1996), 109, HEP-TH 9511104.
\bibitem{AMZ}
E. Abdalla, R. Mohayaee, A. Zadra, HEP-TH 9604063.
%\bibitem{DIV}
%C. Adam, preprint ?, HEP-TH ?
\bibitem{Vary}
J. P. Vary, T. J. Fields, H. J. Pirner, {\em Phys. Rev.} {\bf D53} (1996), 
7231, HEP-PH 9411263.
\bibitem{Ha1}
K. Harada et al, {\em Phys. Rev.} {\bf D52} (1995), 2492, HEP-TH 9509136.
\bibitem{Ell1}
T. Eller, H. C. Pauli, S. J. Brodsky, {\em Phys. Rev.} {\bf D35} (1987), 1493.
\bibitem{Ell2}
T. Eller, H. C. Pauli, {\em Z. Phys.} {\bf C42} (1989), 59.
\bibitem{Els1}
S. Elser, H. C. Pauli, A. C. Kallionatis, HEP-TH 9505069.
\bibitem{Els2}
S. Elser, A. C. Kallionatis, HEP-TH 9601045.
\bibitem{Crew}
D. P. Crewther, C. J. Hamer, {\em Nucl. Phys.} {\bf B170} (1980), 353.
\bibitem{Ham}
C. J. Hamer et al, {\em Nuc. Phys.} {\bf B208} (1982), 413.
\bibitem{CaKe1}
S. R. Carson, R. D. Kenway, {\em Ann. Phys.} {\bf 166} (1986), 364.
\bibitem{Lang1}
H. Gausterer, C. B. Lang, {\em Phys. Lett.} {\bf B341} (1994), 46.
\bibitem{Lang2}
H. Gausterer, C. B. Lang, {\em Nucl.Phys.} {\bf B455} (1995), 785.
\bibitem{Gatt3}
C. Gattringer, {\em Phys. Rev.} {\bf D53} (1996), 5090. 
\bibitem{PQ1}
R. D. Peccei, H. R. Quinn, {\em Phys. Rev.} {\bf D16} (1977), 1791;
{\em Phys. Rev. Lett.} {\bf 38} (1977), 1440.
\bibitem{SaWe1} 
M. Sadzikowski, P. Wegrzyn, HEP-PH 9605242.
\bibitem{Para1}
M. Paranjape, {\em Phys. Rev.} {\bf D48} (1993), 3892; 4946.
\bibitem{Hoso1}
J. E. Hetrick, Y. Hosotani, S. Ito, HEP-TH 9502113.
\bibitem{DECAY}
C. Adam, {\em Phys. Lett.} {\bf B391} (1997), 395,  HEP-TH 9609154.
\bibitem{THREE}
C. Adam, preprint FSUJ-TPI-16/96,  HEP-TH 9610050.
\bibitem{SCAT}
C. Adam, preprint MIT-CTP-2602 (1997), HEP-TH 9701013. 
\bibitem{Eden}
R. J. Eden, P. V. Landshoff, D. I. Olive, J. C. Polkinghorne,
``The Analytic $S$-Matrix'', Cambridge University Press, Cambridge 1966.


\end{thebibliography}
\end{document}